\documentclass[11pt]{article}
\usepackage[final]{pdfpages}
\usepackage[final]{pdfpages}
\usepackage{amsmath}
\usepackage{slashed}
\usepackage{amssymb}
\catcode`@11
\def\seceqaa{\@addtoreset{equation}{section}
\def\theequation{A\arabic{equation}}}
\def\seceqbb{\@addtoreset{equation}{section}
\def\theequation{B\arabic{equation}}}
\def\seceqcc{\@addtoreset{equation}{section}
\def\theequation{C\arabic{equation}}}
\def\seceqdd{\@addtoreset{equation}{section}
\def\theequation{D\arabic{equation}}}
\def\seceqee{\@addtoreset{equation}{section}
\def\theequation{E\arabic{equation}}}
\def\seceqff{\@addtoreset{equation}{section}
\def\theequation{F\arabic{equation}}}
\def\seceqgg{\@addtoreset{equation}{section}
\def\theequation{G\arabic{equation}}}
\def\seceqhh{\@addtoreset{equation}{section}
\def\theequation{H\arabic{equation}}}
\catcode`@11

\begin{document}
\begin{titlepage}
\begin{center}
{\Large\bf New Insights into Properties of Large-$N$ Holographic Thermal QCD at Finite Gauge Coupling at (the Non-Conformal/Next-to) Leading Order in $N$}
\vskip 0.1in Karunava Sil\footnote{email: krusldph@iitr.ac.in}
and
 Aalok Misra\footnote{e-mail: aalokfph@iitr.ac.in}\\
Department of Physics, Indian Institute of Technology,
Roorkee - 247 667, Uttarakhand, India

 \vskip 0.5 true in
\date{\today}
\end{center}
\thispagestyle{empty}

\begin{abstract}{It is believed that large-$N$ thermal QCD laboratories like strongly coupled QGP (sQGP) require not only a large t'Hooft coupling but also a finite gauge coupling \cite{Natsuume}. Unlike almost all top-down holographic models in the literature, holographic large-$N$ thermal QCD models based on this assumption, therefore necessarily require addressing this limit from M theory. This was initiated in \cite{MQGP} which presented a local M-theory uplift of \cite{metrics}'s string theoretic dual of large-$N$ thermal QCD-like theories at finite gauge/string coupling ($g_s\stackrel{\sim}{<}1$ as part of the `MQGP' limit of \cite{MQGP}). Understanding and classifying the properties of systems like sQGP from a top-down holographic model assuming a finite gauge coupling, has been entirely missing in the literature. In this paper we largely address  the following two non-trivial issues pertaining to the same. First, up to LO in $N$ (the number of $D3$-branes), by calculating the temperature dependence of the thermal (and electrical) conductivity and the consequent deviation from the Wiedemann-Franz law, upon comparison with \cite{WF}, we show that remarkably, the results  qualitatively mimic a 1+1-dimensional Luttinger liquid with impurities. Second, by looking at respectively the scalar, vector and tensor modes of metric perturbations and using  \cite{klebanov quasinormal}'s prescription of constructing appropriate gauge-invariant perturbations, we obtain the non-conformal corrections to the conformal results (but at finite $g_s$) respectively for the speed of sound, the shear mode diffusion constant  and the shear viscosity $\eta$ (and $\frac{\eta}{s}$). The new insight gained is that it turns out  these corrections show a partial universality  in the sense that at NLO in $N$ the same are given by the product of $\frac{(g_s M^2)}{N}\ll1$ and $g_s N_f\sim{\cal O}(1)$, $N_f$ being the number of flavor $D7$-branes and $M$ the number of fractional $D3-$branes = the number of colors = 3 in the IR after the end of a Seiberg duality cascade. On the Math side, using the results of \cite{M.Ionel and M.Min-OO (2008)}, at LO in $N$ we finish our argument of \cite{transport-coefficients} and show that  for a predominantly resolved (resolution $>$ deformation - this paper) or deformed (deformation $>$ resolution - \cite{transport-coefficients}), resolved warped deformed conifold, the local $T^3$ of \cite{MQGP} in the MQGP limit, is the $T^2$-invariant special Lagrangian
three-cycle of \cite{M.Ionel and M.Min-OO (2008)} justifying the construction in \cite{MQGP} of the delocalized Strominger-Yau-Zaslow type IIA mirror of the type IIB background of \cite{metrics}. }
\end{abstract}
\end{titlepage}

\section{Introduction}

The AdS/CFT correspondence or in general the gauge/gravity duality has proved to be a very useful tool in understanding the properties of super Yang-Mills theory at large t'Hooft coupling. According to the correspondence, physics of $\mathcal{N}=4$ $SU(N)$ SYM theory in the large $N$ limit can be obtained from type $IIB$ superstring theory on $AdS_5\times S^5$ geometry, where $AdS_5$ is the five dimensional anti-de Sitter space and $S^5$ is the five sphere. The $\mathcal{N}=4$ $SU(N)$ SYM theory is a conformal field theory which means its gauge coupling does not run with the energy scale. On the other hand  QCD is non-conformal. QCD with $SU(N_c)$ gauge group, where $N_c$ is the number of  quark colors, is an asymptotically free theory so that  the gauge coupling is scale dependent and vanishes logarithmically with large characteristic momentum or with short distance. So to deal  with QCD-like theories using Gauge/Gravity duality we need to generalize the AdS/CFT correspondence and incorporate a  running coupling in the theory.  Building up on the Klebanov-Witten \cite{KW}, Klebanov-Nekrasov \cite{KN}  and  Klebanov-Tseytlin \cite{KT} models,  a logarithmic RG flow just like QCD was obtained in the non-conformal Klebanov-Strassler model \cite{KS} by considering $M$ fractional $D3$ branes along with $N$  $D3$ branes in a conifold geometry  wherein the IR geometry was modified resulting in a deformed conifold.

So far we have not talked about the temperature at all. In fact the AdS/CFT correspondence mentioned above is valid at zero temperature. At finite temperature the situation is different on the gravity side of the correspondence. On the other hand the field theory in question i.e. thermal QCD, is an  IR-confining theory at $T=0$ and becomes non-confining at $T\gg\Lambda_{QCD}$, where $\Lambda_{QCD}$ is the only scale that we have here. It possesses a phase transition from confining phase to a non-confining phase at $T=T_c\approx C \Lambda_{QCD}$, where $C=\mathcal{O}(1)$. At sufficiently high temperature i.e. at $T\gg\Lambda_{QCD} $, the interaction strength $\lambda(T)\ll 1$ and hence the theory is weakly coupled. However the thermal gauge theory we want to understand is not in the weak coupling regime. In particular, to explore the physics of QCD at $T\approx T_c$, we have to take a look at the strongly coupled regime where $\lambda\approx 1$. So we cannot apply  perturbative methods any more. In lattice gauge theory using numerical simulations the equilibrium properties of the strongly coupled hot QCD can be explored. But interesting non-equilibrium properties such as hydrodynamic behavior or the real time dynamics cannot be seen from the equilibrium correlation functions. So the lack of non-perturbative methods to study hot QCD forces us to look for either a different theory/model  or a different limit of a known  theory/model.

At finite temperature the equilibrium or non-equilibrium properties of the Euclidean theory are studied requiring time to have periodicity $\beta\sim\frac{1}{T}$. Thus, at non-zero temperature, the Euclidean space-time looks like a cylinder with the topology $\mathbb{R}^3\times S^1$. The AdS/CFT correspondence tells us that at $T=0$ the $4d$ SYM  theory defined on $\mathbb{R}^4$ is dual to string theory on $5d$ AdS space with $\mathbb{R}^4$ as the boundary of the same. So at zero temperature we can think of the field theory as living on the boundary of AdS space. However, the prime interest is to investigate the finite temperature aspects of the dual field theory from the physics of supergravity. Hence at finite temperature, the space-time of the gravitational description somehow has to be changed such that one gets a geometry of the boundary which is equivalent to $\mathbb{R}^3\times S^1$ and not $\mathbb{R}^4$. In other words one needs to find some bulk geometry which has a boundary with the topology $\mathbb{R}^3\times S^1$. One possible answer is the AdS-BH space-time with the following metric sometimes called black-brane metric given as:
$ds^2=a(r)\Biggl(-g(r)dt^2+d\vec{x}^2\Biggr)+b(r)dr^2$ with Minkowskian signature. Here $r$ is the radial coordinate and $g(r)$, dependent on the horizon radius $r_h$, is a `black-hole function'. By construction, the time coordinate is defined to be periodic with period $\beta$ which is inverse of temperature and is related to the horizon radius $r_h$.

Now, let us go back to the Klebanov-Strasslar model  where the temperature is turned on in the field theory side effected by introducing a black hole in the dual geometry. Interestingly the KS background with a black hole has the geometry equivalent to the AdS-BH spacetime in the large $r$ limit. Moreover the embedding of $D7$-branes in KS model via the holomorphic Ouyang embedding \cite{ouyang} and finally the M-theory uplift of the whole set up keeps the background geometry as required provided we  consider some limiting values of the parameters in the theory. The details about this, based on \cite{metrics}, \cite{ouyang},  will be reviewed in Section {\bf 2}.

{ This paper, apart from providing important evidence validating construction in \cite{MQGP} of delocalized Strominger-Yau-Zaslow (SYZ) mirror of \cite{metrics}'s type IIB holographic dual of large $N$ thermal QCD, we believe, fills in a pair of important gaps in the literature pertaining to a top-down holographic study of large-$N$ thermal QCD}.
\begin{itemize}
\item
First, all such large-$N$ holographic models cater to the large t'Hooft-coupling limit while keeping the gauge coupling vanishingly small. However, in systems such as sQGP, it is believed that not only should the t'Hooft coupling be large, but even { the gauge/string coupling should also be finite} \cite{Natsuume}. A finite gauge coupling would imply a finite string coupling which necessitates addressing the limit from an M-theory point of view. Also, for a realistic thermal QCD computation, the number of colors should be set to three. This can be realized in the IR after the end of a Seiberg duality cascade and in the MQGP limit of (\ref{limits_Dasguptaetal-ii}).  This study was  initiated in \cite{MQGP, transport-coefficients} wherein a large-$N$ limit, referred to as the `MQGP limit' (\ref{limits_Dasguptaetal-ii}), was defined in which the gauge coupling was kept to be slightly less than unity and hence finite. By studying  some transport coefficients  in this paper, we obtain even at the leading order in $N$,  a remarkable result that holographic large-$N$ thermal QCD at finite gauge coupling for $\mu_{\rm Ouyang}\equiv$(Ouyang embedding parameter)$\sim r_h^\alpha,\ \alpha\leq0$ mimics qualitatively $D=1+1$ Luttinger liquid with impurities close to `$\frac{1}{3}$-doping'; for $\alpha=\frac{5}{2}$ one is able to reproduce the expected linear large-$T$ variation of DC electrical conductivity characteristic of most strongly coupled gauge theories with five-dimensional gravity duals with a black hole \cite{SJain_sigma+kappa}.

\item
Second, in the context of top-down holographic models of large-$N$ thermal QCD {\it at finite gauge coupling}, there are no previous results that we are aware of pertaining to evaluation of the non-conformal corrections to hydrodynamical quantities such as the shear viscosity $\eta$ (as well as the shear-viscosity-entropy-density ratio $\frac{\eta}{s}$), shear mode diffusion constant  $D$ and the speed of sound $v_s$.  These {non-conformal corrections at finite gauge coupling, determined for the first time in this paper in the given context}, are particularly relevant in the IR and in fact also encode the scale-dependence of aforementioned physical quantities, and hence are extremely important to be determined for  making direct contact with sQGP. { The main non-trivial insight gained via such computations is the realization that at NLO in $N$ there is a partial universality in these corrections determined by $N_f$ and $M$ apart from $N$.}
\end{itemize}

The following is a section-wise description of the sets of issues addressed and the new insights obtained in this paper.
\begin{itemize}
\item
{\bf Sec. 3 - Identification of sLag in a (predominantly) resolved  conifold up to LO in $N$}: Up to  leading order in $N$ and in the UV-IR interpolating region/UV, using the results of \cite{M.Ionel and M.Min-OO (2008)}, we show that the local $T^3$ of \cite{MQGP} is a $T^2$-invariant special Lagrangian three-cycle  in a  resolved conifold. This, together with the results of \cite{transport-coefficients}, shows that for a (predominantly resolved or deformed) resolved warped deformed conifold, the local $T^3$ of \cite{MQGP} in the MQGP limit of \cite{MQGP}, is the $T^2$-invariant special Lagrangian three-cycle of \cite{M.Ionel and M.Min-OO (2008)}, justifying the construction in \cite{MQGP} of the delocalized SYZ type IIA mirror of the type IIB background of \cite{metrics}. {This was a crucial step missing in \cite{MQGP, transport-coefficients} in construction of the delocalized SYZ mirror of the top-down type IIB holographic dual of large-$N$ thermal QCD of \cite{metrics}}, {\it at finite gauge coupling}.

\item
{\bf Transport Coefficients up to (N)LO in $N$:}
We study some transport coefficients of  large-$N$ thermal QCD leading to evaluation of various transport coefficients up to (next-to) leading order in $N$. This boils down to evaluating various retarded Green's functions, but computed from the gravity dual as prescribed in \cite{son correlator}. In order to study the transport phenomenon from the gravity picture, we need to consider a perturbation of the given modified OKS-BH metric - the type IIB string dual of large-$N$ thermal QCD as given in \cite{metrics}. In response to this perturbation the BH will emit gravitational waves with a long period of damping oscillation. The modes associated with this kind of gravitational radiation are called quasinormal modes. Quasinormal modes are the solutions to the linearized EOMs that one gets by considering fluctuations of gravitational background satisfying specific boundary conditions both at the black hole horizon and at the boundary. At the horizon, the quasinormal modes satisfy a pure incoming-wave boundary condition and at the spatial infinity the perturbative field or some gauge invariant combinations of the fields vanishes, that means it follows the Dirichlet boundary condition at infinity. It was shown in \cite{0302026} that the quasinormal frequency associated with the quasinormal modes defined above in an asymptotically AdS spacetime exactly matches with the pole of the two point correlation function involving operators in the field theory dual to different metric perturbations. Hence evaluating the quasinormal frequency $\omega$ as a function of the special momentum $q$, gives the thermodynamic and hydrodynamic behavior of the plasma.

Analogous to \cite{KS}, the non-conformality in \cite{metrics} is introduced via $M$ number of fractional $D3$-branes, the latter appearing explicitly in $B_2, H_3$ and after construction of a delocalized SYZ type IIA mirror (resulting in mixing of $B_2$ with the metric components after taking a triple T-dual of \cite{metrics}) as well as its local M-theory uplift, also in the metric. {In the context of a (local) M-theory uplift of a top-down holographic thermal QCD dual such as that of \cite{metrics} {\it at finite gauge coupling}, to the best of our knowledge, we estimate for the first time, the non-conformal  corrections appearing at the NLO in $N$ to the speed of sound $v_s$, shear mode diffusion constant  $D$, the shear viscosity $\eta$ and the shear viscosity - entropy density ratio $\frac{\eta}{s}$.} { The main new insight gained by this set of results is  that the non-conformal corrections in all the aforementioned quantities are found to display a partial universality in the sense that at the NLO in $N$ the same are always determined by $\left(\frac{(g_s M^2)(g_s N_f)}{N}\right)$}, $N_f$ being the number of flavor $D7$-branes. Thus, {we see that the same are determined by the product of the very small $\frac{g_sM^2}{N}\ll1$ - part of the MQGP limit (\ref{limits_Dasguptaetal-ii}) - and the finite $g_s N_f\sim {\cal O}(1)$ (also part of (\ref{limits_Dasguptaetal-ii})).}  Of course,  the leading order conformal contributions though at vanishing string coupling and large t'Hooft coupling were (in)directly known in the literature. {It is interesting to see the conformal limit of our results at finite $g_s$ obtained by turning off of $M$ - which encodes the non-conformal contributions -  reduce to the known conformal results for vanishing $g_s$}.

\begin{itemize}

\item
{\bf Sec. 4 - (Thermal and electrical) Conductivity, Wiedemann-Franz law and $D=1+1$ Luttinger liquid at LO in $N$}:
As a thermal gradient corresponding to a gauge field fluctuation also turns on vector modes of metric fluctuations, we consider  turning on simultaneously gauge and vector modes of metric fluctuations, and evaluate the thermal
($\kappa_T$) and electrical ($\sigma$) conductivities, and the Wiedemann-Franz law ($\frac{\kappa_T}{T\sigma}$).  {The new insight gained is that for $\mu_{\rm Ouyang}\equiv$(Ouyang embedding parameter)$\sim r_h^\alpha,\ \alpha\leq0$, the temperature dependence of $\kappa_T, \sigma$ and the consequent deviation from the Wiedemann-Franz law, all point to the remarkable similarity with $D=1+1$ Luttinger liquid with impurities at `$\frac{1}{3}$-doping'}; for $\alpha=\frac{5}{2}$ one is able to reproduce the expected linear large-$T$ variation of DC electrical conductivity for most strongly coupled gauge theories with five-dimensional gravity duals with a black hole \cite{SJain_sigma+kappa}.

\item
{\bf Sec. 5 - Speed of sound}: For the metric fluctuations in the sound channel the corresponding quasinormal frequency is given by $w=\pm v_s q-i\Gamma_s q^2$ with $v_s$ defined as the speed of sound and $\Gamma_s$ as the damping constant of the sound mode. Again for the sound channel the pole of the correlations of longitudinal momentum density gives the same dispersion relation in the conformal limit. From the knowledge of quasinormal modes associated with the scalar  modes of metric perturbations, we have  computed the next-to-leading order correction to the speed of sound ($v_s$) at finite gauge coupling (part of the MQGP limit). Up to LO in $N$, we calculate $v_s$ using four routes:\\
(i) (subsection {\bf 5.1.1}) the poles appearing in the common denominator of the solutions to the individual scalar modes of metric perturbations (the pure gauge solutions and the incoming-wave solutions), \\
(ii) (subsection {\bf 5.1.2}) the poles appearing in the coefficient of the asymptotic value of the time-time component of the scalar metric perturbation in the on-shell surface action,\\
(iii) (subsection {\bf 5.2.1}) the dispersion relation obtained via a Dirichlet boundary condition imposed on an appropriate gauge-invariant combination of perturbations - using the prescription of \cite{klebanov quasinormal} - at the asymptotic boundary, \\
and \\
(iv)  (subsection {\bf 5.2.2}) the pole structure of the retarded Green's function calculated from the  on-shell surface action written out in terms of the same single gauge invariant function.

The third approach (of solving a single second-order differential equation for a single gauge-invariant perturbation using the prescription of \cite{klebanov quasinormal}) is then extended to include the non-conformal corrections to the metric and {obtain for the first time in the context of a top-down large-$N$ holographic thermal QCD at finite gauge coupling uplifted to M theory, an estimate of the non-conformal corrections to $v_s$ up to NLO in $N$}.

{\bf Sec. 6 - Shear mode diffusion constant}: The quasinormal frequency for the vector modes of black brane metric fluctuation reads $\omega=-iD q^2$, where $D$ is the shear mode diffusion constant . This dispersion relation also follows from the pole structure of the correlations of transverse momentum density. From the knowledge of quasinormal modes associated with the vector  modes of metric perturbations obtained by imposing  Dirichlet boundary condition at the asymptotic boundary, on an appropriate gauge-invariant perturbation constructed using the prescription of \cite{klebanov quasinormal}, we have  {computed   for the first time in the context of the same top-down large-$N$ holographic thermal QCD at finite gauge coupling uplifted to M theory, the non-conformal corrections to the shear mode diffusion constant up to NLO in $N$}.

\item
{\bf Sec. 7 - Shear viscosity(-to-entropy density ratio)}: We have also {evaluated for the first time in the context of the aforementioned M-theory uplift corresponding to finite $g_s$, the non-conformal temperature-dependent correction at the NLO in $N$, to the shear viscosity $\eta$ and shear viscosity - entropy density ratio $\frac{\eta}{s}$}  from the two point energy-momentum tensor correlation function corresponding to the tensor mode of metric perturbation.

\end{itemize}

{The results for the NLO (in $N$) corrections are particularly important as they suggest a scale dependance to the above mentioned quantities and hence leads to a non-conformal nature of the field theory in the IR.} We have commented on this issue even  in  section {\bf 8}.
\end{itemize}

The paper is organized as follows. First, we briefly review the supergravity dual background of large $N$ strongly coupled QCD like theories. The whole discussion, for the sake of clear understanding of the reader is presented stepwise through first three subsections in section {\bf 2}. In {\bf 2.1}, the type $IIB$ supergravity background of \cite{metrics} dual to large-$N$ thermal gauge theory which is UV complete and closely resembles thermal QCD,  is briefly reviewed. In {\bf 2.2}, the 'MQGP Limit' of \cite{MQGP} and its motivation, in particular to address the properties of strongly coupled QGP medium, is briefly reviewed. In {\bf 2.3}, using the 'MQGP Limit' we review briefly the delocalized SYZ type $IIA$ mirror via three T dualities along a $T^2$-invariant special lagrangian $T^3$ fibered over a large base in a predominantly warped resolved conifold - this serves as a precursor to the material of Sec. {\bf 3}.  In the same sub-section, we discuss it's local uplift to M-theory, where in the large $r$ limit the spacetime is given by $AdS_5\times M_6$. In {\bf 2.4}, following \cite{son correlator} we review the recipe to calculate two-point correlation function with Minkowskian signature. In {\bf 2.5}, following \cite{klebanov quasinormal} the gauge invariant variables for vector, scalar and tensor modes of background metric perturbations are discussed - this will be useful to obtain the results of (sub-)sections {\bf 5.3, 6} and {\bf 7}. In section {\bf 3}, we show that in the MQGP limit of \cite{MQGP}, the local $T^3$ of \cite{MQGP} is the $T^2$-invariant special Lagrangian three-cycle  of a resolved conifold as given in \cite{M.Ionel and M.Min-OO (2008)}. This together with the result reviewd in {\bf 2.3}, shows that in the  MQGP limit, the local $T^3$ of \cite{MQGP} is the $T^2$-invariant sLag of \cite{M.Ionel and M.Min-OO (2008)} for both, a predominantly resolved (resolution $>$ deformation) or predominantly deformed (deformation $>$ resolution), resolved warped deformed conifold. This is important for SYZ mirror construction to work. In section {\bf 4}, we compute the temperature dependance of thermal (electrical) conductivity via Kubo formula at finite temperature and finite baryon density up to LO in $N$. The same and deviations from the Wiedemann-Franz formula, upon comparison with \cite{WF}, mimic remarkably a $D=1+1$ Luttinger liquid with impurities. In section {\bf 5}, through four subsections we present the calculation of speed of sound both at leading order (and NLO) in $N$ in the 'MQGP Limit' in four different ways. We then show that the leading order result as obtained from the quasinormal modes of scalar metric perturbation is consistent with that obtained from the two point correlation function.  In section {\bf 6}, we evaluate the NLO correction to the shear mode diffusion constant  again from the quasinormal modes of the vector metric perturbations. Section {\bf 7}, is devoted to the NLO correction to the shear viscosity $\eta$ and shear viscosity - entropy density ratio $\frac{\eta}{s}$. Section {\bf 8} has a summary of the main results of the paper. The technical details of sections {\bf 3} - {\bf 7} are relegated to eight appendices.

\section{The Background}

In this section, via five sub-sections we will:

\begin{itemize}
\item
provide a short review of the type IIB background of \cite{metrics} which is supposed to provide a UV complete holographic dual of large-$N$ thermal QCD, as well as their precursors in subsection {\bf 2.1},

\item
discuss the 'MQGP' limit of \cite{MQGP} and the motivation for considering the same in subsection {\bf 2.2},

\item
briefly review issues pertaining to construction of delocalized S(trominger) Y(au) Z(aslow) mirror and approximate supersymmetry in subsection {\bf 2.3},

\item
review the recipe of \cite{son correlator} to evaluate Minkowskian-signature space correlators in subsection {\bf 2.4},

\item
briefly discuss the vector, tensor and scalar modes of metric perturbations and construction of gauge-invariant variables in subsection {\bf 2.5}

\end{itemize}

\subsection{Type IIB Dual of Large-$N$ Thermal QCD}

In this subsection, we will discuss a UV complete holographic dual of large-$N$ thermal QCD as given in  Dasgupta-Mia et al \cite{metrics}. As partly mentioned in Sec. {\bf 1}, this was inspired by the zero-temperature Klebanov-Witten model \cite{KW}, the non-conformal Klebanov-Tseytlin model \cite{KT}, its IR completion as given in the Klebanov-Strassler model \cite{KS} and Ouyang's inclusion \cite{ouyang} of flavor in the same \footnote{See \cite{Leo-i} for earlier attempts at studying back-reacted $D3/D7$ geometry at zero temperature; we thank L. Zayas for bringing \cite{Leo-i,Leo-ii} to our attention.}, as well as the non-zero temperature/non-extremal version of \cite{Buchel} (the solution however was not regular as the non-extremality/black hole function and the ten-dimensional warp factor vanished simultaneously at the horizon radius), \cite{Gubser-et-al-finitetemp} (valid only at large temperatures) of the Klebanov-Tseytlin model  and \cite{Leo-ii} (addressing the IR), in the absence of flavors.  \\
 \noindent (a) \underline{Brane construction}

 In order to include fundamental quarks at non-zero temperature in the context of type IIB string theory, to the best of our knowledge, the following model proposed in \cite{metrics} is the closest to a UV complete holographic dual of large-$N$ thermal QCD. The KS model (after a duality cascade) and QCD have similar IR behavior: $SU(M)$ gauge group and IR confinement. However, they differ drastically in the UV as the former yields a logarithmically divergent gauge coupling (in the UV) - Landau pole. This necessitates modification of the UV sector of the KS model apart from inclusion of non-extremality factors. With this in mind and building up on all of the above, the type IIB holographic dual of
 \cite{metrics} was constructed. The setup of \cite{metrics} is summarized below.

 \begin{itemize}
 \item
  From a gauge-theory perspective, the authors of \cite{metrics} considered  $N$ black $D3$-branes placed at the tip of six-dimensional conifold, $M\ D5$-branes wrapping the vanishing two-cycle and $M\ \overline{D5}$-branes  distributed along the resolved two-cycle and placed at the outer boundary  of the IR-UV interpolating region/inner boundary of the UV region.

  \item
  More specifically, the $M\ \overline{D5}$ are distributed around the antipodal point relative to the location of $M\ D5$ branes on the blown-up $S^2$. If  the $D5/\overline{D5}$ separation is given by ${\cal R}_{D5/\overline{D5}}$, then this provides the boundary common to the outer UV-IR interpolating region and the inner UV region. The region $r>{\cal R}_{D5/\overline{D5}}$ is the UV.  In other words, the radial space, in \cite{metrics} is divided into the IR, the IR-UV interpolating region and the UV. To summarize the above:
  \begin{itemize}
  \item
  $r_0$($T=0$), $r_h$($T>0$)$<r<|\mu_{\rm Ouyang}|^{\frac{2}{3}}(T=0), {\cal R}_{D5/\overline{D5}}(T>0)$: the IR/IR-UV interpolating regions with $r\sim\Lambda$: deep IR where the $SU(M)$ gauge theory confines

  \item
  $r>|\mu_{\rm Ouyang}|^{\frac{2}{3}} (T=0), {\cal R}_{D5/\overline{D5}}(T>0)$: the UV region.

  \end{itemize}

\item
$N_f\ D7$-branes, via Ouyang embedding,  are holomorphically embedded in the UV (asymptotically $AdS_5\times T^{1,1}$), the IR-UV interpolating region and dipping into the (confining) IR (up to a certain minimum value of $r$ corresponding to the lightest quark)  and $N_f\ \overline{D7}$-branes present in the UV and the UV-IR interpolating (not the confining IR). This is to ensure turning off of three-form fluxes, constancy of the axion-dilaton modulus and hence conformality and absence of Landau poles in the UV.

\item
The resultant ten-dimensional geometry hence involves a resolved warped deformed conifold. Back-reactions are included, e.g., in the ten-dimensional warp factor. Of course, the gravity dual, as in the Klebanov-Strassler construct, at the end of the Seiberg-duality cascade will have no $D3$-branes and the $D5$-branes are smeared/dissolved over the blown-up $S^3$ and thus replaced by fluxes in the IR.
\end{itemize}

The delocalized S(trominger) Y(au) Z(aslow) type IIA mirror of the aforementioned type IIB background of \cite{metrics} and its M-theory uplift had been obtained in \cite{MQGP,transport-coefficients}, and  newer aspects of the same will be looked into in this paper.

\noindent (b) \underline{Seiberg duality cascade, IR confining $SU(M)$ gauge theory at finite temperature and}\\ \underline{$N_c = N_{\rm eff}(r) + M_{\rm eff}(r)$}

\begin{enumerate}
\item
{\bf IR Confinement after Seiberg Duality Cascade}: Footnote numbered 3 shows that one effectively adds on to the number of $D3$-branes in the UV and hence, one has $SU(N+M)\times SU(N+M)$ color gauge group (implying an asymptotic $AdS_5$) and $SU(N_f)\times SU(N_f)$ flavor gauge group, in the UV: $r\geq {\cal R}_{D5/\overline{D5}}$. It is expected that there will be a partial Higgsing of $SU(N+M)\times SU(N+M)$ to $SU(N+M)\times SU(N)$ at $r={\cal R}_{D5/\overline{D5}}$  \cite{K. Dasgupta et al [2012]}. The two gauge couplings, $g_{SU(N+M)}$ and $g_{SU(N)}$ flow  logarithmically  and oppositely in the IR:
\begin{equation}
\label{RG}
4\pi^2\left(\frac{1}{g_{SU(N+M)}^2} + \frac{1}{g_{SU(N)}^2}\right)e^\phi \sim \pi;\
 4\pi^2\left(\frac{1}{g_{SU(N+M)}^2} - \frac{1}{g_{SU(N)}^2}\right)e^\phi \sim \frac{1}{2\pi\alpha^\prime}\int_{S^2}B_2.
\end{equation}
  Had it not been for $\int_{S^2}B_2$, in the UV, one could have set $g_{SU(M+N)}^2=g_{SU(N)}^2=g_{YM}^2\sim g_s\equiv$ constant (implying conformality) which is the reason for inclusion of $M$ $\overline{D5}$-branes at the common boundary of the UV-IR interpolating and the UV regions, to annul this contribution. In fact, the running also receives a contribution from the $N_f$ flavor $D7$-branes which needs to be annulled via $N_f\ \overline{D7}$-branes. The gauge coupling $g_{SU(N+M)}$ flows towards strong coupling and the $SU(N)$ gauge coupling flows towards weak coupling. Upon application of Seiberg duality, $SU(N+M)_{\rm strong}\stackrel{\rm Seiberg\ Dual}{\longrightarrow}SU(N-(M - N_f))_{\rm weak}$ in the IR;  assuming after repeated Seiberg dualities or duality cascade, $N$ decreases to 0 and there is a finite $M$, {one will be left with $SU(M)$ gauge theory with $N_f$ flavors that confines in the IR - the finite temperature version of the same is what was looked at by \cite{metrics}}.

 \item
{\bf Obtaining $N_c=3$, and Color-Flavor Enhancement of Length Scale in the IR}:  So, in the IR, at the end of the duality cascade, what gets identified with the number of colors $N_c$ is $M$, which in the `MQGP limit' to be discussed below, can be tuned to equal 3. One can identify $N_c$ with $N_{\rm eff}(r) + M_{\rm eff}(r)$, where $N_{\rm eff}(r) = \int_{\rm Base\ of\ Resolved\ Warped\ Deformed\ Conifold}F_5$ and $M_{\rm eff} = \int_{S^3}\tilde{F}_3$ (the $S^3$ being dual to $\ e_\psi\wedge\left(\sin\theta_1 d\theta_1\wedge d\phi_1 - B_1\sin\theta_2\wedge d\phi_2\right)$, wherein $B_1$ is an asymmetry factor defined in \cite{metrics}, and $e_\psi\equiv d\psi + {\rm cos}~\theta_1~d\phi_1 + {\rm cos}~\theta_2~d\phi_2$) where $\tilde{F}_3 (\equiv F_3 - \tau H_3)\propto M(r)\equiv 1 - \frac{e^{\alpha(r-{\cal R}_{D5/\overline{D5}})}}{1 + e^{\alpha(r-{\cal R}_{D5/\overline{D5}})}}, \alpha\gg1$  \cite{IR-UV-desc_Dasgupta_etal}. The effective number $N_{\rm eff}$ of $D3$-branes varies between $N\gg1$ in the UV and 0 in the deep IR, and the effective number $M_{\rm eff}$ of $D5$-branes varies between 0 in the UV and $M$ in the deep IR (i.e., at the end of the duality cacade in the IR). Hence, the number of colors $N_c$ varies between $M$ in the deep IR and a large value [even in the MQGP limit of (\ref{limits_Dasguptaetal-ii}) (for a large value of $N$)] in the UV.  {Hence, at very low energies, the number of colors $N_c$ can be approximated by $M$, which in the MQGP limit is taken to be finite and can hence be taken to be equal to three. } However, in this discussion, the low energy or the IR is relative to the string scale. But these energies which are much less than the string scale, can still be much larger than $T_c$. Therefore, for all practical purposes, as regard the energy scales relevant to QCD, the number of colors can be tuned to three.

 In the IR in the MQGP limit, with the inclusion of terms higher order in $g_s N_f$  in the RR and NS-NS three-form fluxes and the NLO terms in the angular part of the metric, there occurs an IR color-flavor enhancement of the length scale as compared to a Planckian length scale in KS for ${\cal O}(1)$ $M$, thereby showing that quantum corrections will be suppressed. Using \cite{metrics}:
\begin{eqnarray}
\label{NeffMeffNfeff}
& & N_{\rm eff}(r) = N\left[ 1 + \frac{3 g_s M_{\rm eff}^2}{2\pi N}\left(\log r + \frac{3 g_s N_f^{\rm eff}}{2\pi}\left(\log r\right)^2\right)\right],\nonumber\\
& & M_{\rm eff}(r) = M + \frac{3g_s N_f M}{2\pi}\log r + \sum_{m\geq1}\sum_{n\geq1} N_f^m M^n f_{mn}(r),\nonumber\\
& & N^{\rm eff}_f(r) = N_f + \sum_{m\geq1}\sum_{n\geq0} N_f^m M^n g_{mn}(r).
\end{eqnarray}
it was argued in \cite{T_c+Torsion} that  the length scale of the OKS-BH metric in the IR will be given by:
\begin{eqnarray}
\label{length-IR}
& & L_{\rm OKS-BH}\sim\sqrt{M}N_f^{\frac{3}{4}}\sqrt{\left(\sum_{m\geq0}\sum_{n\geq0}N_f^mM^nf_{mn}(\Lambda)\right)}\left(\sum_{l\geq0}\sum_{p\geq0}N_f^lM^p g_{lp}(\Lambda)\right)^{\frac{1}{4}}g_s^{\frac{1}{4}}\sqrt{\alpha^\prime}\nonumber\\
& & \equiv N_f^{\frac{3}{4}}\left.\sqrt{\left(\sum_{m\geq0}\sum_{n\geq0}N_f^mM^nf_{mn}(\Lambda)\right)}\left(\sum_{l\geq0}\sum_{p\geq0}N_f^lM^p g_{lp}(\Lambda)\right)^{\frac{1}{4}} L_{\rm KS}\right|_{\Lambda:\log \Lambda{<}{\frac{2\pi}{3g_sN_f}}},
\end{eqnarray}
which implies that  {in the IR, relative to KS, there is a color-flavor enhancement of the length scale in the OKS-BH metric}. Hence,  in the IR, even for $N_c^{\rm IR}=M=3$ and $N_f=6$ upon inclusion of of $n,m>1$  terms in
$M_{\rm eff}$ and $N_f^{\rm eff}$ in (\ref{NeffMeffNfeff}), $L_{\rm OKS-BH}\gg L_{\rm KS}(\sim L_{\rm Planck})$ in the MQGP limit involving $g_s\stackrel{\sim}{<}1$, implying that {the stringy corrections are suppressed and one can trust supergravity calculations}. As a reminder one will generate higher powers of $M$ and $N_f$ in the double summation in $M_{\rm eff}$ in (\ref{NeffMeffNfeff}), e.g., from the terms higher order in $g_s N_f$ in the RR and NS-NS three-form fluxes that become relevant for the aforementioned values of $g_s, N_f$.

 \item
  Further, the global  flavor group in the UV-IR interpolating and UV regions, due to presence of $N_f$ $D7$ and $N_f\ \overline{D7}$-branes, is $SU(N_f)\times SU(N_f)$, which is broken in the IR to $SU(N_f)$ as the IR has only $N_f$ $D7$-branes.

\end{enumerate}

Hence, the following features of the type IIB model of \cite{metrics} make it an ideal holographic dual of thermal QCD:

\begin{itemize}
\item
the theory having quarks transforming in the fundamental representation, is UV conformal and IR confining with the required chiral symmetry breaking in the IR and restoration at high temperatures

\item
the theory is UV complete with the gauge coupling remaining finite in the UV (absence of Landau poles)

\item
the theory is not just defined for large temperatures but for low and high temperatures

\item
(as will become evident in Sec. {\bf 3}) with the inclusion of a finite baryon chemical potential, the theory provides a lattice-compatible QCD confinement-deconfinement temperature $T_c$ for the right number of light quark flavors and masses, and is also thermodynamically stable; given the IR proximity of the value of the lattice-compatible $T_c$,  after the end of the Seiberg duality cascade, the number of quark flavors approximately equals $M$ which in the `MQGP' limit of (\ref{limits_Dasguptaetal-ii}) can be tuned to equal 3

\item
in the MQGP limit (\ref{limits_Dasguptaetal-ii}) which requires considering a finite gauge coupling and hence string coupling, the theory was shown in \cite{MQGP} to be holographically renormalizable from an M-theory perspective with the M-theory uplift also being thermodynamically stable.

\end{itemize}


\noindent (d) \underline{Supergravity solution on resolved warped deformed conifold}

The working metric is given by :
\begin{equation}
\label{metric}
ds^2 = \frac{1}{\sqrt{h}}
\left(-g_1 dt^2+dx_1^2+dx_2^2+dx_3^2\right)+\sqrt{h}\biggl[g_2^{-1}dr^2+r^2 d{\cal M}_5^2\biggr].
\end{equation}
 $g_i$'s are black hole functions in modified OKS(Ouyang-Klebanov-Strassler)-BH (Black Hole) background and are assumed to be:
$ g_{1,2}(r,\theta_1,\theta_2)= 1-\frac{r_h^4}{r^4} + {\cal O}\left(\frac{g_sM^2}{N}\right)$
where $r_h$ is the horizon, and the ($\theta_1, \theta_2$) dependence come from the
${\cal O}\left(\frac{g_sM^2}{N}\right)$ corrections. The  $h_i$'s are expected to receive corrections of
${\cal O}\left(\frac{g_sM^2}{N}\right)$ \cite{K. Dasgupta  et al [2012]}. We assume the same to also be true of the `black hole functions' $g_{1,2}$.  The compact five dimensional metric in (\ref{metric}), is given as:
\begin{eqnarray}
\label{RWDC}
& & d{\cal M}_5^2 =  h_1 (d\psi + {\rm cos}~\theta_1~d\phi_1 + {\rm cos}~\theta_2~d\phi_2)^2 +
h_2 (d\theta_1^2 + {\rm sin}^2 \theta_1 ~d\phi_1^2) +   \nonumber\\
&&  + h_4 (h_3 d\theta_2^2 + {\rm sin}^2 \theta_2 ~d\phi_2^2) + h_5~{\rm cos}~\psi \left(d\theta_1 d\theta_2 -
{\rm sin}~\theta_1 {\rm sin}~\theta_2 d\phi_1 d\phi_2\right) + \nonumber\\
&&  + h_5 ~{\rm sin}~\psi \left({\rm sin}~\theta_1~d\theta_2 d\phi_1 +
{\rm sin}~\theta_2~d\theta_1 d\phi_2\right),
\end{eqnarray}
$r\gg a, h_5\sim\frac{({\rm deformation\ parameter})^2}{r^3}\ll  1$ for $r \gg({\rm deformation\ parameter})^{\frac{2}{3}}$, i.e. in the UV/IR-UV interpolating region.  The $h_i$'s appearing in internal metric as well as $M, N_f$ are not constant and up to linear order depend on $g_s, M, N_f$ are given as below:
\begin{eqnarray}
\label{h_i}
& & \hskip -0.45in h_1 = \frac{1}{9} + {\cal O}\left(\frac{g_sM^2}{N}\right),\  h_2 = \frac{1}{6} + {\cal O}\left(\frac{g_sM^2}{N}\right),\ h_4 = h_2 + \frac{a^2}{r^2},\nonumber\\
& & h_3 = 1 + {\cal O}\left(\frac{g_sM^2}{N}\right),\ h_5\neq0,\
 L=\left(4\pi g_s N\right)^{\frac{1}{4}}.
\end{eqnarray}
One sees from (\ref{RWDC}) and (\ref{h_i}) that one has a non-extremal resolved warped deformed conifold involving
an $S^2$-blowup (as $h_4 - h_2 = \frac{a^2}{r^2}$), an $S^3$-blowup (as $h_5\neq0$) and squashing of an $S^2$ (as $h_3$ is not strictly unity). The horizon (being at a finite $r=r_h$) is warped squashed $S^2\times S^3$. In the deep IR, in principle one ends up with a warped squashed $S^2(a)\times S^3(\epsilon),\ \epsilon$ being the deformation parameter. Assuming $\epsilon^{\frac{2}{3}}>a$ and given that $a={\cal O}\left(\frac{g_s M^2}{N}\right)r_h$ \cite{K. Dasgupta  et al [2012]}, in the IR and in the MQGP limit, $N_{\rm eff}(r\in{\rm IR})=\int_{{\rm warped\ squashed}\ S^2(a)\times S^3(\epsilon)}F_5(r\in{\rm IR})\ll   M = \int_{S^3(\epsilon)}F_3(r\in{\rm IR})$; we have a confining $SU(M)$ gauge theory in the IR.

 The warp factor that includes the back-reaction, in the IR is given as:
\begin{eqnarray}
\label{eq:h}
&& \hskip -0.45in h =\frac{L^4}{r^4}\Bigg[1+\frac{3g_sM_{\rm eff}^2}{2\pi N}{\rm log}r\left\{1+\frac{3g_sN^{\rm eff}_f}{2\pi}\left({\rm
log}r+\frac{1}{2}\right)+\frac{g_sN^{\rm eff}_f}{4\pi}{\rm log}\left({\rm sin}\frac{\theta_1}{2}
{\rm sin}\frac{\theta_2}{2}\right)\right\}\Biggr],
\end{eqnarray}
where, in principle, $M_{\rm eff}/N_f^{\rm eff}$ are not necessarily the same as $M/N_f$; we however will assume that up to ${\cal O}\left(\frac{g_sM^2}{N}\right)$, they are. Proper UV behavior requires \cite{K. Dasgupta et al [2012]}:
\begin{eqnarray}
\label{h-large-small-r}
& & h = \frac{L^4}{r^4}\left[1 + \sum_{i=1}\frac{{\cal H}_i\left(\phi_{1,2},\theta_{1,2},\psi\right)}{r^i}\right],\ {\rm large}\ r;
\nonumber\\
& & h = \frac{L^4}{r^4}\left[1 + \sum_{i,j; (i,j)\neq(0,0)}\frac{h_{ij}\left(\phi_{1,2},\theta_{1,2},\psi\right)\log^ir}{r^j}\right],\ {\rm small}\ r.
\end{eqnarray}


  In the IR, up to ${\cal O}(g_s N_f)$ and setting $h_5=0$, the three-forms are as given in \cite{metrics}:
\begin{eqnarray}
\label{three-form-fluxes}
& & \hskip -0.4in (a) {\widetilde F}_3  =  2M { A_1} \left(1 + \frac{3g_sN_f}{2\pi}~{\rm log}~r\right) ~e_\psi \wedge
\frac{1}{2}\left({\rm sin}~\theta_1~ d\theta_1 \wedge d\phi_1-{ B_1}~{\rm sin}~\theta_2~ d\theta_2 \wedge
d\phi_2\right)\nonumber\\
&& \hskip -0.3in -\frac{3g_s MN_f}{4\pi} { A_2}~\frac{dr}{r}\wedge e_\psi \wedge \left({\rm cot}~\frac{\theta_2}{2}~{\rm sin}~\theta_2 ~d\phi_2
- { B_2}~ {\rm cot}~\frac{\theta_1}{2}~{\rm sin}~\theta_1 ~d\phi_1\right)\nonumber \\
&& \hskip -0.3in -\frac{3g_s MN_f}{8\pi}{ A_3} ~{\rm sin}~\theta_1 ~{\rm sin}~\theta_2 \left(
{\rm cot}~\frac{\theta_2}{2}~d\theta_1 +
{ B_3}~ {\rm cot}~\frac{\theta_1}{2}~d\theta_2\right)\wedge d\phi_1 \wedge d\phi_2, \nonumber\\
& & \hskip -0.4in (b) H_3 =  {6g_s { A_4} M}\Biggl(1+\frac{9g_s N_f}{4\pi}~{\rm log}~r+\frac{g_s N_f}{2\pi}
~{\rm log}~{\rm sin}\frac{\theta_1}{2}~
{\rm sin}\frac{\theta_2}{2}\Biggr)\frac{dr}{r}\nonumber \\
&& \hskip -0.3in \wedge \frac{1}{2}\Biggl({\rm sin}~\theta_1~ d\theta_1 \wedge d\phi_1
- { B_4}~{\rm sin}~\theta_2~ d\theta_2 \wedge d\phi_2\Biggr)
+ \frac{3g^2_s M N_f}{8\pi} { A_5} \Biggl(\frac{dr}{r}\wedge e_\psi -\frac{1}{2}de_\psi \Biggr)\nonumber  \\
&&  \wedge \Biggl({\rm cot}~\frac{\theta_2}{2}~d\theta_2
-{ B_5}~{\rm cot}~\frac{\theta_1}{2} ~d\theta_1\Biggr). \nonumber\\
\end{eqnarray}
The asymmetry factors in (\ref{three-form-fluxes}) are given by: $ A_i=1 +{\cal O}\left(\frac{a^2}{r^2}\ {\rm or}\ \frac{a^2\log r}{r}\ {\rm or}\ \frac{a^2\log r}{r^2}\right) + {\cal O}\left(\frac{{\rm deformation\ parameter }^2}{r^3}\right),$ $  B_i = 1 + {\cal O}\left(\frac{a^2\log r}{r}\ {\rm or}\ \frac{a^2\log r}{r^2}\ {\rm or}\ \frac{a^2\log r}{r^3}\right)+{\cal O}\left(\frac{({\rm deformation\ parameter})^2}{r^3}\right)$.    As in the UV, $\frac{({\rm deformation\ parameter})^2}{r^3}\ll  \frac{({\rm resolution\ parameter})^2}{r^2}$, we will assume the same three-form fluxes for $h_5\neq0$.

Further, to ensure UV conformality, it is important to ensure that the axion-dilaton modulus approaches a constant implying a vanishing beta function in the UV. This was discussed in detail in  appendix B of \cite{T_c+Torsion}, wherein in particular, assuming an F-theory uplift involving, locally, an elliptically fibered $K3$, it was shown that UV conformality and the Ouyang embedding are mutually consistent.

\subsection{The `MQGP Limit'}

In \cite{MQGP}, we had considered the following two limits:
\begin{eqnarray}
\label{limits_Dasguptaetal-i}
&   & \hskip -0.17in (i) {\rm weak}(g_s){\rm coupling-large\ t'Hooft\ coupling\ limit}:\nonumber\\
& & \hskip -0.17in g_s\ll  1, g_sN_f\ll  1, \frac{g_sM^2}{N}\ll  1, g_sM\gg1, g_sN\gg1\nonumber\\
& & \hskip -0.17in {\rm effected\ by}: g_s\sim\epsilon^{d}, M\sim\left({\cal O}(1)\epsilon\right)^{-\frac{3d}{2}}, N\sim\left({\cal O}(1)\epsilon\right)^{-19d}, \epsilon\ll  1, d>0
 \end{eqnarray}
(the limit in the first line  though not its realization in the second line, considered in \cite{metrics});
\begin{eqnarray}
\label{limits_Dasguptaetal-ii}
& & \hskip -0.17in (ii) {\rm MQGP\ limit}: \frac{g_sM^2}{N}\ll  1, g_sN\gg1, {\rm finite}\
 g_s, M\ \nonumber\\
& & \hskip -0.17in {\rm effected\ by}:  g_s\sim\epsilon^d, M\sim\left({\cal O}(1)\epsilon\right)^{-\frac{3d}{2}}, N\sim\left({\cal O}(1)\epsilon\right)^{-39d}, \epsilon\lesssim 1, d>0.
\end{eqnarray}

Let us enumerate the motivation for considering the MQGP limit which was discussed in detail in \cite{T_c+Torsion}. There are principally two.
\begin{enumerate}
\item
Unlike the AdS/CFT limit wherein $g_{\rm YM}\rightarrow0, N\rightarrow\infty$ such that $g_{\rm YM}^2N$ is large, for strongly coupled thermal systems like sQGP, what is relevant is $g_{\rm YM}\sim{\cal O}(1)$ and $N_c=3$. From the discussion in the previous paragraphs specially the one in point (c) of sub-section {\bf 2.1}, one sees that in the IR after the Seiberg duality cascade, effectively $N_c=M$ which in the MQGP limit of (\ref{limits_Dasguptaetal-ii})  can be tuned to 3. Further, in the same limit, the string coupling $g_s\stackrel{<}{\sim}1$. The finiteness of the string coupling necessitates addressing the same from an M theory perspective. This is the reason for coining the name: `MQGP limit'. In fact this is the reason why one is required to first construct a type IIA mirror, which was done in \cite{MQGP} a la delocalized Strominger-Yau-Zaslow mirror symmetry, and then take its M-theory uplift.

\item
From the perspective of calculational simplification in supergravity, the following are examples of the same and constitute therefore the second set of reasons for looking at the MQGP limit of (\ref{limits_Dasguptaetal-ii}):
\begin{itemize}
\item
In the UV-IR interpolating region and the UV,
$(M_{\rm eff}, N_{\rm eff}, N_f^{\rm eff})\stackrel{\rm MQGP}{\approx}(M, N, N_f)$
\item
Asymmetry Factors $A_i, B_j$(in three-form fluxes)$\stackrel{MQGP}{\rightarrow}1$  in the UV-IR interpolating region and the UV.

\item
Simplification of ten-dimensional warp factor and non-extremality function in MQGP limit
\end{itemize}
\end{enumerate}

 With ${\cal R}_{D5/\overline{D5}}$ denoting the boundary common to the UV-IR interpolating region and the UV region, $\tilde{F}_{lmn}, H_{lmn}=0$ for $r\geq {\cal R}_{D5/\overline{D5}}$ is required to ensure conformality in the UV.  Near the $\theta_1=\theta_2=0$-branch, assuming: $\theta_{1,2}\rightarrow0$ as $\epsilon^{\gamma_\theta>0}$ and $r\rightarrow {\cal R}_{\rm UV}\rightarrow\infty$ as $\epsilon^{-\gamma_r <0}, \lim_{r\rightarrow\infty}\tilde{F}_{lmn}=0$ and  $\lim_{r\rightarrow\infty}H_{lmn}=0$ for all components except $H_{\theta_1\theta_2\phi_{1,2}}$; in the MQGP limit and near $\theta_{1,2}=\pi/0$-branch, $H_{\theta_1\theta_2\phi_{1,2}}=0/\left.\frac{3 g_s^2MN_f}{8\pi}\right|_{N_f=2,g_s=0.6, M=\left({\cal O}(1)g_s\right)^{-\frac{3}{2}}}\ll  1.$ So, the UV nature too is captured near $\theta_{1,2}=0$-branch in the MQGP limit. This mimics addition of $\overline{D5}$-branes in \cite{metrics} to ensure cancellation of $\tilde{F}_3$.

\subsection{Approximate Supersymmetry, Construction of  the Delocalized SYZ IIA Mirror and Its M-Theory Uplift in the MQGP Limit}

A central issue to \cite{MQGP,transport-coefficients} has been implementation of delocalized mirror symmetry via the Strominger Yau Zaslow prescription according to which the mirror of a Calabi-Yau can be constructed via three T dualities along a special Lagrangian $T^3$ fibered over a large base in the Calabi-Yau. This sub-section is a quick review of precisely this.

{ To implement the quantum mirror symmetry a la S(trominger)Y(au)Z(aslow) \cite{syz}, one needs a special Lagrangian (sLag) $T^3$ fibered over a large base (to nullify contributions from open-string disc instantons with boundaries as non-contractible one-cycles in the sLag). Defining delocalized T-duality coordinates, $(\phi_1,\phi_2,\psi)\rightarrow(x,y,z)$ valued in $T^3(x,y,z)$ \cite{MQGP}:
\begin{equation}
\label{xyz defs}
x = \sqrt{h_2}h^{\frac{1}{4}}sin\langle\theta_1\rangle\langle r\rangle \phi_1,\ y = \sqrt{h_4}h^{\frac{1}{4}}sin\langle\theta_2\rangle\langle r\rangle \phi_2,\ z=\sqrt{h_1}\langle r\rangle h^{\frac{1}{4}}\psi,
\end{equation}
using the results of \cite{M.Ionel and M.Min-OO (2008)} it was shown in \cite{transport-coefficients} that the following conditions are satisfied:
\begin{eqnarray}
\label{sLag-conditions}
& & i^* J \approx 0,\nonumber\\
& & \Im m\left( i^*\Omega\right) \approx 0,\nonumber\\
& & \Re e\left(i^*\Omega\right)\sim{\rm volume \ form}\left(T^3(x,y,z)\right),
\end{eqnarray}
for the $T^2$-invariant sLag of \cite{M.Ionel and M.Min-OO (2008)} for a deformed conifold. It will be shown in the Section {\bf 3} that (\ref{sLag-conditions}) is also satisfied for the $T^2$-invariant sLag of \cite{M.Ionel and M.Min-OO (2008)} for a resolved conifold, implying thus: $\left.i^* J\right|_{RC/DC}\approx0, \Im m\left.\left( i^*\Omega\right)\right|_{RC/DC} \approx 0, \Re e\left.\left(i^*\Omega\right)\right|_{RC/DC}\sim{\rm volume \ form}\left(T^3(x,y,z)\right)$. Hence, if the resolved warped deformed conifold is predominantly either resolved or deformed, the local $T^3$ of (\ref{xyz defs}) is the required sLag to effect SYZ mirror construction.}

{Interestingly, in the `delocalized limit' \cite{M. Becker et al [2004]}  $\psi=\langle\psi\rangle$, under the coordinate transformation:
\begin{equation}
\label{transformation_psi}
\left(\begin{array}{c} sin\theta_2 d\phi_2 \\ d\theta_2\end{array} \right)\rightarrow \left(\begin{array}{cc} cos\langle\psi\rangle & sin\langle\psi\rangle \\
- sin\langle\psi\rangle & cos\langle\psi\rangle
\end{array}\right)\left(\begin{array}{c}
sin\theta_2 d\phi_2\\
d\theta_2
\end{array}
\right),
\end{equation}
and $\psi\rightarrow\psi - \cos\langle{\bar\theta}_2\rangle\phi_2 + \cos\langle\theta_2\rangle\phi_2 - \tan\langle\psi\rangle ln\sin{\bar\theta}_2$, the $h_5$ term becomes $h_5\left[d\theta_1 d\theta_2 - sin\theta_1 sin\theta_2 d\phi_1d\phi_2\right]$, $e_\psi\rightarrow e_\psi$, i.e.,  one introduces an isometry along $\psi$ in addition to the isometries along $\phi_{1,2}$. This clearly is not valid globally - the deformed conifold does not possess a third global isometry}.

{ To enable use of SYZ-mirror duality via three T dualities, one also needs to ensure a large base (implying large complex structures of the aforementioned two two-tori) of the $T^3(x,y,z)$ fibration. This is effected via
\cite{F. Chen et al [2010]}:
\begin{eqnarray}
\label{SYZ-large base}
& & d\psi\rightarrow d\psi + f_1(\theta_1)\cos\theta_1 d\theta_1 + f_2(\theta_2)\cos\theta_2d\theta_2,\nonumber\\
& & d\phi_{1,2}\rightarrow d\phi_{1,2} - f_{1,2}(\theta_{1,2})d\theta_{1,2},
\end{eqnarray}
for appropriately chosen large values of $f_{1,2}(\theta_{1,2})$. The three-form fluxes
 remain invariant. The fact that one can choose such large values of $f_{1,2}(\theta_{1,2})$, was justified in \cite{MQGP}.  The guiding principle is that one requires the metric obtained after SYZ-mirror transformation applied to the non-K\"{a}hler  resolved warped deformed conifold is like a non-K\"{a}hler warped resolved conifold at least locally. Then $G^{IIA}_{\theta_1\theta_2}$ needs to vanish \cite{MQGP}.  This is shown to be true anywhere in the UV in Appendix {\bf C}.}


{
The mirror type IIA metric after performing three T-dualities, first along $x$, then along $y$ and finally along $z$, utilizing the results of \cite{M. Becker et al [2004]} was worked out in \cite{MQGP}. We can get a one-form type IIA potential from the triple T-dual (along $x, y, z$) of the type IIB $F_{1,3,5}$ in \cite{MQGP} and using which the following $D=11$ metric was obtained in \cite{MQGP} ($u\equiv\frac{r_h}{r}$):
\begin{eqnarray}
\label{Mtheory met}
& &\hskip -0.6in   ds^2_{11} = e^{-\frac{2\phi^{IIA}}{3}} \left[g_{tt}dt^2 + g_{\mathbb{R}^3}\left(dx^2 + dy^2 + dz^2\right) +  g_{uu}du^2  +   ds^2_{IIA}({\theta_{1,2},\phi_{1,2},\psi})\right] \nonumber\\
& & \hskip -0.6in+ e^{\frac{4{\phi}^{IIA}}{3}}\Bigl(dx_{11} + A^{F_1}+A^{F_3}+A^{F_5}\Bigr)^2 \equiv\ {\rm Black}\ M3-{\rm Brane}+{\cal O}\left(\left[\frac{g_s M^2 \log N}{N}\right] \left(g_sM\right)N_f\right),\nonumber\\
& & {\rm where}:\nonumber\\
& & g_{uu}=\frac{3^{2/3}(2\sqrt{\pi g_s N})}{u^2(1-u^4)}\left(1  - \frac{3 g_s^2 M^2 N_f \log(N) \log \left(\frac{r_h}{u}\right)}{32 \pi ^2 N}\right)\nonumber\\
& & g_{tt} = \frac{3^{2/3}(u^4-1)r_h^2}{u^2(2\sqrt{\pi g_{s}N})} \left(\frac{3 g_s^2 M^2 N_f \log(N) \log \left(\frac{r_h}{u}\right)}{32 \pi ^2 N}+1\right)\nonumber\\
& & g_{\mathbb{R}^3} =  \frac{3^{2/3}r_h^2}{u^2(2\sqrt{\pi g_{s}N})} \left(\frac{3 g_s^2 M^2 N_f \log(N) \log \left(\frac{r_h}{u}\right)}{32 \pi ^2 N}+1\right).
\end{eqnarray}
Further, in the UV:
\begin{eqnarray}
\label{non-conformal-contribution}
& & G_{{x}{x}}^M = G_{{y}{y}}^M = G_{{z}{z}}^M = \frac{3^{2/3}r_h^2}{g_{s}^{2/3}u^2(2\sqrt{\pi g_{s}N})} \left(\frac{3 g_s^2 M^2 N_f \log(N) \log \left(\frac{r_h}{u}\right)}{32 \pi ^2 N}+1\right)\nonumber\\
& & G_{\phi_1r}^M \sim \nonumber\\
& & \frac{2 {g_s}^{4/3} {N_f}^2 {\sin^2\phi_1} 2\sin\left(\frac{\psi}{2}\right)  \sin ^2({\theta_1}) \left(9 \sin ^2({\theta_1})+6 \cos
   ^2({\theta_1})+4 \cos ({\theta_1})\right) \sqrt[4]{{g_s} N \left(1-\frac{3 {g_s}^2 M^2 {N_f} \log (N)
   \log (r)}{16 \pi ^2 N}\right)} }{3^{5/6} \pi ^{7/4} (\cos
   (2 {\theta_1})-5)^2}\nonumber\\
   & & \times \left(9 {h_5} \sin ({\theta_1})+4 \cos ^2({\theta_1}) \csc ({\theta_2})-2
   \cos ({\theta_1}) \cot ({\theta_2})+6 \sin ^2({\theta_1}) \csc ({\theta_2})\right)\nonumber\\
   & & G_{11\ r}^M\sim \frac{3^{\frac{3}{2}}g_s^{\frac{4}{3}} N_f \sin\phi_1\left( - 8 \cos\theta_1 + 3( - 5 +
   \cos(2\theta_1))\right)\sin\theta_1}{\pi\left( - 5 + \cos(2\theta_1)\right)}.
\end{eqnarray}
In \cite{T_c+Torsion}, we showed explicitly the existence of a local type IIA $SU(3)$ structure and a local $G_2$ structure of the M-theory uplift around $\theta_1\sim\frac{1}{N^{\frac{1}{5}}}, \theta_2\sim\frac{1}{N^{\frac{3}{10}}}$. Near this coordinate patch, one sees that:
\begin{eqnarray}
\label{Gphi1r+G11r}
& & G_{\phi_1r}\sim\frac{10 g_s^{\frac{19}{12}}\sin^2\phi_1 \sin\left(\frac{\psi}{2}\right)N^{\frac{3}{20}}}
{2 3^{\frac{5}{6}}\pi^{\frac{7}{4}}} + {\cal O}\left(\frac{1}{N^{\frac{1}{4}}}\right)\ll1\ {\rm for}\ \psi\sim\frac{1}{N^{\alpha\gg\frac{3}{20}}},\nonumber\\
& & G_{11\ r}\sim{\cal O}\left(\frac{1}{N^{\frac{1}{5}}}\right).
\end{eqnarray}
Thus, in the MQGP limit, around  $\theta_1\sim\frac{1}{N^{\frac{1}{5}}}, \theta_2\sim\frac{1}{N^{\frac{3}{10}}}$, the five-dimensional $M_5(t,x_{1,2,3},u)$ decouples from $M_6(\theta_{1,2},\phi_{1,2},\psi,x_{10})$.

As in Klebanov-Strassler construction, a single T-duality along a direction orthogonal to the $D3$-brane world volume, e.g., $z$ of (\ref{xyz defs}), yields $D4$ branes straddling a pair of $NS5$-branes consisting of world-volume coordinates $(\theta_1,x)$ and $(\theta_2,y)$. Further, T-dualizing along $x$ and then $y$ would yield a Taub-NUT space  from each of the two $NS5$-branes \cite{T-dual-NS5-Taub-NUT-Tong}. The $D7$-branes yield $D6$-branes which get uplifted to Kaluza-Klein monopoles in M-theory \cite{KK-monopoles-A-Sen} which too involve Taub-NUT spaces. Globally, probably the eleven-dimensional uplift would involve a seven-fold of $G_2$-structure, analogous to the uplift of $D5$-branes wrapping a two-cycle in a resolved warped conifold \cite{Dasguptaetal_G2_structure}.

Now,  in the delocalized limit of \cite{M. Becker et al [2004]}, in \cite{transport-coefficients}, e.g., $\left.\int_{C_4(\theta_{1,2},\phi_{1/2},x_{10})}G_4\right|_{\phi_{2/1}=\langle\phi_{2/1}\rangle,\psi=\langle\psi\rangle,\langle r\rangle}$ was estimated to be very large. There is a two-fold reason for the same. First, using the local $T^3$-coordinates of (\ref{xyz defs}), this large flux is estimated in the MQGP limit to be $\left(g_s N\right)^{\frac{1}{4}}$ (as, using (\ref{xyz defs}), $G_{\phi_1\ {\rm or}\ \phi_2\ {\rm or}\ \psi\bullet\bullet\bullet}\sim \left(g_s N\right)^{\frac{1}{4}}G_{x\ {\rm or}\ y\ {\rm or}\ z\bullet\bullet\bullet}$ where the bullets denote directions other than $\phi_1,\phi_2,\psi$). This in the MQGP limit, is large. The second is the following. Now, $G_4 = H\wedge (A^{F_1+F-3+F_5} - dx_{10})$ \cite{MQGP} where $A^{F_1+F_3+F_5}$ is the type IIA one-form gauge field obtained after SYZ mirror construction via triple T dualities on the type IIB $F_{1,3,5}$.  As the $S^2(\theta_1,\phi_1)$ is a vanishing two-sphere, to obtain a finite $\int_{S^2(\theta_1,\phi_1)}B_2$  - that appears in the RG equation (\ref{RG}) - one requires a large $B_2$ (From \cite{metrics} one sees that such a large contribution to $B_2$ is obtained near the $\theta_1=\theta_2=0$ branch.) Therefore, this too contributes to a large $G_4$ via a large $H$.

Locally, the uplift (\ref{Mtheory met}) can hence be thought of as black $M3$-brane metric, which in the UV, can be thought of as black $M5$-branes wrapping a two cycle homologous to:
$n_1 S^2(\theta_1,x_{10}) + n_2 S^2(\theta_2,\phi_{1/2}) + m_1 S^2(\theta_1,\phi_{1/2}) + m_2 S^2(\theta_2,x_{10})$ for some large $n_{1,2},m_{1,2}\in\mathbb{Z}$ \cite{transport-coefficients}.  In the large-$r$ limit, the $D=11$ space-time is a warped product of $AdS_5(\mathbb{R}^{1,3}\times\mathbb{R}_{>0})$ and ${\cal M}_6(\theta_{1,2},\phi_{1,2},\psi,x_{10})$
\begin{equation}
\hskip -0.4in
\label{M_6}
\begin{array}{cc}
&{\cal M}_6(\theta_{1,2},\phi_{1,2},\psi,x_{10})   \longleftarrow   S^1(x_{10}) \\
&\downarrow  \\
{\cal M}_3(\phi_1,\phi_2,\psi) \hskip -0.4in & \longrightarrow  {\cal M}_5(\theta_{1,2},\phi_{1,2},\psi)   \\
&\downarrow  \\
 & \hskip 0.9in {\cal B}_2(\theta_1,\theta_2)  \longleftarrow  [0,1]_{\theta_1}  \\
 & \downarrow  \\
& [0,1]_{\theta_2}
\end{array}.
\end{equation}

{The $D=11$ SUGRA EOMs/Bianchi identity \cite{M.S. Bremer [1999]} were shown in \cite{transport-coefficients} to be  satisfied near the $\theta_{1,2}=0,\pi$-branches in the MQGP limit.}

\subsection{Recipe to find Minkowski Correlators}
Following \cite{son correlator} we briefly review the prescription to find the thermal correlator in Minkowski signature. According to AdS/CFT correspondence, there exists an operator $\mathcal{O}$ in the field theory side dual to a field $\phi$ defined in the bulk of AdS geometry such that on the boundary of the anti-de Sitter space $\phi$ tends to a value $\phi_{0}$ which acts as a source for the operator $\mathcal{O}$. we are interested in calculating the retarded Green's function $G^R$ of the operator $\mathcal{O}$ in Minkowski space.

Our working background (Type $IIB$ or it's M-theory uplift) can be expressed as the following $5d$ metric,
\begin{eqnarray}
\label{5dmetric inu}
ds^2=-g_{tt}(u)dt^2+g_{xx}(u)\Biggl(dx^2+dy^2+dz^2\Biggr)+g_{uu}(u)du^2.
\end{eqnarray}
Here $u$ is the radial coordinate defined as $u=r_h/r$ so that $u=0$ is the boundary and $u=1$ is the horizon of the AdS space. A solution of the linearized field equation for any field $\phi(u,x)$ choosing $q^\mu=(w,q,0,0)$ is given as,
 \begin{equation}
 \phi(u,x)=\int\frac{{d^4}q}{(2\pi)^4}e^{-i wt + i qx}f_{q}(u)\phi_{0}(q)
 \end{equation}
where $ f_{q}(u)$ is normalized to 1 at the boundary and satisfies the incoming wave boundary condition at $u=1$, and $\phi_{0}(q)$ is determined by,
 \begin{equation}
 \phi(u=0,x)=\int\frac{{d^4}q}{(2\pi)^4}e^{-i wt + i qx}\phi_{0}(q).
\end{equation}
If the kinetic term for $\phi(u,x)$ is given by: $\frac{1}{2}\int {d^4}x du A(u)\left(\partial_u\phi(x,u)\right)^2$, then using the equation of motion for $\phi$ it is possible to reduce an on-shell action to the surface terms as,
\begin{equation}
S=\int\frac{{d^4}q}{(2\pi)^4}\phi_{0}(-q)\mathcal{F}(q,u)\phi_{0}(q)|^{u=1}_{u=0}
\end{equation}
where the function
\begin{equation}
\label{F}
\mathcal{F}(q,u) = A(u) f_{\pm q}(u)\partial_{u}f_{\pm q}(u).
\end{equation}
 Finally, the retarded Green's function is given by the formula:
\begin{equation}
G^{R}(q)=-2\mathcal{F}(q,u)|_{u=0}.
\end{equation}

The different retarded Green's functions are defined as
 \begin{equation}
 G^{R\ T}_{\mu\nu,\rho\sigma}(q)=-i\int {d^4}x e^{-i wt + i qx}\theta(t) \langle[T_{\mu\nu}(x),T_{\rho\sigma}(0)]\rangle,
\end{equation}
with $\langle[T_{\mu\nu},T_{\rho\sigma}]\rangle\sim\frac{\delta^2S}{\delta h_{\mu\nu}\delta h_{\rho\sigma}}$ and
 \begin{equation}
G^{R\ J}_{\mu\nu}(q)=-i\int {d^4}x e^{-i wt + i qx}\theta(t) \langle[J_{\mu}(x),J_{\nu}(0)]\rangle
\end{equation}
with $\langle[J_{\mu}(x),J_{\nu}(0)]\rangle\sim\frac{\delta^2S}{\delta A_\mu \delta A_\nu}$,
as the energy-momentum tensor $T_{\mu\nu}(x)$ and the current $J_{\mu}(x)$ couple respectively to the metric and gauge field.

\subsection{Perturbations of the background and the gauge invariant combinations}
In the background of section {\bf 2} (Type $IIB$ and it's M-theory uplift), we consider a small linear fluctuation of the black brane metric of (\ref{5dmetric inu}) as:
\begin{eqnarray}
\label{metric perturbation}
g_{\mu\nu}=g^0_{\mu\nu}+h_{\mu\nu},
\end{eqnarray}
where $g^0_{\mu\nu}$ denotes the background metric. The inverse metric is defined as(up to second order in perturbation)
\begin{equation}
\label{inverse_fluc}
g^{\mu\nu}=g^{(0)\mu\nu}-h^{\mu\nu}+h^{\mu l}h_{l}^{~\nu}.
\end{equation}

For the evaluation of the temperature dependance of thermal conductivity in section {\bf 6} we consider the fluctuation gauge field $A_{\mu}$ also along with the metric perturbation given as:
\begin{eqnarray}
A_{\mu}=A^0_{\mu}+\mathcal{A}_{\mu},
\end{eqnarray}
where in this case we consider the coupling of gauge field fluctuation $\mathcal{A}_{\mu}$ with the background metric perturbation.

Assuming the momenta to be along the $x$-direction, the metric and the gauge field fluctuations can be written as the following Fourier decomposed form:
\begin{eqnarray}
h_{\mu\nu}( x,t,u)=\int \frac{d^{4}q}{(2\pi)^4 }e^{-iwt+iqx} h_{\mu\nu}(q,w,u)\nonumber\\
\mathcal{A_{\mu}}(x,t,u)=\int \frac{d^{4}q}{(2\pi)^4 }e^{-iwt+iqx}\mathcal{A_{\mu}}(q,w,u).
\end{eqnarray}
We will work in the gauges where $h_{u\mu}$ and $\mathcal{A}_{u}$ are both zero for all $\mu$ including $u$.

Based on the the spin of different metric perturbations under $SO(2)$ rotation in $(y,z)$ plane, the same can be classified into three types as follows:
\begin{enumerate}
\item[(i)]
 vector modes: $h_{x y}, h_{t y}\neq 0$ or $h_{x z},  h_{t z}\neq 0$, with all other $h_{\mu \nu}=0$.
\item[(ii)]
Scalar modes: $h_{x x}=h_{y y}=h_{z z}=h_{tt}\neq 0$, $h_{x t}\neq 0$, with all other $h_{\mu \nu}=0$.
\item[(iii)]
Tensor modes: $h_{y z}\neq 0$, with all other $h_{\mu \nu}=0$.
\end{enumerate}

The most important step to calculate the two point correlation function as discussed in section {\bf 2} is to solve the linearized equation of motion for the field in question. In this paper the EOMs for the scalar and vector type metric perturbations are all coupled to each other and hence they are not easy to solve. However following \cite{klebanov quasinormal} one can construct a particular combination of different perturbations which is gauge invariant and all the coupled EOMs can be replaced by a single equation involving the gauge invariant variable. This combination which is invariant under diffeomorphisms: $h_{\mu\nu}\rightarrow h_{\mu\nu}-\nabla_{(\mu}\xi_{\nu)}$ is given as \cite{klebanov quasinormal}:
\begin{eqnarray}
\label{Z-vector mode}
 {\rm Vector\ type}: Z_v = q H_{ty} + w H_{xy}~~~~~~~~~~~~~~~~~~~~~~~~~~~~~~~~~~~~~~~~~~~~~~~~
\end{eqnarray}
\begin{eqnarray}
\label{Z-scalar mode}
\nonumber {\rm Scalar\ type}: Z_s=-q^2(1-u^4)H_{tt}+2wqH_{xt}+w^2H_{xx}~~~~~~~~~~~~~~~~~~\\
~~~~~~~~~~~~~~~~~~~~~+q^2(1-u^4)\left(1+\frac{g_{xx}(-4u^3)}{g^{\prime}_{xx}(1-u^4)}-\frac{w^2}{q^2(1-u^4)}\right)H_{yy}
\end{eqnarray}
\begin{eqnarray}
\label{Z-tensor mode}
{\rm Tensor\ type}: Z_t=H_{yz},~~~~~~~~~~~~~~~~~~~~~~~~~~~~~~~~~~~~~~~~~~~~~~~~~~~~~~~~~~
\end{eqnarray}
where $H_{tt}=-g^{tt}h_{tt}$, $H_{xx}=g^{xx}h_{xx}$, $H_{yy}=g^{xx}h_{yy}$, $H_{xt}=g^{xx}h_{xt}$, $H_{xy}=g^{xx}h_{xy}$. The two second order differential equations corresponding to the EOMs of $Z_v$, $Z_s$ and $Z_t$ are solved and the required quasinormal modes are obtained by imposing Dirichlet  boundary conditions at $u=0$ \cite{klebanov quasinormal}.

\section{The Local $T^3$  is a $T^2$-Invariant sLag}

In \cite{transport-coefficients}, we had shown that the local $T^3$ of \cite{MQGP} used for constructing a delocalized SYZ type IIA mirror of the type IIB string theory construct of \cite{metrics} dual to large-$N$ thermal QCD, in the MQGP limit of \cite{MQGP}, is the $T^2$-invariant special Lagrangian (sLag) three-cycle of \cite{M.Ionel and M.Min-OO (2008)} of a deformed conifold. In this section, we show that  the same $T^3$ is also a the $T^2$-invariant sLag of \cite{M.Ionel and M.Min-OO (2008)} of a resolved conifold. The two results together, hence show the existence of a sLag in the MQGP limit in a predominantly resolved (resolution $>$ deformation) or deformed (deformation $>$ resolution) resolved warped deformed conifold of \cite{metrics}.

From \cite{M.Ionel and M.Min-OO (2008)} we note that the following is a $T^2$-invariant special Lagrangian three-cycle in a resolved conifold:
\begin{eqnarray}
\label{sLagRC-I}
& & \frac{K^\prime}{2}\left(|x|^2-|y|^2\right) + 4 a^2\frac{|\lambda_2|^2}{|\lambda_1|^2+|\lambda_2|^2} = c_1,\nonumber\\
& & \frac{K^\prime}{2}\left(|v|^2-|u|^2\right) + 4 a^2\frac{|\lambda_2|^2}{|\lambda_1|^2+|\lambda_2|^2} = c_2,\nonumber\\
& & \Im m\left(xy\right) = c_3,
\end{eqnarray}
wherein one uses the following complex structure for a resolved conifold \cite{Knauf+Gwyn[2007]}:
\begin{eqnarray}
\label{resolvedconifold-compl-struc}
  x & =&  \left ( 9 a^2 r^4 + r ^6 \right ) ^{1/4} e^{i/2(\psi-\phi_1-\phi_2)}\,\sin\frac{\theta_1}{2}\,\sin\frac{\theta_2}{2}  \nonumber\\
  y & =&  \left ( 9 a^2 r^4 + r ^6 \right ) ^{1/4} e^{i/2(\psi+\phi_1+\phi_2)}\,\cos\frac{\theta_1}{2}\,\cos\frac{\theta_2}{2}  \nonumber\\
  u & =&  \left ( 9 a^2 r^4 + r ^6 \right ) ^{1/4} e^{i/2(\psi+\phi_1-\phi_2)}\,\cos\frac{\theta_1}{2}\,\sin\frac{\theta_2}{2}  \nonumber\\
  v & =&  \left ( 9 a^2 r^4 + r ^6 \right ) ^{1/4} e^{i/2(\psi-\phi_1+\phi_2)}\,\sin\frac{\theta_1}{2}\,\cos\frac{\theta_2}{2}\,
\end{eqnarray}
$[\lambda_1:\lambda_2]$ being the homogeneous coordinates of the blown-up $\mathbb{CP}^1=S^2$; $\frac{\lambda_2}{\lambda_1}=\frac{x}{-u}=\frac{v}{-y}=-e^{-i\phi_1}\tan\frac{\theta_1}{2}$. In (\ref{sLagRC-I}),
$\gamma(r^2)\equiv r^2 K^\prime(r^2)= - 2 a^2 + 4 a^4 N^{-\frac{1}{3}}(r^2) + N^{\frac{1}{3}}(r^2)$, where $N(r^2)\equiv\frac{1}{2}\left(r^4 - 16 a^6 + \sqrt{r^8 - 32 a^6 r^4}\right)$. Defining $\rho\equiv \sqrt{\frac{3}{2}}\sqrt{\gamma}$, upon inversion yields: $r\approx \left(\frac{2}{3}\right)^{\frac{3}{4}}\left(3 a^2 + \rho^2\right)^{\frac{3}{4}}$ and $K^\prime(r^2)=\frac{\sqrt{\frac{3}{2}}\rho^2}{\left(3 a^2 + \rho^2\right)^{\frac{3}{2}}}$. The system of equations (\ref{sLagRC-I}) are solved in appendix A to yield (\ref{sLagRC-II}) - (\ref{sLagRC-IV}), which provides an embedding $\rho=\rho(\psi)$ and hence $\theta_{1,2}=\theta_{1,2}(\psi)$.   As $\theta_1, \theta_2\rightarrow0$ as $\frac{1}{N^{\frac{1}{5}}},\frac{1}{N^{\frac{3}{10}}}$ (whereat an explicit local $SU(3)$-structure of the type IIA mirror and an explicit local $G_2$-structure of the M-theory uplift was obtained in \cite{T_c+Torsion}) and in the UV-IR interpolating region/UV: $r\rightarrow {\cal R}_0 > r_0$, therefore in this domain of $(\theta_1,\theta_2,r)$ choose $c_{1,2}$ as given in (\ref{sLagRC-V}).
The large-$N$ [as given in (\ref{sLagRC-VI})] small-$\psi$ [as given in (\ref{sLagRC-VII}) for an appropriate
$\psi=\langle\psi\rangle$ determined by (\ref{sLagRC-VIII}) which is solved to yield (\ref{sLag-XI})] expansion, is discussed in appendix A.

Hence, using (\ref{sLagRC-VI}) of appendix A, the embedding of $\theta_2$ near $(\theta_1\sim\frac{1}{N^{\frac{1}{5}}},\theta_2\sim\frac{1}{N^{\frac{3}{10}}})$, and in the  UV-IR interpolating region/UV: $r={\cal R}_0\gg a, r_h)$ is given as:
\begin{eqnarray}
\label{sLag-XIII}
& & \theta_2\approx\sqrt{\frac{0.324 {\rho_0}^2 {\sin\psi}^{2/3}}{{c_3}^{2/3}}-\frac{0.096 {\rho_0}^6 {\sin\psi}^2}{{c_3}^2}+0.5}\nonumber\\
& & +\frac{a^2 \left(0.193
   {c_3}^{4/3} {\rho_0}^2 {\sin\psi}^{4/3}+0.096 {\rho_0}^6 {\sin\psi}^{8/3}\right)}{{c_3}^{8/3} \sqrt{\frac{0.324 {\rho_0}^2
   {\sin\psi}^{2/3}}{{c_3}^{2/3}}-\frac{0.096 {\rho_0}^6 {\sin\psi}^2}{{c_3}^2}+0.5}} + {\cal O}\left(a^3\right).
\end{eqnarray}
Similarly, using (\ref{sLagRC-XIV}) of appendix A, yields:
\begin{eqnarray}
\label{sLagRC-XV}
& &\hskip -0.6in \theta_1\approx\frac{a^2 \left(-2.89 {c_3}^{2/3} {\rho_0}^2 {\sin\psi}^{4/3}-8.25 {c_3}^{4/3} {\sin\psi}^{2/3}+2.31 {\rho_0}^4
   {\sin\psi}^2\right)}{{c_3}^2 \sqrt{\frac{52.49 {\rho_0}^2 {\sin\psi}^{2/3}}{{c_3}^{2/3}}-\frac{14.6969 {\rho_0}^6
   {\sin\psi}^2}{{c_3}^2}+81.}}\nonumber\\
   & & \hskip -0.6in +\sqrt{\frac{0.32 {\rho_0}^2 {\sin\psi}^{2/3}}{{c_3}^{2/3}}-\frac{0.091 {\rho_0}^6
   {\sin\psi}^2}{{c_3}^2}+0.5}+O\left(a^3\right).
\end{eqnarray}
As is evident from equations (\ref{sLagRC-VI}) and (\ref{sLagRC-XIV}) of appendix A, the numerical factors are rather cumbersome to be retained as such if one is interested in eventually numerically verifying that (\ref{sLagRC-I}) indeed satisfies (\ref{sLag-conditions}). This is the reason why the exact numerical factors in (\ref{sLagRC-VI}) and (\ref{sLagRC-XIV}) of appendix A have been replaced by corresponding decimals to arrive at(\ref{sLag-XIII}) and (\ref{sLagRC-XV}). It is for the same reason that decimals also appear in (\ref{J1}) - (\ref{sLagRC-XXIII}).

The fundamental two-form  is found to be \cite{Knauf+Gwyn[2007]}
\begin{eqnarray}
J &=& -\frac{\rho}{3}\,d\rho\wedge(d\psi+\cos\theta_1\,d\phi_1+\cos\theta_2\,d\phi_2)\nonumber\\
& & -\frac{\rho^2}{6}\,\sin\theta_1\,d\phi_1\wedge
d\theta_1- \frac{\rho^2+6a^2}{6}\, \sin\theta_2\,d\phi_2\wedge d\theta_2
\end{eqnarray}
and is closed, and the holomorphic three-form is \cite{Knauf+Gwyn[2007]}:
\begin{eqnarray}
\label{Omegares}
& & \hskip -0.5in \Omega = \frac{\rho(\rho^2+6a^2)}{6\sqrt{\rho^2+9a^2}}\,(\cos\psi-i\sin\psi)\,d\rho\wedge\Big[\sin\theta_1\,
  d\theta_2\wedge d\phi_1 - \sin\theta_2\,d\theta_1\wedge d\phi_2
  +i(d\theta_1\wedge d\theta_2-\sin\theta_1\sin\theta_2\,d\phi_1\wedge d\phi_2)\Big] \nonumber\\
&& \hskip -0.5in + \, \frac{\rho^2}{18}\,\sqrt{\rho^2+9a^2}\,(\cos\psi-i\sin\psi)\Big[d\theta_1\wedge
  d\theta_2\wedge(d\psi+\cos\theta_1\,d\phi_1+\cos\theta_2\,d\phi_2) -\sin\theta_1\sin\theta_2\,d\phi_1\wedge
  d\phi_2\wedge d\psi \nonumber\\
& & \hskip -0.5in \qquad -i(\sin\theta_1\,d\theta_2\wedge d\phi_1 - \sin\theta_2\,d\theta_1\wedge d\phi_2)\wedge d\psi
  -i(\sin\theta_1\cos\theta_2\,d\theta_2+\cos\theta_1\sin\theta_2\,d\theta_1)\wedge d\phi_1\wedge d\phi_2\Big]\,.
\end{eqnarray}
So, writing $c_1=\alpha_{c_1}\rho_0^2,$ and to LO in $N,\ d\phi_1=\frac{\beta_{\phi_1}dx}{\left(g_sN\right)^{\frac{1}{4}}\frac{1}{N^{\frac{1}{5}}}},\ d\phi_2=\frac{\beta_{\phi_2}dy}{\left(g_sN\right)^{\frac{1}{4}}\frac{1}{N^{\frac{3}{10}}}},\  d\psi=\frac{\beta_{\psi}dz}{\left(g_sN\right)^{\frac{1}{4}}}$ \cite{MQGP}, one obtains (\ref{sLagRC-XVII}) of Appendix A. This implies:
\begin{eqnarray}
\label{J1}
& & i^*J\approx \left(\frac{0.07\alpha_{c_1}^2\rho_0^2\beta_{\psi}\beta_{\phi_1}}{\sqrt{g_s}\sin\psi N^{\frac{3}{10}}} - \frac{0.007\rho_0^2\beta_{\psi}\beta_{\phi_1}}{\sin\psi\sqrt{g_s}N^{\frac{3}{10}}} - \frac{0.84\beta_{\psi}\beta_{\phi_1}}{N^{\frac{3}{10}}\sqrt{g_s}\sin^{\frac{5}{3}}\psi}\right)dz\wedge dx\nonumber\\
& & + \left(\frac{0.07\alpha_{c_1}^2\rho_0^2\beta_{\psi}\beta_{\phi_2}}{\sqrt{g_s}\sin\psi N^{\frac{3}{10}}} - \frac{0.03\rho_0^2\beta_{\psi}\beta_{\phi_2}}{\sin\psi\sqrt{g_s}} - \frac{0.84\beta_{\psi}\beta_{\phi_2}}{N^{\frac{3}{10}}\sqrt{g_s}\sin^{\frac{5}{3}}\psi}\right)dz\wedge dy.
\end{eqnarray}
Choosing: $\sin^{\frac{2}{3}}\langle\psi\rangle = \frac{\alpha}{\rho_0^2}$ and
\begin{equation}
\label{sLagRC-XVIII}
\alpha_{c_1}: \frac{0.07\alpha_{c_1}^2}{3} - 0.03 - \frac{0.84}{\alpha} = 0,
\end{equation}
i.e.,
\begin{equation}
\label{sLagRC-XIX}
\alpha_{c_1}=6.54\sqrt{0.03+\frac{0.84}{\alpha}},
\end{equation}
one obtains:
\begin{equation}
\label{sLagRC-XX}
i^*J\sim 0.02\frac{\beta_{\psi}\beta_{\phi_1}}{N^{\frac{3}{10}}}dz\wedge dx\approx0.
\end{equation}
Further, the three-forms relevant to evaluation of $i^*\Omega$ - using (\ref{Omegares}) - are collected in  (\ref{sLagRC-XXI}). Using (\ref{sLagRC-XIX}), choose $\alpha$:
\begin{equation}
\label{sLagRC-XXII}
\frac{1}{6}\left(\frac{2\alpha_{c_1}^2}{3^3} - \frac{2^{\frac{4}{3}}}{3\alpha}\right) +\frac{1}{36}\left( N^{\frac{1}{10}}0.21 + \frac{2^{\frac{1}{6}}}{3^{\frac{3}{2}}N^{\frac{1}{10}}}\right)\ll1,
\end{equation}
i.e.,
\begin{equation}
\label{sLagRC-XXIII}
0.016 +\frac{0.304}{\alpha }+0.006 \sqrt[10]{N}+\frac{0.006}{\sqrt[10]{N}}\ll1.
\end{equation}
Now, (\ref{sLagRC-XXIII}) will be satisfied by any $\alpha\gg1$ for a reasonably large $N$ but less than $10^{10}$.
Hence,
\begin{eqnarray}
\label{sLagRC-XXIV}
& & i^*J\approx 0;\nonumber\\
& & \Im m (i^*\Omega)=0,\nonumber\\
& & \Re e (i^*\Omega)\sim {\rm vol}\left(T^3\right),
\end{eqnarray}
implying thus: $\left.i^* J\right|_{RC/DC}\approx0, \Im m\left.\left( i^*\Omega\right)\right|_{RC/DC} \approx 0, \Re e\left.\left(i^*\Omega\right)\right|_{RC/DC}\sim{\rm volume \ form}\left(T^3(x,y,z)\right)$. Hence, if the resolved warped deformed conifold is predominantly either resolved (resolution $>$ deformation) or deformed (deformation $>$ resolution), the local $T^3$ of (\ref{xyz defs}) is the required sLag to effect the delocalized SYZ mirror of the type IIB background of \cite{metrics}, carried out in \cite{transport-coefficients}.

\section{Thermal (Electrical) Conductivity, Deviation from the Wiedemann-Franz Law and $D=1+1$ Luttinger Liquids up to LO in $N$}

In this section we compute the temperature dependance of thermal ($\kappa_T$) and electrical ($\sigma$) conductivities for a gauge theory at finite temperature and density, and hence explore deviation from the Wiedemann-Franz law. Remarkably, we find that the results qualitatively mimic those of a $D=1+1$ Luttinger liquid with impurities.

Finite temperature in the  gauge theory is effected by placing a black hole in the dual bulk gravitational background. To get the finite density in the boundary we consider the theory at non-zero chemical potential. The embedding of $N_f$ $D7$ branes in the background geometry introduces $N_f$ no of flavor fields, all in fundamental representation of the gauge group $U(N_f)$. The $U(1)_B$ subgroup of $U(N_f)$ is identified as the baryon number. Hence the $D7$ brane puts the boundary field theory at finite baryon density or equivalently at finite chemical potential $\mu_C$. In the supergravity description one have a $U(1)$ gauge field $A_{\mu}$ in the worldvolume of the $D7$ brane, dual to the current operator $j_{\mu}$ in the boundary. The nonzero time component $A_t$ of the gauge field has to be turned on to get a finite baryon density $<j_t>$ in the field theory. Here we will consider the $5d$ Einstein-Hilbert action and the $D7$ brane DBI action together, of course after integrating over the three angular directions of the later.

The $D7$ brane DBI action in presence of a $U(1)$ gauge field is given as:
\begin{equation}
\label{DBI}
S_{D7}= T_{D7}\int d^8\xi e^{-\Phi}\sqrt{-\det(g+B+\hat{F})}
\end{equation}
where $g$ is the induced metric on $D7$ brane and $\hat{F}$ is the gauge field strength with the only nonzero component given by $\hat{F_{rt}}=\frac{ce^{\Phi}}{\sqrt{c^2e^{2\Phi}+r^{9/2}}}$ \cite{T_c+Torsion}, where $\Phi$ is the dilaton and $c$ is a constant.

Now the finite chemical potential or equivalently the finite charge density will mix the heat (energy) current and the electric current together. According to the AdS/CFT correspondence for every operator in the boundary field theory, there is a bulk field in the dual gravity theory. Heat current is sourced by the energy momentum tensor $T_{\mu\nu}$ in the boundary and the corresponding field in the gravitational description is the bulk metric $g_{\mu\nu}$. Similarly, as already mentioned, the electric current sourced by the current operator $j_{\mu}$ corresponds to the $U(1)$ bulk gauge field $A_{\mu}$. Hence for the computation of thermal conductivity we consider the following linear fluctuations of both the background metric $g^{(0)}_{\mu\nu}$ and the gauge field $A^{(0)}_{\mu}$ as,\begin{equation}g_{\mu\nu}=g^{(0)}_{\mu\nu}+h_{\mu\nu}  ~~~~~~~~~~~~~~~~  A_{\mu}=A^{(0)}_{\mu}+\mathcal{A_{\mu}},\end{equation}
where $h_{\mu\nu}$ and $\mathcal{A_{\mu}}$ represents the metric and the gauge field fluctuations respectively.
Considering the $y$-component of the gauge field as the only perturbation, it can be shown that only the $(ty)$ and the $(xy)$ component of the metric gets perturbed.
Assuming that the above perturbations depends only on the radial coordinate $u$, time $t$ and spatial coordinate $x$, can be decomposed as the following way,
\begin{equation}
h_{ty}=g^{(0)}_{xx}H_{ty}(u)e^{-iwt+iqx} ~~~~~~~~~~~~~  h_{xy}=g^{(0)}_{xx}H_{xy}(u)e^{-iwt+iqx} ~~~~~~~~~~~~~  \mathcal{A}_{y}=\phi(u)e^{-iwt+iqx}
\end{equation}
Now including the above fluctuations in the DBI action, we perform the three angular integrations on $\phi_1$, $\phi_2$ and $\theta_2$. The integration over two of the three angular variable namely $\phi_1$ and $\phi_2$ gives constant factors.
To perform the $\theta_2$ integration, we first expand the DBI action in (\ref{DBI}) (Taking into account the fluctuations) up to quadratic order in fluctuating fields to get,
\begin{eqnarray}
\hskip -0.4in \sqrt{-\det(g+h+B+\hat{F}+F)}=\sqrt a_1 \left(1 + \frac{a_2 H^2_{ty}(u) + a_3 H_{ty}(u) \phi^{'}(u) + a_4 H^2_{xy}(u) + a_5 \phi^{2}(u) + a_6 \phi^{'}(u)^2}{2 a_1}\right).\nonumber\\
& &
\end{eqnarray}
where $h$ and $F$ represents the fluctuations of the two fields. The coefficients $a_1,a_2,a_3,a_4,a_5,a_6$ are given as,
\begin{eqnarray}
\label{a-coefficients}
& & a_1 = -\frac{\cot ^2(\frac{\theta_2}{2}) \csc ^4(\frac{\theta_2}{2})}{1296 \left(c^2 e^{2 \Phi }+r^{9/2}\right)}r^{9/2}\left\{\Biggl(r^3+2 (5 \mu_{\rm Ouyang} ^2-2 r^3) \cos
   \theta_2+14 \mu_{\rm Ouyang}^2+3 r^3 \cos 2 \theta_2\Biggr)\right\}\nonumber\\
& &  \times \Biggl\{\left(8 \mu_{\rm Ouyang}^2-4 r^3\right) \cos \theta_2+r^3 (\cos 2
   \theta_2+3)\Biggr\}\nonumber\\
   & &  a_2 = a_1\frac{r^{-\frac{1}{2}}\left(c^2 e^{2 \Phi }+r^{9/2}\right)}{\left(r^4-r_h^4\right)}e^{2i(qx-tw)}\nonumber\\
       & &  a_3 = 2a_1\frac{ ce^{\Phi}\sqrt{\left(c^2 e^{2 \Phi }+r^{9/2}\right)}}{r^{\frac{9}{2}}}e^{2i(qx-tw)}\nonumber\\
    & &       a_4 = a_1e^{2i(qx-tw)}\nonumber\\
    & & a_5 =4 a_1\frac{(g_s N\pi)}{r^{\frac{9}{2}}\left(r^4-r_h^4\right)} \left\{iw^2 c^2 e^{2 \Phi }+\sqrt{r} \left(iq^2 \left(r_h^4-r^4\right)+iw^2
   r^4\right)\right\}e^{2i(qx-tw)}\nonumber\\
   & & a_6 = a_1\frac{\left(c^2 e^{2 \Phi }+r^{9/2}\right)}{r^{\frac{9}{2}}}\left(1 - \frac{r_h^4}{r^4}\right)e^{2i(qx-tw)},
\end{eqnarray}
where the coordinate $r$ is related to $u$ as $u=\frac{r_h}{r}$. Upon changing the variable from $r$ to $u$ to the above mentioned variables one see that the coefficients $a_2,a_3,a_4,a_5,a_6$ each after the division by $a_1$ are independent of $\theta_2$ and only depends on $u$. The integration of $\sqrt{a_1}$ over $\theta_2$ gives some function of $u$ say $\mathcal{M}(u)$ given by,
\begin{eqnarray}
\mathcal{M}(u)=\sqrt{\mu_{\rm Ouyang}}\left(\frac{r_h}{u}\right)^{9/2}\sqrt{\frac{1}{c^2e^{2\Phi}+\left(\frac{r_h}{u}\right)^{9/2}}}
\end{eqnarray}
 In this way reducing the dimension from eight to five the DBI action takes the following form:
\begin{eqnarray}
\label{kinetic-fluctuations}
& & \hskip -0.3in S_{D7}=\left(\frac{a_{DBI}T_{D_7}}{g_s}\right)\int du~ d^4x~ \mathcal{M}(u)\left(1+\frac{a_2 H^2_{ty}(u) + a_3 H_{ty}(u) \phi^{'}(u) + a_4 H^2_{xy}(u) + a_5 \phi^{2}(u) + a_6 \phi^{'}(u)^2}{2 a_1}\right)\nonumber\\
& &
\end{eqnarray}
where $a_{DBI}$ includes all the constant terms resulting after the angular integrations; $T_{D7}$ is the tension on the $D7$ brane. We will henceforth be working in a hydrodynamical approximation wherein we will approximate the plane-wave exponentials by unity.

Finally taking into account the Einstein-Hilbert action given as,
\begin{eqnarray}
S_{EH}=a_{EH}\int du d^4x\sqrt{-g_{(5)}}(R-2\lambda),
\end{eqnarray}
where $g_{(5)}$ is the determinant of the $5d$ metric, the total action is given by:
$S_{tot}=S_{\rm EH}+S_{D7}$.

The type $IIB$ metric satisfying the above action $S_{tot}$ has the form:
\begin{equation}
ds^2=g_{tt}dt^2+g_{xx}(dx^2+dy^2+dz^2)+g_{uu}du^2,
\end{equation}
where  the different background metric components, in the UV (as the gauge fluctuation will be solved for, near the UV $u=0$) and to LO in $N$, are given as,
\begin{eqnarray}
\nonumber g_{tt}=\frac{\left(u^4-1\right)r_h^2}{2 u^2 \sqrt{\pi g_s   N }}\\
\nonumber g_{xx}=g_{yy}=g_{zz}=\frac{r_h^2}{2 u^2 \sqrt{\pi g_s   N }}\\
g_{uu}=\frac{2 u^2 \sqrt{\pi g_s   N }}{\left(1-u^4\right)r_h^2}.
\end{eqnarray}
Now from the total action defined above, we can write down the EOMs in the hydrodynamical limit for $H_{ty}$, $H_{xy}$~and $\phi$ and they are given as:
\vskip 0.1in
\noindent {\bf $H_{ty}(u)$ EOM:}
\begin{eqnarray}
& & a_{EH}\sqrt{-g_{(5)}}\left(\mathcal{R}^{(1)}_{ty}-\frac{2}{3}\lambda g_{xx}H_{ty}(u)\right)+\left(\frac{a_{DBI}T_{D_7}}{g_s}\right)\mathcal{M}(u)\left(\frac{a_2(u)}{a_1(u)}H_{ty}(u)
+ \frac{a_3(u)}{2a_1(u)}\phi^{'}(u)\right)=0;\nonumber\\
& &
\end{eqnarray}
{\bf $H_{xy}(u)$ EOM:}
\begin{eqnarray}
& & a_{EH}\sqrt{-g_{(5)}}\left(\mathcal{R}^{(1)}_{xy}-\frac{2}{3}\lambda g_{xx}H_{xy}(u)\right)+\left(\frac{a_{DBI}T_{D_7}}{g_s}\right)\mathcal{M}(u)\left(\frac{a_4}{a_1(u)}H_{xy}(u)\right)=0;\nonumber\\
& &
\end{eqnarray}
{\bf $\phi(u)$ EOM:}
\begin{eqnarray}
& & \hskip -0.3in \frac{d}{du}\left(\frac{a_3(u)}{2 a_1(u)}\mathcal{M}(u)H_{ty}(u)\right) + \frac{d}{du}\left(\frac{a_6(u)}{a_1(u)}\mathcal{M}(u)\right)\phi^{'}(u)
+\left(\frac{a_6(u)}{a_1(u)}\mathcal{M}(u)\right)\phi^{''}(u)-\left(\frac{a_5(u)}{a_1(u)}\mathcal{M}(u)\right)\phi(u)=0,
\nonumber\\
& &
\end{eqnarray}
where $\mathcal{R}^{(1)}_{\mu\nu}$ is the linear ordered perturbation of the Ricci tensor. Now substituting the exact form of $\mathcal{M}(u)$ as well as all of the six coefficients $a_1(u), a_2(u), a_3(u), a_4, a_5(u), a_6(u)$, the above three equations regarding $H_{ty},H_{xy}$ and $\phi$ can be rewritten as (\ref{EOM-Hty}) - (\ref{EOM-phi}) in appendix B which also contains their solutions.

As the pre-factor multiplying $\frac{\phi'(u)}{\phi(u)}$ from (\ref{kinetic-fluctuations}), the $A(u)$ in (\ref{F}) - the coefficient of the kinetic term of $\phi(u)$ - that will appear in the current-current correlator is $\left(\frac{\sqrt{\mu}r_h^{\frac{13}{4}}}{72u^{\frac{17}{4}}}\right)$, to obtain
   a finite $\left\{\lim_{{u}\rightarrow0}\frac{1}{u^{\frac{17}{4}}}\frac{\phi'(u)}{\phi(u)}\right\}$, one needs $\phi(u)\sim e^{{\rm constant}\ u^{\frac{21}{4}}}$. Expanding (\ref{phiu0solution_i}) about $u=0$:
\begin{eqnarray}
\label{large-N-small-u-rh_TN-small-w}
 \phi(u\sim0;q=0) & =  & \frac{\left(i g_s N \pi\right)^{7/8}\omega^{7/4}c_2^\Phi u^{\frac{21}{4}}}{33^{\frac{3}{4}}r_h^{\frac{7}{2}}} -\frac{4cg_sc_1^\Phi\Gamma(\frac{13}{24})u^{\frac{13}{4}}}{63r_h^{\frac{9}{4}}\Gamma(\frac{37}{24})}+\frac{4cg_sc_1^\Phi\Gamma(-\frac{1}{3})u^{\frac{13}{4}}}{63r_h^{\frac{9}{4}}\Gamma(\frac{2}{3})}
\nonumber\\
& & + c_1^\Phi + \frac{c}{r_h^{\frac{9}{4}}}{\cal O}(u^6).
\end{eqnarray}
Now, in terms of a dimensionless ratio: $\kappa\equiv\frac{C}{r_h^{\frac{9}{4}}}$ and choosing $C$ to be $m_{\rho}$ and $r_h$ in units of GeV implying $\kappa\ll1$ {\cite{T_c+Torsion}}. Therefore,
\begin{eqnarray}
\label{large-N-small-u-rh_TN-small-w k-small}
& & \phi(u\sim0;q=0)=\frac{(0.08+0.39\ i)\left(g_s N\right)^{\frac{7}{8}}w^{\frac{7}{4}}c_2^\Phi u^{\frac{21}{4}}}{r_h^{\frac{7}{2}}}+c_1^\Phi+\frac{c}{r_h^{\frac{9}{4}}}{\cal O}(u^6)\nonumber \\
& & \approx c_1^\Phi e^{\frac{(0.08+0.39\ i)g_s^{\frac{7}{8}}N^{\frac{7}{8}}w^{\frac{7}{4}}c_2^\Phi u^{\frac{21}{4}}}{c_1r_h^{\frac{7}{2}}}}.
\end{eqnarray}
Analogous to (the reason given in) Sec. {\bf 3}, in (\ref{large-N-small-u-rh_TN-small-w k-small}), we use decimals. At this point we require to calculate some of the thermodynamic parameters like pressure, energy density, entropy density etc. In particular, pressure and energy density follows from the thermodynamic relations as given by $s=\frac{\partial P}{\partial T}$ and $\epsilon=-P+Ts+\mu_{C}n_q$, where $s$ is called the entropy density and is given as,
\begin{eqnarray}
& &  s = \mathcal{O}(1)r_h^3 = \mathcal{O}(1)\pi^3\left(4\pi g_s N\right)^{3/2}T^3.
\end{eqnarray}
Now the density of Gibbs potential $\Omega$ which is equal to the pressure with a minus sign can be used to find the charge density $n_q$ using the relation $n_q=\frac{\partial\Omega}{\partial\mu_C}$, where $\mu_c$ being the chemical potential is given by
\begin{eqnarray}
\label{mu_C_dimensionless}
& & \mu_C = \frac{\left(2 {\kappa} {g_s}\right)^{\frac{4}{9}} {r_h} \Gamma \left(\frac{5}{18}\right) \Gamma \left(\frac{11}{9}\right)}{\sqrt[18]{\pi } (2
   \pi -{g_s} {N_f} \log |\mu_{\rm Ouyang}|)^{4/9}}-{r_h}\ _2F_1 \left(\frac{11}{9};-\frac{(2 \pi -{g_s} {N_f} \log |\mu_{\rm Ouyang}| )^2}{4
   {\kappa}^2 {g_s}^2 \pi ^2}\right)\nonumber\\
   & & = \frac{36 \pi  {\kappa} {g_s} {r_h} \Gamma \left(\frac{11}{9}\right)}{5 \Gamma \left(\frac{2}{9}\right) (2 \pi -{g_s} {N_f}
   \log |\mu_{\rm Ouyang}| )} + {\cal O}\left(\kappa^{\frac{19}{9}}\right),
\end{eqnarray}
from which we get
\begin{equation}
\label{T-func-mu}
 T=\left(\frac{8}{5}\right)^{4/5}\left(\frac{g_s^{\frac{3}{10}}C^{\frac{4}{5}}}{\left(2\pi - g_s N_f \log |\mu_{\rm Ouyang}|\right)^{4/5}}\right)\left(\frac{\mu^{-\frac{4}{5}}_C}{2\pi^{\frac{7}{10}}\sqrt{N}}\right).
 \end{equation}
Substituting the above result for $T$ in the expression for Gibbs potential and differentiating w.r.t $\mu_C$ we get charge density as
\begin{eqnarray}
n_q=\left(\frac{8}{5}\right)^{\frac{16}{5}}\left(\frac{2}{5}\right)\left(\frac{g_s^{\frac{27}{10}}\pi^{\frac{17}{10}}C^{\frac{16}{5}}\mu_C^{-\frac{21}{5}}}{\sqrt{N}\left(2\pi - g_s N_f \log |\mu_{\rm Ouyang}|\right)^{16/5}}\right).
\end{eqnarray}
Hence,
\begin{eqnarray}
\label{kappaT}
& & \hskip -0.3in \kappa_T = \frac{\left(\epsilon+P\right)^2\sigma}{n_q^2T} = -\left(\frac{\epsilon+P}{n_q}\right)^2\left(\frac{\sqrt{|\mu_{\rm Ouyang}|}r_h^{\frac{13}{4}}}{72 T u^{\frac{17}{4}}}\right)\lim_{{\omega}\rightarrow0}\frac{1}{{\omega}}
\Im m\left.\frac{\phi^\prime(u)}{\phi(u)}\right|_{u=0}\nonumber\\
&& \hskip -0.3in = \frac{9}{200\sqrt{2}}\frac{g_s^{\frac{3}{4}}C^2}{ N^{\frac{5}{4}}\pi^{\frac{7}{4}}\left(2\pi - g_s N_f \log |\mu_{\rm Ouyang}|\right)^2T^{\frac{7}{2}}}\sqrt{|\mu_{\rm Ouyang}|}\left(T\pi \sqrt{4\pi g_s N}\right)^{13/4}\lim_{{\omega}\rightarrow0}\frac{(0.39i)g_s^{7/8}N^{7/8}{\omega}^{3/4}c_2^\Phi}{\left(T\pi\sqrt{4\pi g_s N}\right)^{7/2}},\nonumber\\
& &
\end{eqnarray}
which for $c_2^\Phi\sim -i {\omega}^{-\frac{3}{4}}$ implies:
\begin{eqnarray}
\label{WF}
& & \sigma=(0.39)\frac{\sqrt{|\mu_{\rm Ouyang}|}\left(g_s N\right)^{\frac{3}{4}}}{2^{\frac{1}{4}}T^{\frac{1}{4}}\pi^{\frac{3}{8}}},\nonumber\\
& &  \kappa_T = \frac{9\times 0.39}{200\times2^{3/4}}\frac{\sqrt{|\mu_{\rm Ouyang}|}g_s^{\frac{3}{2}}C^2}{\sqrt{N}\pi^{\frac{17}{8}}T^{\frac{15}{4}}\left(2\pi - g_s N_f \log |\mu_{\rm Ouyang}|\right)^2};\nonumber\\
& &  {\rm Wiedemann-Franz\ law}: \frac{\kappa_T}{\sigma T}=\frac{9}{200\sqrt{2}}\frac{g_s^{\frac{3}{4}}C^2}{N^{\frac{5}{4}}\pi^{\frac{7}{2}}\left(2\pi - g_s N_f \log |\mu_{\rm Ouyang}|\right)^2T^{\frac{9}{2}}}.
\end{eqnarray}

(a) Assuming the Ouyang embedding parameter to depend on the temperature via the horizon radius as $|\mu_{\rm Ouyang}|\sim r_h^\alpha,\ \alpha\leq0$. Then, the temperature dependence of $\sigma, \kappa_T$ and the temperature dependences of the Wiedemann-Franz law in (\ref{WF}), upon comparison with Table 2  of \cite{WF}, qualitatively mimic a $D=1+1$ Luttinger liquid with impurities/doping (close to `$\frac{1}{3}$-filling') in the following sense.

With
\begin{itemize}
\item
$v_i, K_i, i=c$(harge), $s$(pin) being the parameters appearing in the Luttinger liquid Hamiltonian as $\sum_{i=c,s}v_i\left[K_i \left(\partial_x\theta_i\right)^2 + \frac{1}{K_i}\left(\partial_x\phi_i\right)^2\right]$ wherein the spin ($s$) and charge ($c$) densities are $\phi_{s,c}$ and their canonically conjugate fields are $\partial_x\theta_i$,
\item
 $n_s=0,1$ for even and odd $n_c$ respectively where $n_{c,s}$ along with $g,a$ appear in the Umklapp scattering Hamiltonian $\frac{g}{\left(2\pi a\right)^{n_c}}\int \left(e^{i\sqrt{2}\left(n_c\phi_c + n_s\phi_s\right) - i \Delta k x} + {\rm h.c.}\right)$,
\item
 $D$ as a parameter appearing in the two-point correlation function of the impurity field $\eta(x)$ via $\langle \eta(x)\eta(x^\prime)\rangle = D \delta(x-x^\prime)$ with $\eta(x)$ appearing in the back-scattering Hamiltonian due to disorder $\frac{1}{\pi a}\int dx \eta(x) \left[e^{i\sqrt{2}\phi_c}\cos\left(\sqrt{2}\phi_s\right) + {\rm h.c.}\right]$,
\end{itemize}
  the authors of \cite{WF} define the following dimensionless parameters: $\tilde{D} \equiv\frac{\rm Impurity\ scattering\ rate}{\rm Umklapp\ scattering\ rate}=\frac{D a^{2n_c-3}}{g^2\left(\frac{a T}{v_c}\right)^\gamma}, \tilde{\delta}\equiv\frac{\delta}{\tilde{D}^{\frac{1}{\gamma}}}$ where
$\delta\equiv\frac{v_c\Delta k}{\pi T}, \gamma\equiv (n_c^2-1)K_c + (n_s^2-1)K_s - 1$ and dimensionless temperature:
$\tilde{T}\equiv\frac{T}{T_D}$ where $T_D\equiv\frac{v_c}{a}\left(\frac{D a^{2n_c-3}}{g^2}\right)^{\frac{1}{\gamma}}$. One then notes that for $\tilde{\delta}=10, 20$ and for $T>T_D$, $\frac{d\sigma}{dT}, \frac{d\kappa_T}{dT}, \frac{d\left(\frac{\kappa_T}{T \sigma}\right)}{dT} < 0$ which is also reflected in (\ref{WF}).  In $\alpha^\prime=1$-units $[T] = [C^{\frac{4}{9}}]$, where $[..]$ denotes that canonical dimension.  To ensure a constant finite value of
$\frac{\kappa_T}{T\sigma}$ for small temperatures as per \cite{WF}, we assume, in the MQGP limit,  for $T :  \frac{T}{C^{\frac{4}{9}}}<1$, i.e., $T\sim C^{\frac{4}{9}}\epsilon^{\alpha_T>0}, 0<\epsilon<1$ and $N\sim\beta_N\epsilon^{-\alpha_N}$, so that if $0<\frac{9\alpha_T}{2} - \frac{5\alpha_N}{4}\ll1$ then $\lim_{T\rightarrow0}\left(\frac{\kappa_T}{T\sigma}\sim\frac{g_s^{\frac{3}{4}}C^2}{N^{\frac{5}{4}}T^{\frac{9}{2}}\left(2\pi - g_s N_f\log\mu\right)^2}\right)\sim\frac{g_s^{\frac{3}{4}}}{\epsilon^{-\frac{5\alpha_N}{4} + \frac{9\alpha_T}{2}}\left(2\pi - g_sN_f\left\{\frac{\alpha_N}{4} + \alpha_T\right\}\log\epsilon + \frac{g_sN_f}{4}\log(\beta_Ng_s)\right)^2}\neq0$.

(b) For $\alpha$(figuring in $|\mu_{\rm Ouyang}|\sim r_h^\alpha$)$>0$, interestingly for the specific choice of
$\alpha=\frac{5}{2}$ one reproduces the large-$T$ (as $T>C^{\frac{4}{9}}=m_\rho=760 MeV$(\cite{T_c+Torsion})$>T_c=175 MeV$, is considered large) linear behavior of DC electrical conductivity $\sigma\sim T$ characteristic of most strongly coupled gauge theories with a five-dimensional gravity dual with a black hole \cite{SJain_sigma+kappa}. As $\frac{C^2}{T^{\frac{9}{2}}}$ is dimensionless, this yields dimensionally $\kappa_T\sim ({\rm temperature})^2$, though $\kappa_T\sim T^{\frac{5}{2}}$ in the aforementioned large-$T$ limit.

\section{Scalar Metric Perturbation Modes and Speed of Sound in MQGP Limit}

In this section, by considering scalar modes of metric perturbations, we will evaluate the speed of sound, first up to leading order in $N$ four ways: (i) (subsection {\bf 5.1.1}) the poles appearing in the common denominator of the solutions to the individual scalar modes of metric perturbations (the pure gauge solutions and the incoming-wave solutions); (ii) (subsection {\bf 5.1.2}) the poles appearing in the coefficient of the asymptotic value of the square of the time-time component of the scalar metric perturbation in the on-shell surface action; (iii) (subsection {\bf 5.2.1}) the dispersion relation obtained via a Dirichlet boundary condition imposed on an appropriate single gauge-invariant metric perturbation - using the prescription of \cite{klebanov quasinormal} - at the asymptotic boundary; (iv) (subsection {\bf 5.2.2}) the poles appearing in the coefficient of the asymptotic value of the square of the time-time component of the scalar metric perturbation in the on-shell surface action written out in terms of the same single gauge-invariant metric perturbation. The third approach is then extended to include the non-conformal corrections to the metric and obtain an estimate of the corrections to $v_s$ up to NLO in $N$.

Having reduced the  $D=11$ metric as given in (\ref{Mtheory met}) to $M_5(\mathbb{R}^{1,3},u)$, up to leading order in $N$ and considering the non-zero scalar modes of metric perturbations defined in subsection {\bf 2.4}, we get a set of seven differential equations from the Einstein's equation. Defining the dimensionless energy and momentum,
\begin{eqnarray}
 \label{four momentum}
 \omega_3=\frac{w}{\pi T},\ q_3=\frac{q}{\pi T},
 \end{eqnarray}
the set of seven equations are given as:
 \begin{eqnarray}
 \label{7scalar_EOMs}
& &  H_{tt}^{\prime\prime} + \frac{1}{u} \left(-\frac{6}{g} + 5\right) H_{tt}^\prime +
  H_s^{\prime\prime} + \frac{1}{u} \left(-\frac{2}{g} + 1\right) H_s^\prime = 0,\nonumber\\
    & & H_{tt}^{\prime\prime} + \frac{2}{u} \left(-\frac{3}{g} + 1\right) H_{tt}^\prime + \frac{1}{u} \left(-\frac{2}{g} + 1\right) H_s^\prime - \frac{q_3^2}{g} H_{tt}
 + \frac{\omega_3^2}{g^2} H_s + 2 \frac{q_3 \omega_3}{g^2} H_{{x}t} = 0,\nonumber\\
& &   H_s^{\prime\prime} - \frac{3}{u} H_{tt}^\prime - \frac{2}{u} \left(1 + \frac{2}{g}\right) H_s^\prime - \frac{q_3^2}{g} H_{tt} + \frac{\omega_3^2}{g^2} H_s -\frac{4q_3^2}{g} H_{{y}{y}} +\frac{2 \omega_3 q_3}{g^2} H_{{x}t} = 0,\nonumber\\
& &  H_{{y}{y}}^{\prime\prime} - \frac{H_{tt}^\prime}{u} - \frac{H_s^\prime}{u} + \frac{1}{u} \left(-\frac{4}{g} + 1\right) H_{{y}{y}}^\prime + \frac{1}{g^2} \left(\omega_3^2 - g q_3^2\right) H_{{y}{y}} =  0,\nonumber\\
& &  H_{{x}t}^{\prime\prime} - \frac{3}{u} H_{{x}t}^\prime + \frac{2 q_3 \omega_3}{g} H_{{y}{y}} = 0,\nonumber\\
& &  q_3 \left(-g H_{tt}^\prime + 2 u^3 H_{tt}\right) - 2 q_3 g H_{{y}{y}}^\prime +  \omega_3 H_{{x}t}^\prime= 0,\nonumber\\
& &   \omega_3 \left(g H_s^\prime + 2 u^3 H_s\right) +
  q_3 \left(g H_{{x}t}^\prime + 4 u^3 H_{{x}t}\right) = 0
\end{eqnarray}
where we define $H_{tt}=\left(\frac{g_s^{2/3}u^2L^2}{r_h^2g}\right)h_{tt}
$, $H_{{x}{x}}=\left(\frac{g_s^{2/3}u^2L^2}{r_h^2}\right)h_{{x}{x}}$, $H_{{y}{y}}=H_{{z}{z}}=\left(\frac{g_s^{2/3}u^2L^2}{r_h^2g}\right)h_{{y}{y}}$, and $H_s=H_{{x}{x}}+2H_{{y}{y}}$.
The above system of equations can be reduced to the following linearly independent set of four equations
\begin{eqnarray}
\label{4scalar_EOMs}
 H^\prime_{{x}{x}} &=&\frac{3\omega_3^2-2q_3^2u^4}{q_3^2\left(u^4-3\right)}H^\prime_{tt}+\frac{2 u \left(q_3^4 \left(1-u^4\right)^2-\omega_3^2 \left(-2 u^6+6 u^2+\omega_3^2\right)\right)}{q_3^2 \left(u^4-3\right)
   \left(1-u^4\right)^2}H_{{y}{y}}\nonumber\\
    & & +\frac{u \omega_3^2 \left(q_3^2 \left(u^4-1\right)+2 u^6-6 u^2-\omega_3^2\right)}{q_3^2 \left(1-u^4\right)^2 \left(u^4-3\right)}H_{{x}{x}}+\frac{2 u w3 \left(q3^2 \left(u^4-1\right)+2 u^6-6 u^2-w3^2\right)}{q3 \left(1-u^4\right)^2 \left(u^4-3\right)}H_{{x}t}\nonumber\\
     & & +\frac{u \left(q_3^2 \left(u^4-1\right)+2 u^6-6 u^2-\omega_3^2\right)}{\left(u^4-3\right) \left(u^4-1\right)}H_{tt}\nonumber\\
     H^\prime_{{y}{y}}&=&-\frac{q3^2 \left(u^4-3\right)+3 \omega_3^2}{2 q_3^2 \left(u^4-3\right)}H^\prime_{tt}+\frac{u \omega_3^2 \left(q3^2 \left(u^4-1\right)-2 u^6+6 u^2+\omega_3^2\right)}{q_3^2 \left(1-u^4\right)^2 \left(u^4-3\right)}H_{{y}{y}}\nonumber\\
     & & +\frac{u \omega_3^2 \left(-2 u^6+6 u^2+\omega_3^2\right)}{2 q_3^2 \left(1-u^4\right)^2 \left(u^4-3\right)}H_{{x}{x}}+\frac{u \omega_3 \left(-2 u^6+6 u^2+\omega_3^2\right)}{q_3 \left(1-u^4\right)^2 \left(u^4-3\right)}H_{{x}t}\nonumber\\
      & & +\frac{u \left(q_3^2 \left(u^4-1\right)+2 u^6-6 u^2-\omega_3^2\right)}{\left(u^4-3\right) \left(u^4-1\right)}H_{tt}\nonumber\\
       H_{{x}t}^\prime&=&\frac{3 \left(u^4-1\right) \omega_3}{q_3 \left(u^4-3\right)}H_{tt}^\prime-\frac{2 u \omega_3 \left(q_3^2 \left(u^4-1\right)-2 u^6+6 u^2+\omega_3^2\right)}{q_3 \left(u^4-3\right) \left(u^4-1\right)}H_{{y}{y}}\nonumber\\
        & & -\frac{u \omega_3 \left(-2 u^6+6 u^2+\omega_3^2\right)}{q_3 \left(u^4-3\right) \left(u^4-1\right)}H_{{x}{x}}+\frac{2 u \left(2 u^6-6 u^2-\omega_3^2\right)}{\left(u^4-3\right) \left(u^4-1\right)}H_{{x}t}\nonumber\\
         & & -\frac{u q_3  \omega_3}{u^4-3}H_{tt}\nonumber\\
          H_{tt}^{\prime\prime}&=&\frac{u^8+2 u^4+9}{u \left(u^4-3\right) \left(u^4-1\right)}H_{tt}^\prime-\frac{2 \left(q_3^2 \left(u^4+1\right)+2 \omega_3^2\right)}{\left(u^4-3\right) \left(u^4-1\right)}H_{{y}{y}}\nonumber\\  & & -\frac{2 \omega_3^2}{\left(u^4-3\right) \left(u^4-1\right)}H_{{x}{x}}-\frac{4 q_3 \omega_3}{\left(u^4-3\right) \left(u^4-1\right)}H_{{x}t} -\frac{2 q_3^2}{u^4-3}H_{tt}.
\end{eqnarray}
To solve the system of equation (\ref{4scalar_EOMs}) we look for the behavior of the solution near $u=1$. Hence for time being we reconsider equation (\ref{7scalar_EOMs}) and write them as the following system of six first order differential equations
\begin{eqnarray}
& &  H^\prime_{tt}=\frac{1}{g}P_{tt}\nonumber\\
& &  H^\prime_{yy}=-\frac{1}{2 g}P_{tt}+\frac{  u^3}{ g}H_{tt}+\frac{ \omega_3}{2 q_3g }P_{xt}\nonumber\\
& &  H^\prime_{s} =-\frac{2 u^3 }{g}H_s-\frac{4 q_3 u^3}{w_{3} g}H_{xt}-\frac{q_3 }{\omega_3}P_{xt}\nonumber\\
& &  H^\prime_{xt}=P_{xt}\nonumber\\
& &  P^\prime_{xt}=\frac{3}{u}P_{xt}-\frac{2q_3 \omega_3 }{g}H_{yy}\nonumber\\
& &  P^\prime_{xt}=-\frac{2(u^4-2)}{u g}P_{tt}+q_3^2H_{tt}-\frac{q_3(u^4+1)}{u \omega_3}P_{xt}-\frac{2u^2+2u^6+\omega_3^2}{g}
 \left(H_s+\frac{2q_3}{\omega_3}H_{xt}\right).
\end{eqnarray}
In matrix form the above equation can be written as
\begin{eqnarray}
\label{one}
X^\prime=A(u)X
\end{eqnarray}
where $A$ is a $6\times6$ matrix and is singular for all values of $u$. Equation (\ref{one}) can be solved by substituting the ansatz $X=(1-u)^r F(u)$ into the same, where the exponent $r$ can be evaluated from the eigenvalues of the matrix $(1-u)A(u)$ near $u=1$. They are given by $r_1=r_2=0,r_3=-1/2,r_4=i\omega_3/4,r_5=-i\omega_3/4$ and $r_6=1/2$. Two of the eigenvalues namely $r=\mp i\omega_3/4$ represent the incoming/outgiong wave.

\subsection{The Longer Route up to Leading Order in $N$ - Via Solutions of EOMs}

In this subsection, we describe the evaluation of $v_s$, first from the solutions to the EOMs for the scalar metric perturbation modes and then putting the same result on a firmer footing, from a two-point correlation function of energy momentum tensor: $\langle T_{00}T_{00}\rangle$. We limit ourselves, in this subsection, to the leading order in $N$.

\subsubsection{From the Pole Structure of Solutions to $H_{ab}(u)$}

Based on \cite{PSS-scalar}, we give below a discussion on three gauge transformations that preserve $h_{\mu u}=0$, for the black $M3$-brane metric (\ref{Mtheory met}) having integrated out the $M_6$ in the (asymptotic) $AdS_5\times M_6$ in the MQGP limit of \cite{MQGP}. This is then utilized to obtain solutions to the scalar metric perturbation modes' equations of motion (\ref{7scalar_EOMs}) near $u=0$ and thereafter the speed of sound. We verify the result for the speed of sound by also calculating the same from an two-point energy-momentum correlation function.

Demanding that infinitesimal diffeomorphism: $x^\mu\rightarrow x^\mu + \xi^\mu, g_{\mu\nu}\rightarrow g_{\mu\nu} - \nabla_{(\mu}\xi_{\nu)}$ preserves the gauge condition $h_{\mu u}=0$ implies imposing \cite{PSS-scalar}:
\begin{equation}
\partial_{(\mu}\xi_{u)} - 2 \Gamma^\rho_{\mu u}\xi_\rho = 0,
\end{equation}
wherein $\Gamma^\rho_{\mu u}$ is calculated w.r.t. $g_{\mu\nu} = g_{\mu\nu}^{(0)} + h_{\mu\nu}$. There are three residual gauge transformations under which the system of differential equations (\ref{7scalar_EOMs}) remains invariant. They are given in (\ref{GTI}), (\ref{GTII}) and (\ref{GT-III}). Choosing $C_{u}, \tilde{C}_{{x},u}:  \left(C_{u},\frac{\tilde{C}_{t,{x}}}{i}\right)\frac{g_s^{\frac{2}{3}}}{L^2}=1$, the non-zero pure gauge solutions gauge equivalent to $H_{ab}=0$ ($H_{ab}=0, \xi_a=0$), near $u=0$, are given by:
\begin{eqnarray}
\label{H_ab-0_u-0}
& & H_{xx}^{(I)}(0) = - 2 q_3, H_{xx}^{(III)}(0) = 2;\nonumber\\
& & H_{tt}^{(II)}(0) = 2 \omega_3;\nonumber\\
& & H_{xt}^{(I)}(0) = \omega_3, H_{xt}^{(II)}(0) = q_3.
\end{eqnarray}

Writing $H^{\rm inc}_{ab}(u)$ as the incoming solution to the differential equations, the general solution can be written as the following form,
\begin{equation}
\label{Hcomponents}
H_{ab}(u) = a H^{(I)}_{ab}(u) + b H^{(II)}_{ab}(u) + c H^{(III)}_{ab}(u) + d H^{\rm inc}_{ab}(u).
\end{equation}
To determine $H^{\rm inc}_{ab}(u)$, we Solve (\ref{7scalar_EOMs}) near the horizon $u=1$ (this enables solving the fourth, fifth and sixth equations of (\ref{7scalar_EOMs}) independent of the first, second, third and seventh equations), where we have already shown that the same is a regular singular point with exponent of the indicial equation corresponding to the incoming solution given by $-\frac{i\omega_3}{4}$, implying that $H_{ab}^{\rm inc}(u)=(1-u)^{-\frac{i\omega_3}{4}}{\cal H}_{ab}(u)$.  Making double perturbative ansatze:
${\cal H}_{ab}(u) = \sum_{m=0}^\infty\sum_{n=0}^\infty {\cal H}_{ab}^{(m,n)}(u)q_3^m\omega_3^n$, one obtains near u=0 the solutions given in (\ref{hab}).

Upon using $H_{tt}(0)=H_{t}^{(0)}, H_{xt}(0)=H_{xt}^{(0)}, H_s(0)=H_s^{(0)}$ and solving for $a, b, c$ and $d$, the following is the common denominator:
\begin{eqnarray}
\label{pole-speed_s}
 \Omega(\omega_3,q_3) & \equiv & \alpha_{yy}^{(0,0)} + \alpha_{yy}^{(1,0)} q_3 + C_{1yy}^{(2,0)} q_3^2 +
 \alpha_{yy}^{(1,0)} \omega_3 + \left(-\frac{i}{4} + C_{2yy}^{(1,1)} - \frac{2}{9} C_{1yy}^{(1,1)} e^3\right) q_3 \omega_3
 \nonumber\\
& & + \left(C_{1yy}^{(0,2)} +
     C_{2yy}^{(0,2)} + \frac{i}{4} \Sigma_{2yy}^{(0,1)}\right) \omega_3^2,
\end{eqnarray}
where $\alpha_{yy}^{(m,n)}, C_{ayy}^{(m,n)}, a,b=1,2$ are constants appearing in the solutions to ${\cal H}_{ab}^{(m,n)}(u)$ in (\ref{hab}).
Now, (\ref{pole-speed_s}) can be solved for $\omega_3$ and the solution is given in (\ref{pole-speed_s_ii}) in appendix E.
 Assuming $\alpha_{yy}^{(0,0)}\ll1, |\Sigma_{2\ yy}^{(0,1)}|\gg1(i \Sigma_{2\ yy}^{(0,1)}\in\mathbb{R}): \alpha_{yy}^{(0,0)}\Sigma_{2\ yy}^{(0,1)}<1; \alpha_{yy}^{(1,0)} = - |\alpha_{yy}^{(1,0)}|$, consistent with the constraints such as (\ref{constraints_I}) and (\ref{pole-speed_s_ii}) of appendix E, implies the roots (\ref{root1-i}) and (\ref{root2-i}) as given in appendix E. In the same appendix, it is shown that:
 \begin{eqnarray}
 \label{pole-speed_s_iii}
 \omega_3\approx \pm q_3\left(1 + i \frac{\alpha_{yy}^{(00)}\Sigma_{2\ yy}^{(0,1)}}{2\left(\alpha_{yy}^{(1,0)}\right)^2}\right)\equiv\pm v_s q_3.
 \end{eqnarray}
One can show that one can consistently choose $ \frac{\alpha_{yy}^{(00)}\left(i\Sigma_{2\ yy}^{(0,1)}\right)}{2\left(\alpha_{yy}^{(1,0)}\right)^2} = \frac{1}{\sqrt{3}} - 1$ to yield $v_s=\frac{1}{\sqrt{3}}$.


\subsubsection{Via Two-Point Correlation Function $\langle T_{00} T_{00}\rangle$ using ON-Shell Reduction of Action and LO EOM's Solutions}

To put the results of {\bf 5.1.1} on a sound footing, we will now looking at the evaluation of the two-point correlation function $\langle T_{00}T_{00}\rangle$ from the on-shell action having dimensionally reduced $M$ theory on $M_5\times M_6$ in the MQGP limit to $M_5$, which asmptotically is $AdS_5$.

On-shellness dictates that:
$R^{(0)}=\frac{10}{3}\Lambda$ under the metric perturbation given in (\ref{metric perturbation}). The pure gravitational part of the $5d$ action along with the Gibbons-Hawking York surface term \cite{Liu+Tseytlin} and a counter term (required to regularize the action) is given by:
\begin{eqnarray}
\label{full action}
\int_0^1 du \int {d^4}x \sqrt{-g}(R - 2 \Lambda)+\int {d^4}x\sqrt{-g_B}~2K+a\int {d^4}x \sqrt{-g_B}
\end{eqnarray}
where $\Lambda$ is a cosmological constant term, $g_B^{\mu\nu}$ is the pull-back metric on the boundary of AdS space and $K$ is the extrinsic curvature. For the given metric in this paper the cosmological constant is $\lambda=-\frac{6 g_s^{2/3}}{L^2}$, also we choose $a=-\frac{6 g^{1/3}_s}{L}$ to make the action in equation (\ref{full action}) finite. On-shell, the bilinear part of the above action, in the limit $q_3\rightarrow0, \omega_3\rightarrow0$, reduces to the following surface term:
\begin{eqnarray}
\label{on-shell-surface-action}
& & \int {d^4}x\Biggl[\frac{1}{4}\Biggl(H_{tt}^2+8H_{{x}t}^2+2H_{{x}{x}}H_{tt}+4H_{{y}{y}}H_{tt}+4H_{{x}{x}}H_{{y}{y}}-H_{{x}{x}}^2\Biggr)
\nonumber\\
& & -\frac{1}{2\epsilon^3}\Biggl(H_{{x}t}^2+H_{{y}{y}}^2+H_{{x}{x}}H_{tt}+H_{{y}{y}}H_{tt}+2H_{{x}{x}}H_{{y}{y}}\Biggr)^\prime~\Biggr].
\end{eqnarray}

The equations of motion imply that $H_{tt}^\prime(u=0) = H_s^\prime(u=0) = H_{xt}^\prime(u=0) = H_{yy}^\prime(u=0) = 0$, and we will further assume that
\begin{eqnarray}
& & \left(\begin{array}{c}H_{yy}(u=0)\\ H_{s}(u=0)\end{array}\right) = \left(\begin{array}{cc} -\beta_{yt} & -\beta_{yx} \\ -\beta_{st} & -\beta_{sx} \end{array}\right)\left(\begin{array}{c} H_{tt}(u=0)\\ H_{xx}(u=0)\end{array}\right).
\end{eqnarray}
So, the relevant two-point correlation function involving $T_{00}$ will require finding out the coefficient of $\left(H^{(0)}_{t}\right)^2$ upon substitution of (\ref{GTI}) - (\ref{GTIII}) and (\ref{hab}) along with the values of $a, b, c, d$ with the common denominator $\Omega(\omega_3,q_3)$ of (\ref{pole-speed_s}). As the generic form of this two-point function in the hydrodynamical limit \cite{hyrdodynamical_limit} : $\omega_3\rightarrow0, q_3\rightarrow 0: \frac{\omega_3}{q_3}=\alpha\equiv$ constant - is expected to be of the form: $\frac{q_3^2}{\omega_3^2 - v_s^2 q_3^2}$, we isolate these terms and work up to leading order in $\Sigma_{2yy}^{(0,1)}$. We find from (\ref{on-shell-surface-action}) the following  coefficients of $(H_t^{(0)})^2$ coming from the $H^2$-like terms and $HH^{\prime}$-like terms:
\begin{eqnarray}
\label{T00T00}
& & \hskip -0.7in H^2\ {\rm terms}:\nonumber\\
  & &\hskip -0.7in -\frac{i \Sigma_{2yy}^{01} \Biggr(\alpha^4 \left(\beta_{st}^2+\beta_{st} (2-8 \beta_{yt})+12 \beta_{yt}^2-1\right)-\alpha ^2
   \left(\beta_{st}^2+\beta_{st} (56 \beta_{yt}+2)+12 \beta_{yt}^2-1\right)-32 \beta_{yt} (\pi \beta_{yt}-2)\Biggl)}{16 \left(\alpha
   ^2-1\right)};\nonumber\\
& & \hskip -0.7in \left.\frac{\left(HH^\prime\right)^{{\cal O}(u^0)}}{u^3}\right|_{u=\epsilon} =\nonumber\\
 & & \hskip -0.7in-\frac{i \alpha ^2  \Sigma_{2yy}^{01}\beta_{yt} \Biggr(\alpha ^2 ((8+\pi ) \beta_{st}-2 ((\pi -6) \beta_{yt}+1))+(16+\pi ) \beta_{st}-12
   \beta_{yt}+\pi ^2 \beta_{yt}+14 \pi  \beta_{yt}-2 \pi -22\Biggl)}{16 \left(\alpha ^2-1\right)} ;\nonumber\\
& & \hskip -0.7in\left.\frac{\left(HH^\prime\right)^{{\cal O}(u)}}{u^3}\right|_{u=\epsilon}=-\frac{1}{16 \left(\alpha
   ^2-1\right)}\Biggl\{i \alpha ^2 \Sigma_{2yy}^{01} \Biggl(\alpha ^2 \beta_{yt} ((16+\pi ) \beta_{st}-(\pi -20)\beta_{yt}+2)-2 \alpha  (2 \beta_{st}+\pi
   \beta_{yt}-2)\nonumber\\
 & &\hskip -0.7in+\beta_{yt} \left((\pi -24)\beta_{st}+\left(-20-3 \pi +\pi ^2\right) \beta_{yt}-2 \pi +6\right)\Biggr)\Biggr\};\nonumber\\
   & &\hskip -0.7in \left.\frac{\left(HH^\prime\right)^{{\cal O}(u^2)}}{u^3}\right|_{u=\epsilon}-\frac{1}{32 \left(\alpha ^2-1\right)}\Biggr\{i \alpha ^2  \Sigma_{2yy}^{01}\beta_{yt} \Biggl(2 \alpha ^2 ((\pi -24) \beta_{st}+10\beta_{yt})-(\pi -36) \alpha  (2  \beta_{st}+\pi
   \beta_{yt}-2)\nonumber\\
   & &\hskip -0.7in +2 \left((\pi -24)  \beta_{st}+\left(-10-24 \pi +\pi ^2\right) \beta_{yt}-2 \pi +48\right)\Biggr)\Biggl\};\nonumber\\
   & &\hskip -0.7in \left.\frac{\left(HH^\prime\right)^{{\cal O}(u^3)}}{u^3}\right|_{u=\epsilon}=\nonumber\\
   & & \hskip -0.7in\frac{1}{160 \left(\alpha ^2-1\right)^2}\Biggl\{i \Sigma_{2yy}^{01} \Biggl(5 \alpha ^6 \left(24 \beta_{st}^2+ \beta_{st} (8-2 (\pi -2) \beta_{yt})+\beta_{yt} (2-\pi
   \beta_{yt})\right)-40 \alpha ^5 \beta_{yt} (2  \beta_{st}+\pi \beta_{yt}-2)\nonumber\\
  & & \hskip -0.7in+2 \alpha ^4 \left(60  \beta_{st}^2+ \beta_{st} (6 (15 \pi
   -8) \beta_{yt}-200)-5 \left((\pi -11) \pi \beta_{yt}^2+(22-4 \pi ) \beta_{yt}+4\right)\right)+40 \alpha ^3\beta_{yt} (2 \beta_{st}+\pi
   \beta_{yt}-2)\nonumber\\
  & & \hskip -0.7in +\alpha ^2 \left(2  \beta_{st} ((38+35 \pi )\beta_{yt}-60)+\pi  (70 \pi -233) \beta_{yt}^2+(466-280 \pi )
   \beta_{yt}+280\right)+128 \beta_{yt} (\pi  \beta_{yt}-2)\Biggr)\Biggr\}.
   \end{eqnarray}
   From (\ref{T00T00}), we see that for $\beta_{yt}=0,\beta_{st}=1$, the first line in (\ref{on-shell-surface-action}) yields
   a contribution: $i\alpha^2\Sigma_{2yy}^{01}\frac{q_3^2}{\left(\omega_3^2 - v_s^2 q_3^2\right)}$ and
   from the second line in (\ref{on-shell-surface-action}), there is no required contribution from $\frac{\left(HH^\prime\right)^{{\cal O}(u^{0,1,2})}}{u^3}$ and $\frac{\left(HH^\prime\right)^{{\cal O}(u^3)}}{u^3}$ terms yield:
   $i \alpha^2\Sigma_{2yy}^{01}\frac{q_3^2}{\left(\omega_3^2 - v_s^2 q_3^2\right)}$.

\subsection{The Shorter Route - Use of Gauge-Invariant Variable}

In this subsection, we carry on the same calculation as we did in the last subsection for the speed of sound up to leading order in $N$ via a different approach. This time following \cite{klebanov quasinormal}, we first obtain the EOM for appropriate gauge-invariant variable corresponding to the non zero scalar modes of metric perturbations as defined in (\ref{Z-scalar mode}) and then compute the quasinormal modes, hence the speed of sound $v_s$ by solving that EOM in the hydrodynamic approximation. we have also calculate the two point correlation function of energy momentum tensor using the above solution for the gauge invariant variable. Latter following the same approach we compute the next to leading order correction to speed of sound by using the metric components as given in (\ref{Mtheory met}) corrected up to NLO in $N$.
\subsubsection{From the solution of Gauge Invariant Variable up to Leading Order in $N$}

Going back to (\ref{4scalar_EOMs}) we see that the four linearly independent equations using the following gauge invariant combination of perturbations namely,
\begin{eqnarray}
\label{gauge invariant}
Z_s(u)=2 q_3 \omega_3 H_{{x}t}+\omega_3^2 H_{{x}{x}}+H_{{y}{y}}\left[q_3^2
   \left(u^4+1\right)-\omega_3^2\right]-q_3^2 \left(1-u^4\right) H_{tt},
\end{eqnarray}
can be written as a single second order differential equation involving $Z_s(u)$:
\begin{eqnarray}
\label{single EOM}
& &  Z_s^{\prime\prime}(u)-\frac{q_3^2 \left(7 u^8-8 u^4+9\right)-3 \left(u^4+3\right) \omega_3^2}{u \left(u^4-1\right) \left(q_3^2 \left(u^4-3\right)+3
  \omega_3^2\right)}Z_s^\prime(u)\nonumber\\
  & & +\frac{q_3^4 \left(u^8-4 u^4+3\right)+2 q_3^2 \left(8 u^{10}-8 u^6+2 u^4 \omega_3^2-3 \omega_3^2\right)+3 \omega_3^4}{\left(1-u^4\right)^2
   \left(q_3^2 \left(u^4-3\right)+3 \omega_3^2\right)}Z_s(u)=0.
\end{eqnarray}
The above equation can be solve by considering an ansatz $Z_s(u)=(1-u)^rF(u)$ where $F(u)$ is regular near the horizon $u=1$. We have already obtained the value of exponent $r$ at the end of section {\bf 5} and it is given by $\pm \frac{i\omega_3}{4}$. we choose the negative sign here as it represents an incoming wave. The evaluation of the function $F(u)$ can be done perturbatively using hydrodynamic approximation, given as: $\omega_3\ll1$, $q_3\ll1$. For analytic solution the momentum has to be light-like, means $\omega_3$ and $q_3$ would be of the same order. Hence we can rescale $\omega_3$ and $q_3$ by a same parameter $\lambda$ as: $\omega_3\rightarrow\lambda \omega_3$, $q_3\rightarrow\lambda q_3$ and expand equation (\ref{single EOM}) to first order in $\lambda$, where the limit $\lambda\ll 1$ ensure that we are working in the hydrodynamic regime.
We choose the following series expansion of $F(u)$ for small frequency and momentum as:
\begin{eqnarray}
\label{regular F}
F(u)=F_0(u)+\omega_3F_1(u)+\mathcal{O}(\omega_3^2,q_3^2,\omega_3q_3).
\end{eqnarray}
Plugging in the equation (\ref{regular F}) into the equation (\ref{single EOM}) one can get an equation involving $F_0(u)$ only:
\begin{eqnarray}
 u \left(u^4-1\right) \left(q_3^2 \left(u^4-3\right)+3 \omega_3^2\right) F_0^{\prime\prime}+ \left(q_3^2 \left(-7 u^8+8
   u^4-9\right)+3 \left(u^4+3\right) \omega_3^2\right)F_0^\prime+16 q_3^2 u^7 F_0=0.\nonumber\\
   & &
\end{eqnarray}
A solution to the above equation is given by,
\begin{eqnarray}
F_0(u)=\frac{c_1 \left(q_3^2 \left(u^4+1\right)-3 \omega_3^2\right)}{17 q_3^2-3 \omega_3^2}+\frac{c_2 \left(q_3^2 \left(u^4+1\right)-3
   \omega_3^2\right) \left(-\frac{2 q_3^2-3 \omega_3^2}{q_3^2 \left(u^4+1\right)-3 \omega_3^2}-\frac{1}{4} \log
   \left(u^4-1\right)\right)}{17 q_3^2-3 \omega_3^2}.\nonumber\\
   & &
\end{eqnarray}
For the regularity of $F_0(u)$ near the horizon $u=1$, we choose the constant $c_2$ to be equal to zero.
Using this solution for $F_0(u)$, another equation for $F_1(u)$ can be found from (\ref{single EOM}),
\begin{eqnarray}
& & u \left(u^4-1\right)\left\{17 q_3^4
   \left(u^4-3\right)-3 q_3^2\omega_3^2 \left(u^4-20\right) -9 \omega_3^4\right\}F_1^{\prime\prime}\nonumber\\
    & & + \left\{-17 q_3^4 \left(7 u^8-8
   u^4+9\right)+3 q_3^2 \omega_3^2\left(7 u^8+9 u^4+60\right) -9 \omega_3^4 \left(u^4+3\right)\right\}F_1^{\prime}\nonumber\\
   & & +16  u^7q_3^2 \left(17
  q_3^2-3 \omega_3^2\right) F_1+16 i  u^7 q_3^2\left(2 q_3^2-3 \omega_3^2\right)c_1=0.
\end{eqnarray}
A general solution is given as:
\begin{eqnarray}
& & F_1(u)=-\frac{c_1 i \left(2 q_3^2-3 \omega_3^2\right)}{17 q_3^2-3 \omega_3^2}+\frac{c_2 \left(q_3^2 \left(u^4+1\right)-3
  w3^2\right)}{17 q_3^2-3 \omega_3^2}\nonumber\\
  & & +\frac{c_3 \left(q_3^2 \left(u^4+1\right)-3 \omega_3^2\right) \left(-\frac{2 q_3^2-3
   \omega_3^2}{q_3^2 \left(u^4+1\right)-3 \omega_3^2}-\frac{1}{4} \log \left(u^4-1\right)\right)}{17 q_3^2-3 \omega_3^2}.
\end{eqnarray}
Again demanding the regularity of the above solution near the horizon, we put $c_3$ to zero. Also imposing a boundary condition $F_1(u=1)=0$, we determine the constant $c_2$ to be equal to $ic_1$. With this the final expression of $Z_s(u)$ is given as:
\begin{eqnarray}
\label{solution single eom}
Z_s(u)=c_1(1-u^4)^{-i\omega_3/4}\left(\frac{q_3^2 \left(u^4+1\right)-3 \omega_3^2}{17 q_3^2-3 \omega_3^2}-\frac{i q_3^2\omega_3 \left(1-u^4\right)}{17 q_3^2-3 \omega_3^2}\right).
\end{eqnarray}
Imposing the Dirichlet boundary condition $Z(u=0)=0$ we get the quasinormal frequency,
\begin{eqnarray}
\omega_3=\pm\frac{q_3}{\sqrt{3}}-\frac{i q_3^2}{6}+\mathcal{O}.
\end{eqnarray}
Using (\ref{four momentum}), we get the following dispersion relation:
\begin{eqnarray}
w=\pm\frac{q}{\sqrt{3}}-\frac{i q^2}{6\pi T}.
\end{eqnarray}
Comparing this with the dispersion relation corresponding to the sound wave mode,
\begin{eqnarray}
w=\pm q v_s - i\Gamma_s q^2
\end{eqnarray}
where $v_s$ is the speed of sound and $\Gamma_s$ is the attenuation constant, we get their exact values.

\subsubsection{Via Two-Point Correlation Function $\langle T_{00} T_{00}\rangle$- Using the Solution of EOM involving Gauge Invariant Variable}

The relevant part of the bilinear surface term of the full action (\ref{full action}) as given in (\ref{on-shell-surface-action}) can be rewritten in terms of the gauge invariant variable $Z_s(u)$ as:
\begin{eqnarray}
\label{bilinear action2}
S^{(2)}_{\epsilon}=\lim_{u\rightarrow 0}\int \frac{dw dq}{2\pi^2} A(\omega_3,q_3,u)Z_s^\prime(u,q)Z_s(u,-q).
\end{eqnarray}
Using the equations of motion (\ref{4scalar_EOMs}) along with (\ref{on-shell-surface-action}), we find the function $A(\omega_3,q_3,u)$ as:
\begin{eqnarray}
A(\omega_3,q_3,u)=\frac{3}{u^3 \left(q_3^2 \left(u^4-3\right)+3 \omega_3^2\right)^2}
\end{eqnarray}
For the computation of two point function we need the solution of equation (\ref{single EOM}) as given in equation (\ref{solution single eom}), where the constant $c_1$ is determined by the boundary condition
\begin{eqnarray}
Z_s(u=0)=-H_{tt}^0 q_3^2+2 H_{{x}t}^0 q_3 \omega_3+H_{{x}{x}}^0 \omega_3^2+H_{{y}{y}}^0 \left(q_3^2-\omega_3^2\right),
\end{eqnarray}where we define $H_{ab}(u=0)=H_{ab}^0$.
Now putting the above expression of $A(\omega_3,q_3,u)$ and the solution $Z_s(u)$ back in equation (\ref{bilinear action2}) one get the two point correlator $G_{tt,tt}$ as:
\begin{eqnarray}
\nonumber G_{tt,tt}=\frac{\delta^2S_{\epsilon}^{(2)}}{\delta H_{tt}^{(0)}(k)\delta H_{tt}^{(0)}(-k)}\\
=\frac{8 q^6}{3 \left(q^2-3 w^2\right) \left(q^2-w^2\right)^2}
\end{eqnarray}
Hence the pole structure of the Green's function gives the correct value of the speed of sound wave, $v_s=\frac{1}{\sqrt{3}}$ propagating through hot plasma. The above value of speed of sound also matches exactly with the value that we have already got from the solution of hydrodynamic equations, thus provides a quantitative checks of the validity of Gauge/Gravity duality.
\subsection{From the solution of Gauge Invariant Variable - Going up to NLO in $N$ in the MQGP Limit}

Considering the Next-to-Leading Order corrections in $N$ of the metric components as given in (\ref{Mtheory met}), and using the gauge invariant combination given in (\ref{Z-vector mode}), (\ref{Z-scalar mode}) and (\ref{Z-tensor mode}), the Einstein equation can be expressed in terms of a single equation of the form $Z_i^{\prime\prime}(u) = m_i(u) Z_i^\prime(u) + l_i(u) Z_i(u)$, where, $i=s({\rm calar})$, $v({\rm ector})$, $t({\rm ensor })$.

In {\bf 5.3.1}, we first evaluate $v_s$ including the non-conformal contribution to the M-theory metric evaluated at a finite $r$ and large $N$, i.e., $\log\left(\frac{r}{\sqrt{\alpha^\prime}}\right)<\log N$, thereby dropping $\log r \log N$  as compared to $\left(\log N\right)^2$. Then, in {\bf 5.3.2}, we attempt a full-blown non-conformal estimate of $v_s$ up to NLO in $N$ by working at an $r: \log\left(\frac{r}{\sqrt{\alpha^\prime}}\right)\sim \log N$. It turns out, unlike the former, the horizon becomes an irregular singular point for the latter.  We set $\alpha^\prime$ to unity throughout. Given that in both, {\bf 5.3.1} and {\bf 5.3.2}, we are interested in numerics, exact numerical factors in all expressions will be replaced by their decimal equivalents.

\subsubsection{Dropping $\log r \log N$ As Compared to $\left(\log N\right)^2$}

Including the NLO terms, the EOM for the gauge invariant variable $Z_s(u)$ - given by (\ref{Z-scalar mode}) -  can be rewritten as:
\begin{equation}
\label{EOM_vs_i}
(u-1)^2Z_s^{\prime\prime}(u) + (u-1)P(u-1) Z_s^\prime(u) + Q(u-1) Z_s(u) = 0,
\end{equation}
in which $P(u-1) = \sum_{n=0}^\infty p_n(u-1)^n$ and $Q(u-1) = \sum_{m=0}^\infty q_n (u-1)^n$  wherein, up to ${\cal O}\left(\frac{1}{N}\right)$, $p_n, q_n$ are worked out in (\ref{pn+qn_up_to_2nd_order}).  The horizon $u=1$ being a regular singular point of (\ref{EOM_vs_i}), the Frobenius method then dictates that the incoming-wave solution is given by:
\begin{equation}
\label{solution-i}
Z_s(u) = \left(1 - u \right)^{\frac{3 {g_s}^2 M^2 {N_f}  \log (N) \left(8 {q_3}^2 {\omega_3}^2 \log (N)+\left({\omega_3}^2+4\right) \left(10 {q_3}^2-27
   {\omega_3}^2\right)\right)}{2048 \pi ^2 N q_3^2 \omega_3 \left(-1\right)^{3/2}}-\frac{i {\omega_3}}{4}}\left(1 + \sum_{m=1}a_m (u - 1)^m\right),
\end{equation}
where $a_{1,2}$ are given in (\ref{a1a2-sound}).  Following \cite{klebanov quasinormal},  imposing Dirichlet boundary condition $Z_s(u=0)=0$ and going up to second order in powers of $(u-1)$ in (\ref{solution-i}) and considering in the hydrodynamical limit $\omega_3^nq_3^m:m+n=2$ one obtains:
\begin{equation}
\label{vs+Gammas-second_order}
\omega_3 = -\frac{2 {q_3}}{\sqrt{3}}-\frac{9 i {q_3}^2}{32},
\end{equation}
which yields a result for the speed of sound similar to, though not identical to, (\ref{v_s 1})  for $n=0,1$.

To get the LO or conformal result for the speed of sound $v_s = \frac{1}{\sqrt{3}}$, let us go to the fourth order in (\ref{solution-i}). For this, up to ${\cal O}\left(\frac{1}{N}\right)$, $p_n, q_n$ are worked out in (\ref{pn+qn_up_to_fourth_order}).

We will not quote the expressions for $a_3$ and $a_4$ because they are too cumbersome. Substituting the expressions for $a_{1,2,3,4}$ into $Z_s(u)$ and implementing the Dirichlet boundary condition: $Z_s(u=0)=0$, in the hydrodynamical limit, going up to ${\cal O}(\omega_3^4)$ one sees that one can write the Dirichlet boundary condition as a quartic: $a \omega_3^4 + b \omega_3^3 + c \omega_3^2 + f \omega_3 + g = 0$ where $a, b, c, d, f, g$ are given in (\ref{a+b+c+f+g}).
One of the four roots yields:
\begin{equation}
\label{dispersion}
\omega_3 \approx 0.46 q_3 - 0.31 i q_3^2,
\end{equation}
with no ${\cal O}\left(\frac{1}{N}\right)$-corrections! The coefficient of $q_3$ is not too different from the conformal value of $\frac{1}{\sqrt{3}}\approx 0.58$. We expect the leading order term in the coefficient of $q_3$ to converge to $\frac{1}{\sqrt{3}}$. Also, the coefficient of $q^2$ term turns out to be $\frac{0.31}{\pi}$ which is not terribly far from the conformal result of $\frac{0.17}{\pi}$. We are certain that the inclusion of higher order terms in (\ref{solution-i}) will ensure that we get a perfect match with the conformal result. The reason we do obtain the NLO non-conformal contribution to $v_s$ is that at the very outset, we have neglected the non-conformal $\log r$-contributions by working at a large but finite $r$, but such that $\frac{\frac{r}{\sqrt{\alpha^\prime}}}{N}\ll1.$ We will see how to obtain the non-conformal contribution with the inclusion of the same in {\bf 5.3.2} below.

\subsubsection{Retaining $\log r \log N$ and $\left(\log N\right)^2$ Terms}

Constructing a $Z_s(u)$ given by (\ref{Z-scalar mode}) and retaining the non-conformal $\log r \log N$-contribution as well as $\left(\log N\right)^2$ terms, one sees one obtains (\ref{Z-EOM}) as the equation of motion for $Z_s(u)$.
The horizon $u=1$ due to inclusion of the non-conformal corrections to the metric,  becomes an irregular singular point. One then tries the ansatz: $Z_s(u) = e^{S(u)}$ near $u=1$ \cite{Bender_Orzag}. Assuming that $\left(S^{\prime}\right)^2\gg S^{\prime\prime}(u)$ near $u=1$ the differential equation (\ref{Z-EOM}), which could written as $Z_s^{\prime\prime}(u) = m(u)Z_s^\prime + l(u) Z_s(u)$ can be approximated by:
\begin{equation}
\label{S_EOM-text}
\left(S^\prime\right)^2 - m(u) S^\prime(u) - l(u) \approx 0.
\end{equation}
A solution to (\ref{S_EOM-text}) is given in (\ref{S-solution}) in Appendix D. Taking first the MQGP limit,  integrating with respect to $u$, the solution (\ref{S-solution})  will reflect the singular nature of $Z_s(u)$'s equation of motion (\ref{S-solution}) via
\begin{equation}
\label{pole-soln-Z}
Z_s(u)\sim \left(1 - u \right)^{-\frac{1}{2} + \frac{15 {g_s}^2 M^2 {N_f} {\omega_3}^2 \log \left(\frac{1}{N}\right)}{256 \pi ^2 N \left(2 {q_3}^2-3{\omega_3}^2\right)}}F(u),
\end{equation}
 where $F(u)$ is regular in $u$ and its equation of motion, around $u=0$, is given by (\ref{F-EOM}) whose solution is given in (\ref{F-eom-solution}). One notes from (\ref{F-eom-solution}) that $F(u\sim0) = c_1$. This needs to be improved upon by including the sub-leading terms in $u$ in $F'(u)$ which is discussed in detail in Appendix D.

 For $Z_s(u=0)=0$ to obtain $\omega = \omega(q)$ to determine the speed of sound, one requires $F(u=0)=0$. From (\ref{solution-improved-F-EOM}), this can be effected by requiring
\begin{equation}
\hskip -0.5in\frac{225 {g_s}^4 {N_f}^2 {\omega_3}^2 \log ^2(N) M^4+4800 {g_s}^2 N
   {N_f} \pi ^2 \left(4 {q_3}^2-5 {\omega_3}^2\right) \log (N) M^2+139264 N^2 \pi ^4 \left(2 {q_3}^2-3 {\omega_3}^2\right)}{128 N \pi ^2
   \left(15 {g_s}^2 {N_f} \left(8 {q_3}^2-11 {\omega_3}^2\right) \log (N) M^2+896 N \pi ^2 \left(2 {q_3}^2-3
   {\omega_3}^2\right)\right)} = - n\in\mathbb{Z}^-
   \end{equation}
   or
   \begin{equation}
   \label{v_s 1}
\omega = {q_3} \left(\frac{\sqrt{14 {n}+17}}{\sqrt{21 {n}+\frac{51}{2}}}+\frac{5 (2 {n}+5) {g_s}^2 M^2 {N_f} \log N}{128 \pi ^2 \sqrt{14 {n}+17} \sqrt{84 {n}+102} N}\right),
\end{equation}
implying the following estimate of the speed of sound:
\begin{equation}
\label{v_s 2}
v_s\approx  \frac{\sqrt{14 {n}+17}}{\sqrt{21 {n}+\frac{51}{2}}}+\frac{5 (2 {n}+5) {g_s}^2 M^2 {N_f} \log
   N}{128 \pi ^2 \sqrt{14 {n}+17} \sqrt{84 {n}+102} N}.
\end{equation}
Given that (\ref{S-solution}) is an approximate solution to (\ref{S_EOM-text}), one expects to obtain an expression for $v_s$ from an $M3$-brane uplift\footnote{For a $p$-brane solution, to LO in $N$, one expects $v_s=\frac{1}{\sqrt{p}}$ \cite{Herzog-vs}.}, to be of the form $v_s \approx \frac{{\cal O}(1)}{\sqrt{3}} + {\cal O}\left(\frac{g_sM^2}{N}\right)$,
and (\ref{v_s 1}) is exactly of this form for $n=0,1$.

\section{Vector Mode Perturbations and Shear Mode Diffusion Constant  up to NLO in $N$ in the MQGP Limit}

The equations of motion for the vector perturbation modes up next-to-leading order in $N$, can be reduced to the following single equation of motion in terms of a gauge-invariant variable $Z_v(u)$ (given by (\ref{Z-vector mode})):
\begin{equation}
\label{vector-modes-ZEOM-text}
Z_v^{\prime\prime}(u) - m_v(u) Z_v^\prime(u) - l_v(u) Z_v(u) = 0,
\end{equation}
where $m_v(u), l_v(u)$ are given in (\ref{m+l-vec-definitions}).
The horizon $u=1$ is a regular singular point of (\ref{vector-modes-ZEOM-text}) and the root of the indicial equation corresponding to the incoming-wave solution is given by:
\begin{equation}
\label{root-solution-incoming-wave}
-\frac{i {\omega_3}}{4} + \frac{3 i {g_s}^2 M^2 {N_f} {\omega_3} \log ^2(N)}{256 \pi ^2 N}.
\end{equation}
(a) Using the Frobenius method, taking the solution about $u=1$ to be:
\begin{equation}
\label{solution1-text}
Z_v(u) = (1 - u)^{-\frac{i {\omega_3}}{4} + \frac{3 i {g_s}^2 M^2 {N_f} {\omega_3} \log ^2(N)}{256 \pi ^2 N}}\left(1 + \sum_{n=1}^\infty a_n (u - 1)^n\right),
\end{equation}
by truncating the infinite series in (\ref{solution1-text}) to ${\cal O}((u-1)^2)$ one obtains in (\ref{a1a2}) of Appendix F, values for $a_1, a_2$.

The Dirichlet boundary condition $Z(u=0)=0$ in the hydrodynamical limit retaining therefore terms only up to ${\cal O}(\omega_3^mq_3^n):\ m+n=4$, reduces to:
$a \omega_3^4 + b \omega_3^3 + c \omega_3^2 + f \omega_3 + g = 0$ where $a, b, c, d, f, g$ are given in (\ref{a b c d f g-i}). Analogous to {\bf 5.3}, once again as we are interested in numerics, exact numerical factors in all expressions will be replaced by their decimal equivalents for most part of this section.

One of the four roots of  $Z_v(u=0)=0$ is:
\begin{equation}
\label{root-at-second-order}
\omega_3 = -8.18 i + \frac{0.14 i g_s^2 M^2 N_f(\log N)^2}{N} + \left(-0.005 i - \frac{0.002 i g_s^2 M^2 N_f (\log N)^2}{N}\right)q_3^2 + {\cal O}(q_3^3).
\end{equation}

(b) Using the Frobenius method and going up to ${\cal O}((u-1)^3)$ in (\ref{solution1-text}), one obtains in (\ref{NLOa3}) values of $a_3$.

The Dirichlet condition $Z_v(u=0)=0$ reduces to $a \omega_3^4 + b \omega_3^3 + c \omega_3^2 + f \omega_3 + g = 0$
where $a, b, c, d, f, g$ are given in (\ref{a b c d f g-ii}).
One of the four roots of the quartic in $\omega_3$ is:
\begin{equation}
\label{root-at-third-order}
\omega_3 =  \left(- 0.73 i + \frac{0.003 i g_s^2 M^2 N_f (\log N)^2}{N}\right)q_3^2 + {\cal O}(q_3^3).
\end{equation}
The leading order coefficient of $q_3^2$ is not terribly far off the correct value $-\frac{i}{4}$ already at the third order in the infinite series (\ref{solution1-text}).

(c) Let us look at (\ref{solution1-text}) up to the fourth order. One finds in (\ref{a_4}) the value of $a_4$.
In the hydrodynamical limit the Dirichlet boundary condition $Z_v(u=0)=0$ reduces to $a \omega_3^4 + b \omega_3^3 + c \omega_3^2 + f \omega_3 + g = 0$
where $a, b, c, d, f, g$ are given in (\ref{a b c d f g-iii}).
Incredibly, one of the roots of the quartic equation in $\omega_3$ is:
\begin{eqnarray}
\label{root-at-fourth-order}
& & \omega_3 = \left( - \frac{i}{4} + \frac{3 i g_s^2 M^2 N_f \log N\left(5 + 2 \log N\right)}{512 \pi^2 N}\right)q_3^2 + {\cal O}\left(q_3^3\right).
\end{eqnarray}
Hence, the leading order (in $N$) yields a diffusion constant of the shear mode $D = \frac{1}{4\pi T}$, exactly the conformal result! Including the non-conformal corrections which appear at NLO in $N$, one obtains:
\begin{equation}
\label{D}
D = \frac{1}{\pi T}\left(\frac{1}{4} - \frac{3  g_s^2 M^2 N_f \log N\left(5 + 2 \log N\right)}{512 \pi^2 N}\right).
\end{equation}
We conjecture that all terms in (\ref{solution1-text}) at fifth order and higher, do not contribute to the Dirichlet boundary condition up to the required order in the hydrodynamical limit.

\begin{figure}
 \begin{center}
 \includegraphics[scale=0.8]
 {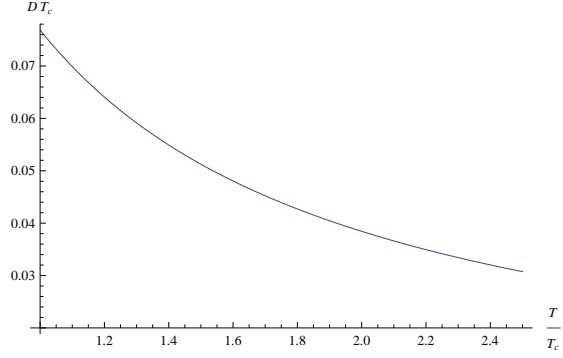}
 \end{center}
 \caption{$D T_c$ vs. $\frac{T}{T_c}$ for $T\geq T_c$}
\end{figure}

The variation of the shear mode diffusion constant with temperature is shown in Figure 1 for $N_f=3$, $M=3$, $g_s=0.9$, $N=100$. As the lowest order conformal result we obtain exactly $\frac{1}{4\pi T}$ as obtained in \cite{Starinets-Quasinormal spectrum}, for the black brane metric of the form (\ref{5dmetric inu}).

\section{NLO Corrections in N to $\eta$ and $\frac{\eta}{s}$}

Using the non-conformal $M_5(\mathbb{R}^{1,3},u)$ metric components of (\ref{Mtheory met}), we now evaluate the non-conformal $\frac{g_s M^2}{N}$-corrections to the shear viscosity $\eta$ by considering the EOM for the tensor mode of metric fluctuations up to NLO in $N$,  and also estimate the same for the shear viscosity - entropy density ratio $\frac{\eta}{s}$.

The EOM for the tensor mode of metric fluctuation, using (\ref{Z-tensor mode}),  is given as under:
\begin{eqnarray}
\label{EOM-tensor}
& & Z_t^{\prime\prime}(u) + Z_t^\prime(u) \left(-\frac{15 {g_s}^2 M^2 {N_f} \log {N}}{64 \pi ^2 N u}+\frac{u^4+3}{u \left(u^4-1\right)}\right)+Z_t(u) \Biggl[\frac{{q_3}^2
   \left(u^4-1\right)+{\omega_3}^2}{\left(u^4-1\right)^2}\nonumber\\
   & & -\frac{3 \left({q_3}^2 u^4-{q_3}^2+{\omega_3}^2\right) \left({g_s}^2 M^2 {N_f} \log
   ^2{N}+2 {g_s}^2 M^2 {N_f} \log {N} \log \left(\frac{2 \pi ^{3/2} \sqrt{{g_s}} T}{u}\right)\right)}{32 \pi
   ^2 N \left(u^4-1\right)^2}\Biggr]=0.\nonumber\\
   & &
   \end{eqnarray}
Realizing the horizon is a regular singular point, one makes the following double perturbative ansatz in $\omega_3$ and $q_3$:z
\begin{equation}
\label{ansatz-solution-tensor}
\hskip -0.4in Z_t(u) = \left(1 - u\right)^{-i {\omega_3} \left(\frac{1}{4}-\frac{3  {g_s}^2 M^2 {N_f} \log (N) \log r_h}{128 \pi ^2 N}\right)}\left(z_{00}(u) + \omega_3 z_{01}(u) + q_3 z_{10}(u) + {\cal O}(q_3^m\omega_3^n)_{m+n\geq2}\right).
\end{equation}
Using equations (\ref{w3}) - (\ref{constant}) in Appendix G,
\begin{eqnarray}
\label{Phi}
& & Z_t(u) = -\frac{i}{3072 \pi ^2 N}(1-u)^{-i {\omega_3} \left(\frac{1}{4}-\frac{3  {g_s}^2 M^2 {N_f} \log (N) \log r_h}{128 \pi ^2 N}\right)}\nonumber\\
    & & \times\Biggl(-3 {g_s}^2 M^2 {N_f} u \log {N} \left(4 c_2 \left(2 u^2+3 u+6\right) {\omega_3} \log \left(2
   \pi ^{3/2} \sqrt{{g_s}} T\right)+15 i c_5 {q_3} u^3 (1-4 \log (u))\right)\nonumber\\
   & & -6 c_2 {g_s}^2 M^2 {N_f} u \left(2 u^2+3 u+6\right) {\omega_3} \left(\log
   {N}\right)^2 +128 \pi ^2 N \biggl[6 i \left(c_5 {q_3} u^4+4 c_3 {q_3}+4 c_4 {\omega_3}\right)\nonumber\\
   & & +c_2 \left(2 u^3 {\omega_3}+3 u^2 {\omega_3}+6 u
   {\omega_3}+24 i\right)\biggr]\Biggr).
\end{eqnarray}
Setting $q_3=0$ one obtains (\ref{Phi'overPhi}) wherein the ${\cal O}(u^3\omega_3)$ term, without worrying about overall numerical multiplicative constants, is given by:
   \begin{equation}
   \label{w3u3}
   \frac{i}{4}-\frac{3 i {g_s}^2 M^2 {N_f} \log (N) \log r_h}{128 \pi ^2 N}
   \end{equation}
   Using arguments of \cite{transport-coefficients},  setting $\kappa_{11}^2=1$, the coefficient of the kinetic term of $Z_t(u)$ near $u=0$ and near   $\theta_1=\frac{\alpha_{\theta_1}}{N^{\frac{1}{5}}}$ (whereat an explicit local $SU(3)$-structure of the type IIA mirror and an explicit local $G_2$-structure of the M-theory uplift was obtained in \cite{T_c+Torsion}) is
 \begin{equation}
 \label{kinetic-term-fluctuation}
 \frac{r_h^4}{g_s^2u^3}\int d\theta_1\cot ^3\theta_1 \sin\theta_1 f_1(\theta_1)\delta\left(\theta_1 - \frac{\alpha_\theta}{N^{\frac{1}{5}}}\right)\sim \frac{r_h^4}{g_s^2u^3}\frac{N^{\frac{2}{5}}}{\alpha_N\alpha_{\theta_1}^2},
 \end{equation}
 where $f_1(\theta_1) = \frac{\cot\theta_1}{\alpha_N},\ f_2(\theta_2) = -\alpha_N\cot\theta_2$ \cite{T_c+Torsion}. The exact result for the temperature, assuming the resolution to be larger than the deformation in the resolved warped deformed conifold in the type IIB background of \cite{metrics} \footnote{Refer to appendix {\bf H} for details pertaining to this assumption.} in the MQGP limit, and utilizing the IR-valued warp factor $h(r,\theta_1,\theta_2)$, is given as under:
\begin{eqnarray}
\label{T}
& & T = \frac{\partial_rG^{\cal M}_{00}}{4\pi\sqrt{G^{\cal M}_{00}G^{\cal M}_{rr}}}\nonumber\\
& & = {r_h} \left[\frac{1}{2 \pi ^{3/2} \sqrt{{g_s} N}}-\frac{3 {g_s}^{\frac{3}{2}} M^2 {N_f} \log ({r_h}) \left(-\log
   {N}+12 \log ({r_h})+\frac{8 \pi}{g_s N_f} +6-\log (16)\right)}{64 \pi ^{7/2} N^{3/2}} \right]\nonumber\\
   & & + a^2 \left(\frac{3}{4 \pi ^{3/2} \sqrt{{g_s}} \sqrt{N} {r_h}}-\frac{9 {g_s}^{3/2} M^2 {N_f} \log ({r_h})
   \left(\frac{8 \pi }{{g_s} {N_f}}-\log (N)+12 \log ({r_h})+6-2 \log (4)\right)}{128 \pi ^{7/2} N^{3/2}
   {r_h}}\right).\nonumber\\
   & &
\end{eqnarray}
Now, we will assume a non-zero `bare' resolution parameter $\sim\alpha$ in the resolution parameter $a(r_h)$ and substitute $a = \left(\alpha + \gamma\frac{g_s M^2}{N} + \beta \frac{g_s M^2}{N}\log r_h\right)r_h$ \cite{K. Dasgupta  et al [2012]} into (\ref{T}).  One can hence calculate the shear viscosity $\eta$:
   \begin{eqnarray}
   \label{eta}
   & &\hskip -0.8in \eta = \Upsilon\frac{N^{\frac{2}{5}}}{g_s^2\alpha_N\alpha_{\theta_1}^2}\lim_{\omega_3\rightarrow0}\left(\frac{1}{\omega_3 T}\lim_{u\rightarrow0}\left[\frac{r_h^4}{u^3}\Im m \left(\frac{Z_t^\prime(u)}{Z_t(u)}\right)\right]\right).
   \end{eqnarray}
   where $\Upsilon$ is an overall multiplicative constant.

For the purpose of comparison of $\frac{\eta}{s}$ with lattice/RHIC data for QGP and consequently be able to express
$r_h$ in terms of $\tilde{t}\equiv \frac{T}{T_c}-1$, let us revisit our calculation of $T_c$ as given in \cite{T_c+Torsion} with the difference that unlike \cite{T_c+Torsion} wherein we had assumed a constant dilaton profile, this time around we will take:
\begin{eqnarray}
\label{dilaton}
& (a) & T>0:\nonumber\\
& & e^{-\Phi} = \frac{1}{g_s} - \frac{N_f}{8\pi}\log(r^6 + a^2 r^4) - \frac{N_f}{2\pi}\log\left(\sin\frac{\theta_1}{2}\sin\frac{\theta_2}{2}\right),\ r<{\cal R}_{D5/\overline{D5}},\nonumber\\
& & e^{-\Phi} = \frac{1}{g_s},\ r>{\cal R}_{D5/\overline{D5}};\nonumber\\
& (b) & T=0:\nonumber\\
& & e^{-\Phi} = \frac{1}{g_s} - \frac{3 N_f}{4\pi}\log r - \frac{N_f}{2\pi}\log\left(\sin\frac{\theta_1}{2}\sin\frac{\theta_2}{2}\right),\ r<\left|\mu_{\rm Ouyang}\right|^{\frac{2}{3}},\nonumber\\
& & e^{-\Phi} = \frac{1}{g_s},\ r>\left|\mu_{\rm Ouyang}\right|^{\frac{2}{3}}.
\end{eqnarray}
Hence, setting the Newtonian constant to unity,  performing a large-N expansion and then a large $r_\Lambda$-expansion, one obtains for the thermal background ($r_h=0$) for which $r\in[r_0,r_\Lambda]$ where $r_0$ and $r_\Lambda$ are respectively the IR and UV cut-offs:
\begin{eqnarray}
\label{V_1}
& & V_1 = -\frac{1}{2}\int_{r=r_0}^{r_\Lambda} d^5x\sqrt{-g}e^{-2\Phi}\left(R - 2\Lambda\right) - \int_{r=r_\Lambda}\sqrt{-h}e^{-2\Phi}K
\nonumber\\
& & = \frac{3  {r_\Lambda}^4}{2{N^{5/4}} \sqrt{2} \pi ^{5/4} {g_s}^{13/4}} -\frac{1}{32 \left(\sqrt{2} \pi ^{9/4} {g_s}^{13/4}\right)}\Biggl\{\frac{1}{{N^{5/4}}} \Biggl(4
   \pi  {g_s}^2 \log ^2 N \left({|\mu_{\rm Ouyang}|^{\frac{8}{3}}}-{r_0}^4\right)
   \nonumber\\
   & & - {g_s} \log N
   \Biggl[\left({r_0}^4-{|\mu_{\rm Ouyang}|^{\frac{8}{3}}}\right) (-12 {g_s} {N_f} \log ({|\mu_{\rm Ouyang}|^{\frac{2}{3}}})+{g_s} {N_f} (3+16 \pi  \log (4))+16 \pi )\nonumber\\
   & & -12
   {g_s} {N_f} {r_0}^4 \log \left(\frac{{r_0}}{{|\mu_{\rm Ouyang}|^{\frac{2}{3}}}}\right)\Biggr]-16 \pi  \left({|\mu_{\rm Ouyang}|^{\frac{8}{3}}}-2
   {r_h}^4\right)\Biggr)\Biggr\}\nonumber\\
& & + {\cal O}\left(\frac{1}{{r_\Lambda}^2}\right).
\end{eqnarray}
Similarly, for the black hole background, for which $r\in[r_h,r_\Lambda]$ one obtains:
\begin{eqnarray}
\label{V_2}
& & V_2 = -\frac{1}{2}\int_{r=r_h}^{r_\Lambda} d^5x\sqrt{-g}e^{-2\Phi}\left(R - 2\Lambda\right) - \int_{r=r_\Lambda}\sqrt{-h}e^{-2\Phi}K
\nonumber\\
& & = \frac{3 \left(\frac{1}{N}\right)^{5/4} {r_\Lambda}^4}{2 \sqrt{2} \pi ^{5/4} {g_s}^{13/4}}+\frac{9 a^2 \left(\frac{1}{N}\right)^{5/4}
   {r_\Lambda}^2}{4 \sqrt{2} \pi ^{5/4} {g_s}^{13/4}}\nonumber\\
   & & +\frac{1}{32 \sqrt{2} \pi ^{9/4} {g_s}^{13/4}
   {{\cal R}^2_{D5/\overline{D5}}}}\Biggl\{\frac{1}{N^{5/4}} \Biggl[6 \pi  a^2 {g_s}^2 \log
   ^2\left(\frac{1}{N}\right) \left({r_h}^4-{{\cal R}^4_{D5/\overline{D5}}}\right)\nonumber\\
   & & - 3 a^2 {g_s} \log N \Biggl(6 {g_s} {N_f}
   \left({r_h}^4-{{\cal R}^4_{D5/\overline{D5}}}\right) \log ({{\cal R}_{D5/\overline{D5}}})\nonumber\\
   & & +\left({{\cal R}^2_{D5/\overline{D5}}}-{r_h}^2\right) \Biggl[{g_s} {N_f} \left({{\cal R}^2_{D5/\overline{D5}}} (8 \pi
    \log (4)-9)+{r_h}^2 (8 \pi  \log (4)-3)\right)\nonumber\\
    & & +8 \pi  \left({{\cal R}^2_{D5/\overline{D5}}}+{r_h}^2\right)\Biggr]\Biggr)+8 \pi  \left(3 a^2
   \left({{\cal R}^4_{D5/\overline{D5}}}-{r_h}^4\right)-4 {{\cal R}^2_{D5/\overline{D5}}} {r_h}^4\right)\Biggr]\Biggr\}\nonumber\\
   & & + {\cal O}\left(\frac{1}{{r_\Lambda^2}}\right).
\end{eqnarray}
Now, in the $r_\Lambda\rightarrow\infty$-limit, realizing:
\begin{eqnarray}
\label{CT-i}
& & \left.\sqrt{-h^{\rm Thermal}}\right|_{r=r_\Lambda} = \frac{{r_\Lambda}^4}{4 \pi  {g_s} N},\nonumber\\
& & \left.\sqrt{-h^{\rm BH}}\right|_{r=r_\Lambda} = \frac{{r_\Lambda}^4-3 a^2 {r_\Lambda}^2}{8 \sqrt{2} \pi ^{3/4} {g_s}^{3/4} N^{3/4}},
\end{eqnarray}
one sees that the required counter term required to be added to $V_2-V_1$ (required later) is:
\begin{eqnarray}
\label{CT-ii}
& & \int_{r=r_\Lambda}\left(-\frac{3 \sqrt[4]{\frac{1}{N}} \left(\sqrt[4]{\pi } \sqrt[4]{{g_s}} \sqrt[4]{N} \sqrt{-h^{\rm Thermal}}-2 \sqrt{2}
   \sqrt{-h^{\rm BH}}\right)}{\sqrt{2 \pi } {g_s}^{5/2} \sqrt[4]{N}}\right).
\end{eqnarray}
Therefore,
\begin{eqnarray}
\label{V_2-V_1-UVfinite}
& & \hskip -0.3in (V_2 - V_1)^{\rm UV-finite} = \nonumber\\
& & \frac{1}{32 \sqrt{2} \pi ^{9/4} {g_s}^{13/4}}\left(\frac{1}{N}\right)^{5/4} \Biggl(3 {g_s}^2 {\log N} {N_f} \left(9 a^4-{r_h}^4\right) (2 \log (a)+\log (3))\nonumber\\
& & -9 a^4
   \left({g_s}^2 {\log N} (2 \pi  {\log N}+{N_f} (8 \pi  \log (4)-9))+8 \pi  {g_s} {\log N}-8 \pi \right)-18 a^2
   {g_s}^2 {\log N} {N_f} {r_h}^2\nonumber\\
   & & +4 \pi  {g_s}^2 {\log N}^2 \left(|\mu_{\rm Ouyang}|^{8/3}-{r_0}^4\right)-{g_s} {\log N}
   \Biggl[\left({r_0}^4-|\mu_{\rm Ouyang}|^{8/3}\right) (-8 {g_s} {N_f} \log (|\mu_{\rm Ouyang}|)\nonumber\\
   & & +{g_s} {N_f} (3+16 \pi  \log (4))+16 \pi ) -12
   {g_s} {N_f} {r_0}^4 \log \left(\frac{{r_0}}{|\mu_{\rm Ouyang}|^{2/3}}\right)\Biggr]-16 \pi  \left(|\mu_{\rm Ouyang}|^{8/3}-2
   {r_h}^4\right)\nonumber\\
   & & -{r_h}^4 \left({g_s}^2 {\log N} ({N_f} (3-8 \pi  \log (4))-2 \pi  {\log N}) -8 \pi  {g_s}
   {\log N}+40 \pi \right)\Biggr).
\end{eqnarray}
Now assuming
${\cal R}_{D5/\overline{D5}} = \sqrt{3}a$ (to be justified via a finite temperature $0^{++}$ glueball mass calculation via the WKB quantization method in \cite{Sil+Misra-glueball}), $|\mu_{\rm Ouyang}|^{\frac{2}{3}}=\delta r_0$ and assuming an IR-valued $r_h, r_0$, $(V_2 - V_1)^{\rm UV-finite} =0$ yields:
\begin{eqnarray*}
& & \hskip -0.6in 2 \pi  {g_s}^2 {\log N}^2 \left(\left(1-9 \alpha ^4\right) {rh}^4+2 \left(\delta ^{8/3}-1\right) {r_0}^4\right)+6 {g_s}^2
   {\log N} {N_f} \left(\left(9 \alpha ^4-1\right) {r_h}^4 \log ({rh})-2 \left(\delta ^{8/3}-1\right) {r0}^4 \log
   ({r_0})\right)\nonumber\\
   & & \hskip -0.6in -8 \pi  \left(\left(1-9 \alpha ^4\right) {r_h}^4+2 \delta ^{8/3} {r0}^4\right) = 0,
   \end{eqnarray*}
whose solution is given by:
\begin{eqnarray}
\label{V2-V1=0}
& & \hskip -0.4in \frac{\sqrt[4]{\frac{2}{3}} \sqrt[4]{-\frac{\left(9 \alpha ^4-1\right) {rh}^4 \left(\pi  \left({g_s}^2 {\log N}^2-4\right)-3
   {g_s}^2 {\log N} {N_f} \log ({rh})\right)}{\left(\delta ^{8/3}-1\right) {g_s}^2 {\log N}
   {N_f}}}}{\sqrt[4]{{\cal PL}\left(-\frac{\left(9 \alpha ^4-1\right) {rh}^4 \left(2^{\frac{1}{2} \left(\delta ^{8/3}-1\right)} \exp
   \left(\frac{\pi  \left(\delta ^{8/3} \left(4-{g_s}^2 {\log N}^2\right)+{g_s}^2 {\log N}^2\right)}{3 {g_s}^2 {\log N}
   {N_f}}\right)\right)^{\frac{4}{\delta ^{8/3}-1}} \left(\pi  \left({g_s}^2 {\log N}^2-4\right)-3 {g_s}^2 {\log N} {N_f}
   \log ({rh})\right)}{6 \left(\delta ^{8/3}-1\right) {g_s}^2 {\log N} {N_f}}\right)}},\nonumber\\
   & &
\end{eqnarray}
where ${\cal PL}$ is the `ProductLog' function. This yields:
\begin{equation}
\label{r0rh}
r_0  =  r_h\sqrt[4]{\left|\frac{9 \alpha ^4-1}{2(\delta^{\frac{8}{3}}-1)}\right|} + {\cal O}\left(\frac{1}{\log N}\right).
\end{equation}
For subsequent calculations and comparison with RHIC data, we will be defining: $\tilde{t}\equiv \frac{T}{T_c}-1$.
Now, as we will show in \cite{Sil+Misra-glueball}, the lightest $0^{++}$ scalar glueball mass is given by:
\begin{equation}
\label{glueball-mass-i}
m_{\rm glueball} \approx \frac{4 r_0}{L^2}.
\end{equation}
Now, lattice calculations for $0^{++}$ scalar glueball masses at zero temperature (zero temperature being relevant as the IR cut-off is required at zero temperature whereas $r_h$ provides the same for the black hole background)
\cite{SU3_lattice_glueball_masses}, yield the lightest mass to be around $1,700 MeV$. From (\ref{glueball-mass-i}), replacing $\frac{r_0}{L^2}$ by $\frac{m_{\rm glueball}}{4}$ we obtain:
\begin{equation}
\label{Tc}
T_c = \left.\frac{m_{\rm glueball}\left(1 + \frac{3\alpha^2}{2}\right)}{2^{\frac{7}{4}}\pi\sqrt[4]{\left|\frac{9 \alpha ^4-1}{2(\delta^{\frac{8}{3}}-1)}\right|}}\right|_{\alpha=0.6,\delta=1.02}
=179 MeV.
\end{equation}

As the expressions in the following will become very cumbersome to deal with and to type, specially for the purpose of comparison with RHIC QGP data, we will henceforth deal only with numerical expressions setting $g_s=0.9, N=100, M=3, N_f=2, \alpha=0.6, \delta=1.02$.

We now discuss the $\frac{1}{N}$-corrections to the entropy density $s$ by estimating the same from the $D=11$ supergravity action result of \cite{MQGP}, and hence work out the $\frac{1}{N}$ corrections to $\frac{\eta}{s}$.
The UV-finite part of the $D=11$ supergravity action, given by the Gibbons-Hawking-York (GHY) surface action $S_{\rm GHY}$ from \cite{MQGP} (without worrying about overall multiplicative constants) is \cite{MQGP}:
\begin{equation}
\label{K_MQGP_i}
\left.\int_{r={\cal R}_{\rm UV}\equiv{\rm UV\ cut-off}}K\sqrt{h}\right|_{\theta_{1,2}\sim0}\sim\frac{\cot^2\theta_1 f_2(\theta_2)}{g_s^{\frac{11}{4}}N^{\frac{3}{4}}\left(\sin^2\theta_1 + \sin^2\theta_2\right)}\left(\frac{1}{T}\right),
\end{equation}
 ($K$ being the extrinsic curvature and $h$ being the determinant of the pull-back of the $D=11$ metric on to $r={\cal R}_{\rm UV}$). Further, assume that what appears in (\ref{K_MQGP_i}) is $f_1(\theta_1)$. Now, unlike the scaling given in (\ref{limits_Dasguptaetal-ii}) used in \cite{MQGP}, we will be using:
 $\theta_{1,2}\rightarrow0$ as $\theta_1=\frac{\alpha_{\theta_1}}{N^{\frac{1}{5}}},\ \theta_2=\frac{\alpha_{\theta_2}}{N^{\frac{3}{10}}} (N\sim10^8)$ - as used in \cite{T_c+Torsion} (to discuss a local $SU(3)$ structure of the type IIA delocalized SYZ mirror and a local $G_2$ structure of its M-theory uplift), as well as this paper. This can be used to evaluate $ S^{\rm UV-finite}_{GHY}$ and the entropy density: $s =  - T \frac{\partial S^{\rm UV-finite}_{\rm GHY}}{\partial T} - S^{\rm UV-finite}_{\rm GHY}$. This yields:
\begin{eqnarray}
\label{eta_over_s-i}
& & \hskip -0.8in \frac{\eta}{s} =  {\cal O}(1)\times\nonumber\\
& & \hskip -0.8in \left[\frac{\frac{1}{4 \pi }-0.00051 \log ({r_h})}{1 - 0.064 \gamma + 0.004 \gamma^2 + \sum_{n=1}^4 a_n(\beta,\gamma)\log^n r_h + \frac{\sum_{n=0}^4 b_n(\beta,\gamma)\log^n r_h}{\sum_{n=0}^2c_n(\beta,\gamma)\log^nr_h}}\right],
\end{eqnarray}
\noindent where $a_n,b_n,c_n$ are known functions of $\beta$ and $\gamma$, and there is freedom to choose the ${\cal O}(1)$ constant. We will impose two conditions, as per RHIC QGP data, on $\beta$ and $\gamma$ and the ${\cal O}(1)$ constant: $\left.\frac{\eta}{s}\right|_{T=T_c}=0.1,$ and
$\left.\frac{d\left(\frac{\eta}{s}\right)}{d\tilde{t}}\right|_{\tilde{t}>0}>0$. Numerically, one sees that setting $(\beta,\gamma)=(4,4)$ and consequently $r_h=\frac{35546.9 ({\tilde{t}}+1)}{{\cal PL}(2706.3 ({\tilde{t}}+1))}$ where ${\cal P L}$ is the ``ProductLog" function, and the ${\cal O}(1)$ constant equal to $5.8$, fits the bill. Hence,
{\footnotesize
\begin{eqnarray}
\label{eta_over_s-ii}
& & \hskip -0.7in\frac{\eta}{s} = 5.8\Biggl[\frac{{9.18\times 10^{-8}} \log ^3({r_h})-1.6\times 10^{-5} \log
   ^2({r_h})+2.7\times 10^{-4} \log ({r_h})+1.7\times 10^{-3}}{-{2.5\times 10^{-7}} \log
   ^6({r_h})+\frac{9 \log ^5({r_h})}{10^6}-\frac{\log ^4({r_h})}{10^4}+3.1\times 10^{-4}
   \log ^3({r_h})+0.002 \log^2({r_h})+3.6\times 10^{-3} \log ({r_h})+0.047}\Biggr].\nonumber\\
   & &
\end{eqnarray}}
The graphical variation of $\frac{\eta}{s}\left(N_f=3,M=3,g_s=0.9,N=100\right)$ vs. $\tilde{t}=\frac{T - T_c}{T_c}$ is shown in the following graph in Figure 2, and the RHIC data plot from \cite{eta-over-s-RHIC}\footnote{One of us (KS) thanks R. Lacey to permit us to reproduce the graph in Figure 3 from their paper \cite{eta-over-s-RHIC}.}, is shown in Figure 3.

\begin{figure}
 \begin{center}
 \includegraphics[scale=0.8]
 {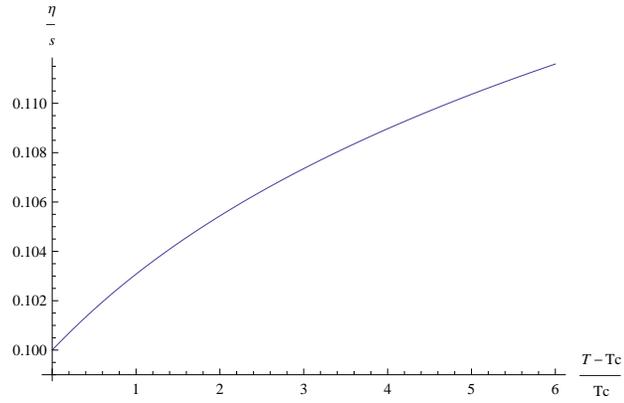}
 \end{center}
 \caption{$\frac{\eta}{s}$ vs. $\frac{T-T_c}{T_c}$ for $T\geq T_c$ assuming $\left.\frac{\eta}{s}\right|_{T=T_c}=0.1$}
\end{figure}

\begin{figure}
 \begin{center}
 \includegraphics[scale=0.5]
 {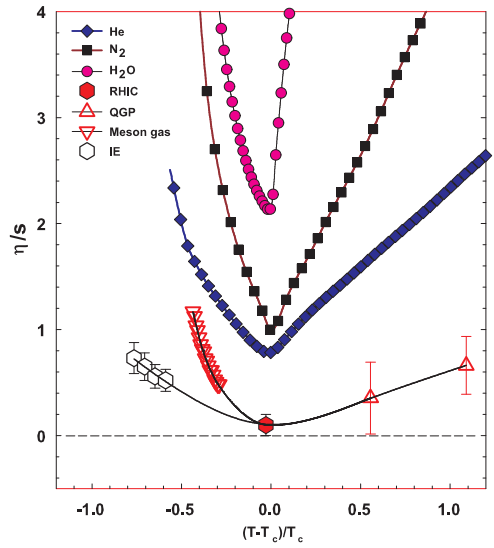}
 \end{center}
 \caption{$\frac{\eta}{s}$ vs. $\frac{T-T_c}{T_c}$ reproduced from \cite{eta-over-s-RHIC}.}
\end{figure}

We draw a third graph in which the plots of Figures 2 and 3 are drawn on the same graph.
\begin{figure}
 \begin{center}
 \includegraphics[scale=0.8]
 {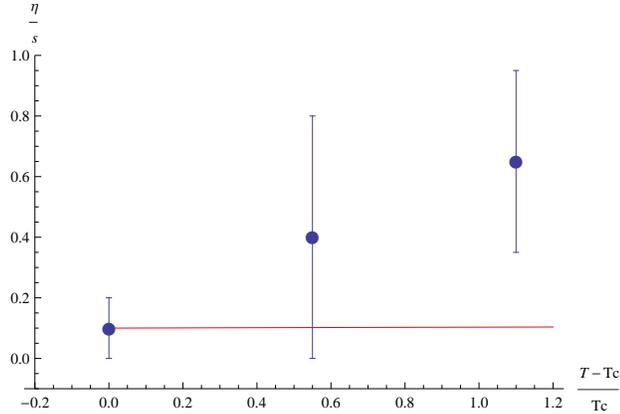}
 \end{center}
 \caption{Combined plots of Figures 2 and 3: the graph in red is from Figure 2 (our calculations) and the set of three points with error bars are from Figure 3 (RHIC QGP data from \cite{eta-over-s-RHIC}).}
\end{figure}
The combined plots in Figure 4 make the comparison of our results with those of RHIC data in \cite{eta-over-s-RHIC}, very clear. We conclude the following:
 \begin{itemize}
 \item
 $\frac{\eta}{s}(T=T_c)= 0.1$, and $\left.\frac{d\left(\frac{\eta}{s}\right)}{dT}\right|_{T>T_c}>0$ - this is clear from Figure 2.

 \item
 The numerical values, unlike \cite{eta-over-s-RHIC}, remain close to the value at $T=T_c$. In other words, unlike Figure 3 quoted from \cite{eta-over-s-RHIC}, in Figure 4, $\frac{\eta}{s}$ is found to be a much more slowly varying function of $\tilde{t}=\frac{T - T_c}{T_c}$.  Also, $\frac{d^2\left(\frac{\eta}{s}\right)}{d\tilde{t}^2}<0$ in Figure 2  and $\frac{d^2\left(\frac{\eta}{s}\right)}{d\tilde{t}^2}>0$ in Figure 3. The error bars appearing in Figure 3 as shown more clearly in Figure 4,  for $\frac{T - T_c}{T_c}\in[0,1.1]$ - the range covered in \cite{eta-over-s-RHIC} - permit our deviations from \cite{eta-over-s-RHIC} at least for $\frac{T - T_c}{T_c}\in[0,0.6]$.

\end{itemize}

\section{Summary and New Insights into (Transport) Properties of Large $N$ Thermal QCD at Finite Gauge Coupling}

A realistic computation pertaining to  thermal QCD systems such as sQGP, require a finite gauge coupling and not just a large t'Hooft coupling \cite{Natsuume}, and the number of colors $N_c$ equal to three. Such computations, missing in the literature, were initiated in \cite{MQGP,transport-coefficients}.  Further, computations quantifying the IR non-conformality in physical quantities pertaining  to large-$N$ thermal QCD at finite gauge coupling that appear at the NLO in $N$ in the corresponding holographic description in string \cite{metrics}/M-theory \cite{MQGP}, have been missing thus far in the literature. In this paper, at finite gauge coupling with $N_c=M_{r\in{\rm IR}}=3$ as part of the MQGP limit (\ref{limits_Dasguptaetal-ii}), we have addressed a Math issue and {obtained new insights into some transport properties at LO in $N$, and  non-conformal corrections appearing at the NLO in $N$ in a variety of hydrodynamical quantities crucial to characterizing thermal QCD - like systems at finite gauge coupling such as sQGP}.

In this paper we have discussed identification of the local $T^3$ of \cite{MQGP} (used for construction of the delocalized SYZ type IIA mirror in \cite{MQGP} of \cite{metrics}'s type IIB holographic dual of large-$N$ thermal QCD) with a special Lagrangian three-cycle, as well as a number of different issues relevant to the transport properties of large-$N$ thermal QCD at finite gauge coupling often inclusive of the non-conformal $\frac{1}{N}$ temperature-dependent corrections, in the context of gauge/gravity duality. For the latter set of issues, the calculations from the gravitational description involves scalar, vector and tensor modes of the asymptotically $AdS_5$ metric perturbations. In particular, solving the Einstein's equation involving gauge invariant combination of different perturbations we obtain the quasinormal frequencies. The speed of sound follows from the quasinormal frequency corresponding to scalar modes of metric perturbations while the diffusion constant of the shear mode is obtained from the quasinormal frequency corresponding to the vector modes of metric perturbation.

Before summarizing our main results, we would first summarize the assumptions made to arrive at the results.
\begin{enumerate}
\item
The three-form fluxes of (\ref{three-form-fluxes}) obtained in \cite{metrics} for a resolved warped conifold, is also valid in the UV-IR interpolating region and the UV for a resolved warped deformed conifold which is predominantly (warped and) resolved. The reason, as stated below (\ref{three-form-fluxes}), has to do with the fact that the corrections due to the resolution parameter $a$ appear as $\frac{a^2}{r^2}$ whereas those due to the deformation parameter $\epsilon$ appear as $\frac{\epsilon^2}{r^3}$. In the UV-IR interpolating region, assuming $a,\epsilon^{\frac{3}{2}}<r_h$  and in the UV, the latter is sub-dominant as compared to the former.

\item
\begin{enumerate}
\item
To ensure UV conformality for $r>{\cal R}_{D5/\overline{D5}}$, as explained in the last paragraph of {\bf 2.2}, one requires a vanishing $D5$-brane and $D7$-brane charges in the UV. This requires that $F_3$ and $B_3$ vanish in the UV which, as explained in {\bf 2}, is effected by including $M\ \overline{D5}$-branes which like $M\ D5$-branes, wrap the vanishing $S^2$, but are distributed at the antipodal points of the resolved $S^2$ relative to the $D5$-branes at $r={\cal R}_{D5/\overline{D5}}$.

This is implemented in our calculations by assuming that $r\rightarrow r_\Lambda$(UV cut-off)$\rightarrow\infty$ as $\epsilon^{-\gamma_r}$ near the coordinate patch $\theta_{1,2}=0$ effected by $\theta_{1,2}\rightarrow\epsilon^{\gamma_{\theta_{1,2}}}$ for $\epsilon\ll1$. In the MQGP limit of (\ref{limits_Dasguptaetal-ii}), one can then show that $\lim_{r\rightarrow\infty}(F_3,H_3)=0$.

One can do a better job, as suggested in \cite{IR-UV-desc_Dasgupta_etal}, by working with the following ansatze for ${\widetilde F}_3$:
\begin{eqnarray}
\label{Better_F3tilde}
&& \hskip -0.3in {\widetilde F}_3  = \left({a}_o - {3 \over 2\pi r^{g_sN_f}} \right)
\sum_\alpha{2M(r)c_\alpha\over r^{\epsilon_{(\alpha)}}}
\left({\rm sin}~\theta_1~ d\theta_1 \wedge d\phi_1-
\sum_\alpha{f_\alpha \over r^{\epsilon_{(\alpha)}}}~{\rm sin}~\theta_2~ d\theta_2 \wedge
d\phi_2\right)\nonumber\\
&&\hskip -0.3in \wedge~ {e_\psi\over 2}-\sum_\alpha{3g_s M(r)N_f d_\alpha\over 4\pi r^{\epsilon_{(\alpha)}}}
~{dr}\wedge e_\psi \wedge \left({\rm cot}~{\theta_2 \over 2}~{\rm sin}~\theta_2 ~d\phi_2
- \sum_\alpha{g_\alpha \over r^{\epsilon_{(\alpha)}}}~
{\rm cot}~{\theta_1 \over 2}~{\rm sin}~\theta_1 ~d\phi_1\right)\nonumber \\
&& \hskip -0.3in -\sum_\alpha{3g_s M(r) N_f e_\alpha\over 8\pi r^{\epsilon_{(\alpha)}}}
~{\rm sin}~\theta_1 ~{\rm sin}~\theta_2 \left({\rm cot}~{\theta_2 \over 2}~d\theta_1 +
\sum_\alpha{h_\alpha \over r^{\epsilon_{(\alpha)}}}~
{\rm cot}~{\theta_1 \over 2}~d\theta_2\right)\wedge d\phi_1 \wedge d\phi_2\label{brend}
\end{eqnarray}
where $M(r)\equiv 1 - \frac{e^{\alpha(r-{\cal R}_{D5/\overline{D5}})}}{1 + e^{\alpha(r-{\cal R}_{D5/\overline{D5}})}}, \alpha\gg1$, and $a_o = 1 + {3\over 2\pi}$ and ($c_\alpha, \epsilon_{\alpha}, h_\alpha$) are constants. Further investigation using (\ref{Better_F3tilde}) however will be deferred to a later work.

\item
Further, given that the number $N_f$ of flavor $D7$-branes appears in the expression of the dilaton in (\ref{dilaton}), to ensure a constant axion-dilaton modulus in the UV required for UV conformality, as explained in {\bf 2}, one adds an equal number of $\overline{D7}$-branes in the UV and the UV-IR interpolating region and not the IR.

This is implemented in our calculations by assuming that $\frac{3\gamma_r}{4}=\gamma_\theta$ ensuring that \\ $\lim_{r\rightarrow\infty}e^{-\Phi}(\theta_{1,2}\rightarrow0)=\frac{1}{g_s}$.
\end{enumerate}

\item
The functions $h_i$ of (\ref{h_i}) appearing in the resolved warped deformed conifold metric (\ref{RWDC}) along with (\ref{metric}) are assumed to receive corrections of ${\cal O}\left(\frac{g_s M^2}{N}\right)$.

\item
The $D=10$ warp factor $h(r,\theta_{1,2},\phi_{1,2})$, as stated in (\ref{eq:h}) and (\ref{h-large-small-r}), is assumed to be:
\begin{eqnarray}
&& \hskip -0.45in h =\frac{L^4}{r^4}\Bigg[1+\frac{3g_sM_{\rm eff}^2}{2\pi N}{\rm log}r\left\{1+\frac{3g_sN^{\rm eff}_f}{2\pi}\left({\rm
log}r+\frac{1}{2}\right)+\frac{g_sN^{\rm eff}_f}{4\pi}{\rm log}\left({\rm sin}\frac{\theta_1}{2}
{\rm sin}\frac{\theta_2}{2}\right)\right\}\Biggr],\ {\rm in\ the\ IR};\nonumber\\
& & \hskip -0.45in h = \frac{L^4}{r^4}\left[1 + \sum_{i=1}\frac{{\cal H}_i\left(\phi_{1,2},\theta_{1,2},\psi\right)}{r^i}\right],\ {\rm in\ the\ UV},
\end{eqnarray}
where, in principle, $M_{\rm eff}/N_f^{\rm eff}$ are not necessarily the same as $M/N_f$; we however assume that up to ${\cal O}\left(\frac{g_sM^2}{N}\right)$, they are. We also assume that ${\cal H}_i\left(\phi_{1,2},\theta_{1,2},\psi\right)={\cal O}\left(\frac{g_s M^2}{N}\right)$.

\end{enumerate}

The following provides a summary of the new results obtained in this paper as well as the new insights into the Physics of strongly coupled thermal QCD laboratories like sQGP gained therefrom.

\begin{itemize}
\item
{\bf Legitimacy of the local $T^3$ of \cite{MQGP} for effecting delocalized SYZ mirror transformation}: In the MQGP limit, in the UV (as well as the UV-IR interpolating) region(s): $r\gg r_h$, we have shown that the local $T^3$ defined in \cite{MQGP} is the same as the $T^2$-invariant special Lagrangian (sLag) three-cycle of \cite{M.Ionel and M.Min-OO (2008)} in a resolved conifold. Earlier in \cite{transport-coefficients}, it was already shown that  in the MQGP limit the aforementioned $T^3$ is also the $T^2$-invariant sLag of \cite{M.Ionel and M.Min-OO (2008)} in a deformed conifold. Together, { the new insight gained is that  the local $T^3$ defined in \cite{MQGP}  used for constructing the type IIA delocalized Strominger-Yau-Zaslow(SYZ) mirror of \cite{metrics}'s type IIB holographic dual of large-$N$ thermal QCD, in the MQGP limit, is shown to be a bonafide special Lagrangian three-cycle necessary to construct the required type IIA SYZ mirror}. This is valid for both, a predominantly resolved (resolution $>$ deformation - this paper) or a predominantly deformed (deformation $>$ resolution - \cite{transport-coefficients}) resolved warped deformed conifold. Though we limited ourselves to the LO in $N$ for this purpose, but the same can also be shown to be true at the NLO in $N$ - the computations will become extremely cumbersome though. This was crucial in justifying the construction of the SYZ type IIA mirror in \cite{MQGP} of the \cite{metrics}'s type IIB holographic dual of large-$N$ thermal QCD.

\item
{\bf $\kappa_T, \sigma$, Wiedemann-Franz law at LO in $N$ and  $D=1+1$ Luttinger Liquid with impurities}: As gauge fluctuations are tied  to vector modes of metric fluctuations, by solving the coupled set of equations for both, we obtained the temperature dependence of the thermal and electrical conductivities as well as looked at whether the Wiedemann-Franz law was satisfied. {This revealed a remarkable insight into the properties of large-$N$ thermal QCD at finite gauge coupling} namely that { the type IIB holographic dual of large-$N$ thermal QCD with a temperature-dependent Ouyang embedding parameter: $|\mu_{\rm Ouyang}|\sim r_h^{\alpha\leq0}$, effectively  qualitatively mimicked a $D=1+1$ Luttinger liquid with impurities/doping}. It will be extremely interesting to explore this unexpected duality, further. For $\alpha=\frac{5}{2}$, one is able to reproduce the usual linear large-temperature dependence of DC electrical conductivity for most strongly coupled systems with five-dimensional gravity duals with  a black hole \cite{SJain_sigma+kappa}.

\item
{\bf The non-conformal/NLO-in-$N$ corrections to Transport Coefficients}: For ease of readability and convenience of the reader, the main results pertaining to obtaining the non-conformal temperature-dependent ${\cal O}\left(\frac{(g_sM^2)(g_sN_f)}{N}\right)$ corrections to $v_s$ (the speed of sound), $D$ (shear mode diffusion constant ), $\eta$ (shear viscosity) and $\frac{\eta}{s}$ (shear-viscosity-entropy density ratio) are summarized in Table 1 below.
\begin{table}[h]
\begin{tabular}{|c|c|c|}\hline
S. No. & Quantity & Expression up to ${\cal O}\left(\frac{(g_sM^2)(g_s N_f)}{N}\right)$  \\ \hline
1 & $v_s$ & $v_s\approx  \frac{\sqrt{14 {n}+17}}{\sqrt{21 {n}+\frac{51}{2}}}+\frac{5 (2 {n}+5) {g_s}^2 M^2 {N_f} \log
   N}{128 \pi ^2 \sqrt{14 {n}+17} \sqrt{84 {n}+102} N}, n\in\mathbb{Z}^+\cup{0}$ \\
   & & $\downarrow\ n=0,1$ \\
   & & $\frac{{\cal O}(1)}{\sqrt{3}} + {\cal O}\left(\frac{g_sM^2}{N}\right)$  \\ \hline
   2 & $D$ & $\frac{1}{\pi T}\left(\frac{1}{4} - \frac{3  g_s^2 M^2 N_f \log N\left(5 + 2 \log N\right)}{512 \pi^2 N}\right)$
   \\  \hline
   3 & $\frac{\eta}{s}$ & See (\ref{eta_over_s-i}) and (\ref{eta_over_s-ii}) \\ \hline
   \end{tabular}
   \caption{Summary of local non-nonformal ${\cal O}\left(\frac{g_sM^2}{N}\right)$ corrections   to $v_s, D, \frac{\eta}{s}$}
   \end{table}

We showed that in the spirit of  gauge/gravity duality, the leading order result of speed of sound from the quasinormal modes can be reproduced from (a) the pole of the common denominator that appears in the solutions to the scalar modes of metric perturbations, (b) the pole of the retarded Green's function corresponding to the energy momentum tensor two-point correlation function $\langle T_{00}T_{00}\rangle$ using the on-shell surface action written in terms of the metric perturbation modes, (c) imposing Dirichlet boundary condition on the solution to the EOM of an appropriate single gauge-invariant perturbation and (d) $\langle T_{00}T_{00}\rangle$-computation using the on-shell surface action written in terms of this gauge-invariant perturbation. The leading order result for the diffusion constant of the shear mode as well as the ratio of shear viscosity-to-entropy density ratio were already discussed in \cite{transport-coefficients}.

{The non-trivial insight thus gained at LO in $N$ into the transport properties of holographic large-$N$ thermal QCD at finite gauge coupling is that the LO-in-$N$ conformal result for finite $g_s$ as obtained in this paper, matches the LO-in-$N$ conformal result for vanishing $g_s$ as is expected/known in the literature for a $p$-brane for $p=3$.}

 { The non-conformal corrections in all the aforementioned quantities, start appearing at  ${\cal O}\left(\frac{(g_s M^2)(g_s N_f)}{N}\right)$}, $N_f$ being the number of flavor $D7$-branes. Thus, at NLO in $N$, { the new insight gained is that  there is a partial universality in the non-conformal corrections to the transport coefficients in the sense that the same are determined by the product of the very small $\frac{g_sM^2}{N}\ll1$ - part of the MQGP limit (\ref{limits_Dasguptaetal-ii}) - and the finite $g_s N_f\sim {\cal O}(1)$ (also part of (\ref{limits_Dasguptaetal-ii})).} The NLO-corrected results in this paper reflect the non-conformality of the field theory in the IR. As discussed in section {\bf 2} that in the Klebanov-Strassler backgroud \cite{KS} the number of $D3$ branes $N$ decreases with  decreasing (the non-compact radial coordinate) $r$, which according to AdS/CFT dictionary, behaves as an energy scale. This decrease in $N$ is due to a series of repeated Seiberg dualities, where in the extreme IR, at the end of this duality cascade the number of fractional $D3$ branes $M$ which is taken to be finite in the 'MQGP Limit' gets identified with the number of colors in the theory. In other words, the number of $D3$ branes $N$ exhibits a scale dependance due to the duality cascade. Hence from the NLO-corrected expressions of the shear mode diffusion constant and the viscosity, we conclude that these quantities also exhibit a scale dependance through $N$; the appearance of $M$ in the NLO-in-$N$ corrections to the transport coefficients appearing as $\frac{(g_s M^2)(g_s N_f)}{N}$ signals the non-conformality of the field theory in the IR. This is because of the following reason. In the KS picture the presence of finite number $M$ of fractional $D3$ branes makes the field theory non-conformal in the IR while in the UV the presence of $\overline{D5}$ branes cancels the effects of the  $D5$-branes and restore the conformality in the UV. Now at large $r$ the effective number $N_{\rm eff}$ of (\ref{NeffMeffNfeff}), is so large that the NLO term can be neglected and we will be left with the leading order conformal results. But in the IR region the NLO terms have to be considered due to small value of $N_{\rm eff}$ - this is rather nicely captured, e.g., by the non-conformal/NLO corrections to $\eta$ (See e.g. Table 1.)

\end{itemize}

We compared our results for $\frac{\eta}{s}$ with the QGP-related RHIC data for $T\geq T_c$ in Section {\bf 7}. Let us also make some remarks as regard comparison of some of our results with some well-cited bottom-up holographic QCD models like \cite{Kiritsis_et_al} (as well as references therein) and the more recent \cite{Veneziano-i} based on the Veneziano's QCD model. As regard the speed of sound, like \cite{Kiritsis_et_al}, for $T>T_c$ (which is the temperature range in which we calculated the speed of sound in Section {\bf 5}) the speed of sound approaches a constant value; the difference however is that the NLO non-conformal corrections in our results pushes the value to slightly above $\frac{1}{\sqrt{3}}$ - our LO result and the saturation value in \cite{Kiritsis_et_al}. Upon comparison with some of the results of \cite{Veneziano-ii} which works with the finite temperature version of \cite{Veneziano-i}, one sees that the authors of the same work in the limit: $N_f\rightarrow\infty, N_c\rightarrow\infty: \frac{N_f}{N_c}\equiv$ fixed and $g_{\rm YM}^2 N_c\equiv$ fixed, which is very different from the MQGP limit of (\ref{limits_Dasguptaetal-ii}). A similarity however pertaining to the QCD phase diagram in the same and our results of \cite{T_c+Torsion} is that $\mu_C(T=T_c)$(for $N_f=2$)$\approx0$.

\section*{Acknowledgements}

One of us (KS) is supported by a senior research fellowship (SRF) from the Ministry of Human Resource and Development (MHRD).  One of us (AM) would like to thank N. Evans for useful discussions and MPI (Werner Heisenberg Institute) for Physics, Munich, and AEI (MPI for Gravitational Physics), Golm, for their hospitality where part of this work was completed. Some of the results of this paper were presented in seminars given by one of us (AM) at U. Southampton, MPI Munich and AEI Golm. We also thank P. Pandey for participating in, as part of his Masters project, the material discussed in Sec. {\bf 4}.

\appendix

\section{Details of Local $T^3$ Being a  $T^2$-Invariant sLag  in a Resolved Conifold in the MQGP Limit}
\setcounter{equation}{0} \seceqaa

The system of equations (\ref{sLagRC-I}) are solved to yield (\ref{sLagRC-II}).
\begin{eqnarray}
\label{sLagRC-II}
& & \hskip -0.7in \cos^2\frac{\theta_1}{2} = \nonumber\\
& & \hskip -0.7in \frac{4 a^2 \sqrt{3 a^2+\rho ^2} \left(-3 {c_1}-3 {c_2}+8 \rho ^2\right)+\rho ^2 \left(-4 {c_1} \sqrt{3 a^2+\rho ^2}-4 {c_2} \sqrt{3 a^2+\rho
   ^2}+\sqrt{6} \sqrt{9 a^2 \rho ^4+\rho ^6}\right)+96 a^4 \sqrt{3 a^2+\rho ^2}}{32 a^2 \rho ^2 \sqrt{3 a^2+\rho ^2}+2 \sqrt{6} \rho ^2 \sqrt{9 a^2 \rho ^4+\rho
   ^6}+96 a^4 \sqrt{3 a^2+\rho ^2}},\nonumber\\
   & & \hskip -0.7in \cos^2\frac{\theta_2}{2} = \frac{\rho ^2 \left(-4 {c_1} \sqrt{3 a^2+\rho ^2}+4 {c_2} \sqrt{3 a^2+\rho ^2}+\sqrt{6} \sqrt{9 a^2 \rho ^4+\rho ^6}\right)-12 a^2 \sqrt{3 a^2+\rho ^2}
   ({c_1}-{c_2})}{2 \sqrt{6} \rho ^2 \sqrt{9 a^2 \rho ^4+\rho ^6}}.
\end{eqnarray}
  Substituting (\ref{sLagRC-II}) into the third equation of (\ref{sLagRC-I}), one obtains:
\begin{eqnarray}
\label{sLagRC-III}
& & \frac{1}{9} \left(a^2 {\sin \psi}^2 \left(-4 \left(16 \sqrt{6}-27\right) a^2 ({c_1}+{c_2})+\left(8 \sqrt{6}-27\right) {c_1}^2+16 \sqrt{6} {c_1}
   {c_2}+\left(8 \sqrt{6}-27\right) {c_2}^2\right)-9 {c_3}^2\right)\nonumber\\
   & & -\frac{{\sin \psi}^2 \left(32 \sqrt{6} a^2 ({c_1}-{c_2})^2
   ({c_1}+{c_2})-4 \sqrt{6} \left({c_1}^2-{c_2}^2\right)^2\right)}{9 \sqrt{6} \rho ^2} +\frac{9}{16} a^2 \rho ^4 {\sin \psi}^2\nonumber\\
   & &   -\frac{1}{3} \rho ^2
   {\sin \psi}^2 \left(8 a^4-4 a^2 ({c_1}+{c_2})+{c_1}^2+{c_2}^2\right) +\frac{\rho ^6 {\sin \psi}^2}{16} + {\cal O}\left(\frac{a^2c_1^4\sin^2\psi}{\rho^4}\right) = 0.
 \end{eqnarray}
This obtains:
\begin{equation}
\label{sLagRC-IV}
\rho^2 = -3 a^2+\frac{4\ 2^{2/3} {\sin \psi}^{2/3} \left({c_1}^2+{c_2}^2\right)}{9 {c_3}^{2/3}}+\frac{2 \sqrt[3]{2} {c_3}^{2/3}}{{\sin \psi}^{2/3}} + {\cal O}\left(\frac{a^2c_1^2\sin^{\frac{4}{3}}\psi}{c_3^{\frac{4}{3}}}\right).
\end{equation}
As $\theta_1,\theta_2\rightarrow0$ as $\frac{1}{N^{\frac{1}{5}}},\frac{1}{N^{\frac{3}{10}}}$ (whereat an explicit local $SU(3)$-structure of the type IIA mirror and an explicit local $G_2$-structure of the M-theory uplift was obtained in \cite{T_c+Torsion}) and in the UV-IR interpolating region/UV: $r\rightarrow {\cal R}_0$, therefore in this domain of $(\theta_1,\theta_2,r)$ choose:
\begin{equation}
\label{sLagRC-V}
c_1\sim {\cal R}_0^2,\ c_2\sim\frac{{\cal R}_0^2}{N^{\frac{2}{5}}}.
\end{equation}
Hence, making a large-$N$ expansion:
{\footnotesize
\begin{eqnarray}
\label{sLagRC-VI}
& & \hskip -1.2in \cos^2\theta_2 = \nonumber\\
& & \hskip -1.2in \frac{1}{4
   \left(-3 a^2+\frac{4\ 2^{2/3} {\rho_0}^4 {\sin \psi}^{2/3}}{9 {c_3}^{2/3}}+\frac{2 \sqrt[3]{2} {c_3}^{2/3}}{{\sin \psi}^{2/3}}\right)
   \sqrt{\frac{19683 a^6 {c_3}^2 {\sin \psi}^2+54\ 2^{2/3} {c_3}^{4/3} {\rho_0}^4 {\sin \psi}^{8/3} \left(16 {\rho_0}^4-81
   a^4\right)-243 \sqrt[3]{2} {c_3}^{8/3} {\sin \psi}^{4/3} \left(81 a^4-16 {\rho_0}^4\right)+5832 {c_3}^4+128 {\rho_0}^{12}
   {\sin \psi}^4}{{c_3}^2 {\sin \psi}^2}}}\nonumber\\
   & & \hskip -1.2in \times\Biggl\{9 \sqrt{3} \Biggl(\frac{1}{243 {c_3}^{2/3} {\sin \psi}^{2/3}}\Biggl\{2 \left(-27 a^2 {c_3}^{2/3} {\sin \psi}^{2/3}+18 \sqrt[3]{2} {c_3}^{4/3}+4\ 2^{2/3} {\rho_0}^4
   {\sin \psi}^{4/3}\right)\nonumber\\
    & & \hskip -0.6in\Biggl[\sqrt{3} \sqrt{\frac{19683 a^6 {c_3}^2 {\sin \psi}^2+54\ 2^{2/3} {c_3}^{4/3} {\rho_0}^4 {\sin \psi}^{8/3}
   \left(16 {\rho_0}^4-81 a^4\right)-243 \sqrt[3]{2} {c_3}^{8/3} {\sin \psi}^{4/3} \left(81 a^4-16 {\rho_0}^4\right)+5832 {c_3}^4+128
   {\rho_0}^{12} {\sin \psi}^4}{{c_3}^2 {\sin \psi}^2}}\nonumber\\
   & & \hskip -0.8in - 18\ 2^{2/3} {\rho_0}^2 \sqrt{\frac{2 \sqrt[3]{2} {\rho_0}^4
   {\sin \psi}^{2/3}}{{c_3}^{2/3}}+\frac{9 {c_3}^{2/3}}{{\sin \psi}^{2/3}}}\Biggr]\Biggr\}-4\ 2^{2/3} a^2
   {\rho_0}^2 \sqrt{\frac{2 \sqrt[3]{2} {\rho_0}^4 {\sin \psi}^{2/3}}{{c_3}^{2/3}}+\frac{9 {c_3}^{2/3}}{{\sin \psi}^{2/3}}}\Biggr)\Biggr\} + {\cal O}\left(\frac{1}{N^{\frac{2}{5}}}\right).
\end{eqnarray}}
Making subsequently a small-$\psi$ expansion:
\begin{equation}
\label{sLagRC-VII}
\cos^2\theta_2 = \frac{0.003 {\sin\psi}^2 \left(32 {\rho_0}^6-405 a^4 {\rho_0}^2\right)}{{c_3}^2}-\frac{0.386 a^2 {\rho_0}^2
   {\sin\psi}^{4/3}}{{c_3}^{4/3}}-\frac{0.324 {\rho_0}^2 {\sin\psi}^{2/3}}{{c_3}^{2/3}}+0.5
   + {\cal O}\left(\frac{a^2\rho_0^6\sin^{\frac{8}{3}}\psi}{c_3^{\frac{8}{3}}}\right).
   \end{equation}
For (\ref{sLagRC-VII}) to be a valid embedding near $\theta_1=\frac{1}{N^{\frac{1}{5}}}, \theta_2=\frac{1}{N^{\frac{3}{10}}}$ (whereat an explicit local $SU(3)$-structure of the type IIA mirror and an explicit local $G_2$-structure of the M-theory uplift was obtained in \cite{T_c+Torsion}) in the UV, $\psi$ is near $\langle\psi\rangle$ determined by:
\begin{equation}
\label{sLagRC-VIII}
0.5 - 0.32 \xi + \frac{a^2(-0.386 - 0.82 a^4 + 0.216\rho_0^4)\xi^2}{\rho_0^4} + \left(0.091 - \frac{1.148 a^4}{c_3\rho_0^4}\right) = 1
\end{equation}
where $\xi\equiv\frac{\rho_0^2\sin^{\frac{2}{3}}\langle\psi\rangle}{c_3^{\frac{2}{3}}}$. This is solved to yield:
\begin{equation}
\label{sLagRC-IX}
\xi  =  2.416 - 0.993 (a \rho_0)^2 + 0.304 (a \rho_0)^4 - 0.034 (a \rho_0)^6 + {\cal O}\left((a \rho_0)^8\right).
\end{equation}
One sees that (\ref{sLagRC-IX}) can be satisfied by requiring:
\begin{equation}
\label{sLargeRC-X}
\xi\sim 1.9;\ a^2 \rho_0^2 \sim 0.8,
\end{equation}
implying:
\begin{equation}
\label{sLag-XI}
\sin\langle\psi\rangle\sim \frac{1.9^{\frac{3}{2}}c_3}{\rho_0^3},\ a\sim\frac{0.9}{\rho_0}.
\end{equation}

Similarly,
\begin{eqnarray}
\label{sLagRC-XIV}
& &  \cos^2\theta_1 = \frac{1}{2} + \frac{\left(3 \sqrt{6}-16\right) a^2 {\sin\psi}^{4/3} \left(4 a^2-{\rho_0}^2\right)}{12\ 2^{2/3} {c_3}^{4/3}}+\frac{{\sin\psi}^{2/3} \left(4
   a^2-{\rho_0}^2\right)}{2^{5/6} \sqrt{3} {c_3}^{2/3}}\nonumber\\
   & & -\frac{{\sin\psi}^2 \left(4 a^2-{\rho_0}^2\right) \left(3 \left(576-391
   \sqrt{6}\right) a^4+32 \sqrt{6} {\rho_0}^4\right)}{864 {c_3}^2} + {\cal O}\left({\sin\psi}^{7/3}\right).
\end{eqnarray}

From (\ref{sLagRC-IV}), (\ref{sLagRC-VII}) and (\ref{sLagRC-XIV}):
\begin{eqnarray}
\label{sLagRC-XVI}
& & \rho d\rho\approx \left(- \frac{2^{\frac{4}{3}}}{3\sin^{\frac{5}{3}}\psi} + \frac{2^{\frac{5}{3}}(c_1^2+c_2^2)}{3^3c_3^{\frac{2}{3}}}\sin^{\frac{1}{3}}\psi\right)d\psi;\nonumber\\
& & -\sin 2\theta_1d\theta_1\approx-2\sin\theta_1d\theta_1\approx \left(-\frac{2}{3}\frac{\rho_0^2}{2^{\frac{5}{6}}\sqrt{3}\sin^{\frac{1}{3}}\psi c_3^{\frac{2}{3}}} + \frac{4a^2(16 - 3\sqrt{6})\rho_0^2\sin^{\frac{1}{3}}\psi}{36 2^{\frac{2}{3}}c_3^{\frac{4}{3}}}\right)d\psi\nonumber\\
& & \approx -2\left(\frac{2^{\frac{1}{6}}}{3^{\frac{3}{2}}\sin\psi} + {\cal O}\left(\frac{a^2}{\sin\psi\rho_0^2}\right)\right)d\psi;\nonumber\\
& & -\sin 2\theta_2d\theta_2\approx-2\sin\theta_2d\theta_2\approx\left(\frac{0.64}\rho_0^2{3c_3^{\frac{2}{3}}\sin^{\frac{1}{3}}\psi} - \frac{1.56\rho_0^2\sin^{\frac{1}{3}}\psi a^2}{3c_3^{\frac{4}{3}}}\right)d\psi\nonumber\\
& & -2\left(\frac{0.64}{3\sin\psi} + {\cal O}\left(\frac{a^2}{\sin\psi\rho_0^2}\right)\right)d\psi.
\end{eqnarray}
So, writing $c_1=\alpha_{c_1}\rho_0^2, d\phi_1=\frac{\beta_{\phi_1}dx}{\left(g_sN\right)^{\frac{1}{4}}\frac{1}{N^{\frac{1}{5}}}} d\phi_2=\frac{\beta_{\phi_2}dy}{\left(g_sN\right)^{\frac{1}{4}}\frac{1}{N^{\frac{3}{10}}}}, d\psi=\frac{\beta_{\psi}dz}{\left(g_sN\right)^{\frac{1}{4}}}$:
\begin{eqnarray}
\label{sLagRC-XVII}
& & \frac{\rho}{3}d\rho\wedge \cos\theta_1d\phi_1\approx\left(\frac{0.07\alpha_{c_1}^2\rho_0^2}{\sin\psi} - \frac{0.84}{\sin^{\frac{5}{3}}\psi}\right)N^{\frac{3}{10}}\beta_{\psi}\beta_{\phi_1}\frac{dz\wedge dx}{\sqrt{g_s}};\nonumber\\
& & \frac{\rho}{3}d\rho\wedge \cos\theta_2d\phi_2\approx\left(\frac{0.07\alpha_{c_1}^2\rho_0^2}{\sin\psi} - \frac{0.84}{\sin^{\frac{5}{3}}\psi}\right)\beta_{\psi}\beta_{\phi_2}\frac{dz\wedge dy}{\sqrt{g_s}};\nonumber\\
& & \frac{\rho^2}{6}\sin\theta_1d\phi_1\wedge d\theta_1\approx-\frac{0.007\rho^2_\Lambda\beta_{\psi}\beta_{\phi_1}}{\sin\psi}\frac{dz\wedge dx}{\sqrt{g_s}};\nonumber\\
& & \frac{(\rho^2+6a^2)}{6}\sin\theta_2d\phi_2\wedge d\phi_2\sim-\frac{\rho_0^2}{6}\frac{0.64}{3\sin\psi}d\phi_2\wedge d\psi\nonumber\\
& & =-\frac{0.03\rho_0^2\beta_{\psi}\beta_{\phi_2}}{\sin\psi}\frac{dz\wedge dy}{\sqrt{g_s}}.
\end{eqnarray}
This implies:
\begin{eqnarray}
\label{J}
& & i^*J\approx \left(\frac{0.07\alpha_{c_1}^2\rho_0^2\beta_{\psi}\beta_{\phi_1}}{\sqrt{g_s}\sin\psi N^{\frac{3}{10}}} - \frac{0.007\rho_0^2\beta_{\psi}\beta_{\phi_1}}{\sin\psi\sqrt{g_s}N^{\frac{3}{10}}} - \frac{0.84\beta_{\psi}\beta_{\phi_1}}{N^{\frac{3}{10}}\sqrt{g_s}\sin^{\frac{5}{3}}\psi}\right)dz\wedge dx\nonumber\\
& & + \left(\frac{0.07\alpha_{c_1}^2\rho_0^2\beta_{\psi}\beta_{\phi_2}}{\sqrt{g_s}\sin\psi N^{\frac{3}{10}}} - \frac{0.03\rho_0^2\beta_{\psi}\beta_{\phi_2}}{\sin\psi\sqrt{g_s}} - \frac{0.84\beta_{\psi}\beta_{\phi_2}}{N^{\frac{3}{10}}\sqrt{g_s}\sin^{\frac{5}{3}}\psi}\right)dz\wedge dy.
\end{eqnarray}

Further,
\begin{eqnarray}
\label{sLagRC-XXI}
& & -i\frac{\rho^2}{6}d\rho(\cos\psi - i \sin\psi)\wedge\sin\theta_1d\phi_1\wedge\sin\theta_2d\phi_2
\approx -i\frac{\rho_0^6\cos\psi}{18\left(g_sN\right)^{\frac{3}{4}}}\frac{\left(\frac{2\alpha_{c_1}^2}{3^3} - \frac{2^{\frac{4}{3}}}{3\alpha}\right)}{\alpha^{\frac{3}{2}}}\beta_{\psi}\beta_{\phi_1}\beta_{\phi_2}dz\wedge dx\wedge dy;\nonumber\\
& & -\frac{\rho^3}{18}(\cos\psi - i \sin\psi)\sin\theta_1d\phi_1\wedge\sin\theta_2d\phi_2\wedge d\psi\approx
-\frac{\rho_0^3\cos\psi\beta_{\psi}\beta_{\phi_1}\beta_{\phi_2}}{\left(g_sN\right)^{\frac{3}{4}}}dx\wedge dy\wedge dz;\nonumber\\
& & -i\frac{\rho^3}{18}(\cos\psi - i\sin\psi)(\sin\theta_1\cos\theta_2d\theta_2 + \sin\theta_2\cos\theta_1d\theta_1)\wedge d\phi_1\wedge d\phi_2\nonumber\\
& & \approx -i\frac{\rho_0^6\cos\psi}{36\left(g_sN\right)^{\frac{3}{4}}}\frac{\beta_{\psi}\beta_{\phi_1}\beta_{\phi_2}}{\alpha^{\frac{3}{2}}}
\left(N^{\frac{1}{10}}0.21 + \frac{2^{\frac{1}{6}}}{3^{\frac{3}{2}}N^{\frac{1}{10}}}\right)dz\wedge dx\wedge dy.
\end{eqnarray}

\section{EOMs for (Vector Mode) Metric  and Gauge Fluctuations,  and Their Solutions near $u=0$}
\setcounter{equation}{0} \seceqbb

(A)~~EOM for $H_{ty}$
\begin{eqnarray}
\label{EOM-Hty}
& & \frac{e^{2 i (q x-t w)}\sqrt{\mu } \sqrt{c^2 e^{2 \Phi }+\left(\frac{r_h}{u}\right)^{9/2}} r_h^2  \left(H_{ty}(u) \sqrt{c^2 e^{2 \Phi }+\left(\frac{r_h}{u}\right)^{9/2}}-c
   e^{\Phi } u^4 \phi '(u)+c e^{\Phi } \phi '(u)\right)}{36\sqrt{c^2 e^{2 \Phi }
   u^4+\left(\frac{r_h^9}{u}\right)^{1/2}} r_h \left(u^4-1\right)}-\nonumber\\
   & & \frac{e^{i (q x-t w)}
   \sqrt{\frac{r_h^8}{L^6 u^{10} g_s^{10/3}}} \Biggl(g_1(u) r_h^2 \left(3 H_{ty}'(u)-u H_{ty}''(u)\right)+L^4 q^2 u H_{ty}(u)+L^4 q u w
   H_{xy}(u)\Biggr)}{2 L^4 u}=0.\nonumber\\
   & &
\end{eqnarray}
(B)~~EOM for $H_{xy}$
\begin{eqnarray}
\label{EOM-Hxy}
& & \hskip -0.7in \frac{e^{i (q x-2 t w)}}{36 L^4 u^4 g_1(u)}\Biggl[L^4 {H_{xy}}(u)\Biggl\{\sqrt{\mu } {r_h}^3 g_1(u) e^{i q x} \sqrt{\frac{{r_h}}{u}} \sqrt{\frac{{r_h}^4}{c^2 e^{2 \Phi } u^4+\left(\frac{r_h^9}{u}\right)^{1/2}}}+18 u^4 w^2 e^{i t w} \sqrt{\frac{r_h^8}{L^6 u^{10}
   g_s^{10/3}}}\Biggr\}\nonumber\\
   & & \hskip -0.7in +18 L^4 q u^4 w {H_{ty}}(u) e^{i t w} \sqrt{\frac{r_h^8}{L^6 u^{10} g_s^{10/3}}}+18 u^3 g_1(u) r_h^2 e^{i t w} \Biggl(u g_1(u)
   {H_{xy}}''(u)+\left(g_1(u)-4\right) {H_{xy}}'(u)\Biggr) \sqrt{\frac{r_h^8}{L^6 u^{10} g_s^{10/3}}}\Biggr].\nonumber\\
   & &
\end{eqnarray}
(C) For $\mu=u$ and $\nu=y$
\begin{eqnarray}
-a_{EH}\frac{i e^{i (q x-t w)} \sqrt{\frac{r_h^8}{L^6 u^{10} g_s^{10/3}}} \left(q g_1(u) H_{xy}'(u)+w H_{ty}'(u)\right)}{2 g_1(u)}=0.
\end{eqnarray}
(D)~~EOM for $\phi$:
\begin{eqnarray}
\label{EOM-phi}
& & \hskip -0.7in \Biggl(\frac{c \sqrt{\mu } r_h^5 \left(\frac{{r_h}}{u}\right)^{11/4} \sqrt{c^2 e^{2 \Phi }+\left(\frac{{r_h}}{u}\right)^{9/2}} \left(c^6 e^{6 \Phi
   } u^{13}+3 c^4 e^{4 \Phi } r_h^4 u^9 \sqrt{\frac{{r_h}}{u}}+3 c^2 e^{2 \Phi } {r_h}^9 u^4+{r_h}^{13}
   \sqrt{\frac{{r_h}}{u}}\right) e^{\Phi +2 i q x-2 i t w}}{18 \left(c^2 e^{2 \Phi } u^4+{r_h}^4 \sqrt{\frac{{r_h}}{u}}\right)^5
   \left(\frac{r_h^5}{c^2 e^{2 \Phi } u^5 \sqrt{\frac{{r_h}}{u}}+{r_h}^5}\right)^{3/2}}\Biggr)H_{ty}
 \nonumber\\
 & &  \hskip -0.7in -\Biggl(\frac{c \sqrt{\mu } {r_h}^6 \left(\frac{{r_h}}{u}\right)^{7/4} \sqrt{c^2 e^{2 \phi }+\left(\frac{{r_h}}{u}\right)^{9/2}} \left(c^6 e^{6 \Phi
   } u^{13}+3 c^4 e^{4 \Phi } {r_h}^4 u^9 \sqrt{\frac{{r_h}}{u}}+3 c^2 e^{2 \Phi } {r_h}^9 u^4+{r_h}^{13}
   \sqrt{\frac{{r_h}}{u}}\right) e^{\Phi +2 i q x-2 i t w}}{36 \left(c^2 e^{2 \Phi } u^4+{r_h}^4 \sqrt{\frac{{r_h}}{u}}\right)^5
   \left(\frac{{r_h}^5}{c^2 e^{2 \Phi } u^5 \sqrt{\frac{{r_h}}{u}}+{r_h}^5}\right)^{3/2}}\Biggr)H^{'}_{ty}
  \nonumber \\
  & & \hskip -0.7in +\Biggl(\frac{\sqrt{\mu } {r_h}^6 \left(u^4-1\right) \left(\frac{{r_h}}{u}\right)^{7/4} \left(c^6 e^{6 \Phi } u^{13}+3 c^4 e^{4 \Phi } {r_h}^4 u^9
   \sqrt{\frac{{r_h}}{u}}+3 c^2 e^{2 \Phi } {r_h}^9 u^4+{r_h}^{13} \sqrt{\frac{{r_h}}{u}}\right) e^{2 i (q x-t w)}}{36 \left(c^2 e^{2 \Phi
   } u^5+{r_h}^4 u \sqrt{\frac{{r_h}}{u}}\right)^4 \left(\frac{{r_h}^5}{c^2 e^{2 \Phi } u^5
   \sqrt{\frac{{r_h}}{u}}+{r_h}^5}\right)^{3/2}}\Biggr)\phi^{''}
   \nonumber \\
   & & \hskip -0.7in -\Biggl[\frac{\sqrt{\mu } {r_h}^6 \left(\frac{{r_h}}{u}\right)^{7/4}e^{2 i (q x-t w)}}{144 u^5 \left(c^2 e^{2 \Phi } u^4+{r_h}^4 \sqrt{\frac{{r_h}}{u}}\right)^4
   \left(\frac{{r_h}^5}{c^2 e^{2 \Phi } u^5 \sqrt{\frac{{r_h}}{u}}+{r_h}^5}\right)^{3/2}} \nonumber\\
   & & \hskip -0.4in \left\{-8 c^6 e^{6 \Phi } u^{13} \left(u^4+1\right)-3 c^4 e^{4 \Phi }  u^9
   \left(5 u^4+11\right) \sqrt{\frac{r_h^9}{u}}-6c^2e^{2 \Phi }{r_h}^9 u^4 \left(u^4+7\right)+ \left(u^4-17\right)
   \sqrt{\frac{r_h^{27}}{u}}\right\} \Biggr]\phi^{'}
   \nonumber \\
   & & \hskip -0.7in +\Biggl(\frac{\pi  {g_s} \sqrt{\mu } N u \left(\frac{{r_h}}{u}\right)^{3/4} \sqrt{\frac{{r_h}^5}{c^2 e^{2 \Phi } u^5
   \sqrt{\frac{{r_h}}{u}}+{r_h}^5}} e^{2 i (q x-t w)} \left({i w}^2 \left(c^2 e^{2 \Phi } u^4+{r_h}^4
   \sqrt{\frac{{r_h}}{u}}\right)+{iq}^2 {r_h}^4 \left(u^4-1\right) \sqrt{\frac{{r_h}}{u}}\right)}{9 {r_h}^6 \left(u^4-1\right)}\Biggr)\phi = 0.\nonumber\\
   & &
\end{eqnarray}

The $H_{ty}(u)$ EOM, setting $q=0$ and near $u=0$ is:
\begin{eqnarray}
\label{Htyq0u0}
& & \frac{1}{36} \sqrt{\mu } {r_h}^{13/4} u^{7/4} {H_{ty}}(u)-\frac{{r_h}^6 \left(u {H_{ty}}''(u)-3 {H_{ty}}'(u)\right)}{2 {g_s}^{5/3} L^7}=0,
\end{eqnarray}
whose solution is given by:
\begin{eqnarray}
\label{Htyq0u0solution}
& & H_{ty}(u) =  \frac{4 2^{2/11} c_1 {g_s}^{40/33} L^{56/11} \mu ^{4/11} u^2 \Gamma \left(-\frac{5}{11}\right) I_{-\frac{16}{11}}\left(\frac{4 \sqrt{2} {g_s}^{5/6}
   L^{7/2} \sqrt[4]{\mu } u^{11/8}}{33 {r_h}^{11/8}}\right)}{33 33^{5/11} {r_h}^2}\nonumber\\
   & & -\frac{64 (-1)^{5/11} 2^{2/11} c_2 {g_s}^{40/33} L^{56/11} \mu
   ^{4/11} u^2 \Gamma \left(\frac{16}{11}\right) I_{\frac{16}{11}}\left(\frac{4 \sqrt{2} {g_s}^{5/6} L^{7/2} \sqrt[4]{\mu } u^{11/8}}{33
   {r_h}^{11/8}}\right)}{363 33^{5/11} {r_h}^2}\nonumber\\
   & & = \kappa_1 + \kappa_2 u^{\frac{11}{4}} + \gamma\kappa_2 u^4 + ....,
\end{eqnarray}
where:
\begin{eqnarray}
\label{Htynearu0}
& & \kappa_1 \equiv c_1;\nonumber\\
& & \kappa_2 \equiv -\frac{64 \sqrt{2} \pi ^{7/4} c_1 {g_s}^{41/12} \sqrt{\mu } N^{7/4}}{495 {r_h}^{11/4}};\nonumber\\
& & \gamma \equiv \frac{320 (-1)^{5/11} 2^{21/22} \pi ^{35/44} c_2 {g_s}^{205/132} \mu ^{5/22} N^{35/44} \Gamma \left(\frac{16}{11}\right)}{121 33^{10/11} c_1 {r_h}^{5/4}
   \Gamma \left(\frac{27}{11}\right)}.
\end{eqnarray}
The $H_{xy}$ EOM near $u=0$ is:
\begin{eqnarray}
\label{HxyEOM-i}
& & 18 r_h^2 \left(u {H_{xy}}''(u)-3 {H_{xy}}'(u)\right) \sqrt{\frac{r_h^8}{L^6 g_s^{10/3}}}+L^4 \sqrt{\mu } {r_h}^{13/4} u^{7/4} {H_{xy}}(u)=0,
\end{eqnarray}
whose solution is given by:
\begin{eqnarray}
& & H_{xy}(u)=\frac{64 2^{2/11} c_2 L^{56/11} \mu ^{4/11} {r_h}^{26/11} u^2 \Gamma \left(\frac{16}{11}\right) g_s^{40/33} J_{\frac{16}{11}}\left(\frac{4 \sqrt{2} L^{7/2}
   \sqrt[4]{\mu } {r_h}^{13/8} u^{11/8} g_s^{5/6}}{33 r_h^3}\right)}{363 33^{5/11} r_h^{48/11}}\nonumber\\
   & & +\frac{4 2^{2/11} c_1 L^{56/11} \mu ^{4/11} {r_h}^{26/11}
   u^2 \Gamma \left(-\frac{5}{11}\right) g_s^{40/33} J_{-\frac{16}{11}}\left(\frac{4 \sqrt{2} L^{7/2} \sqrt[4]{\mu } {r_h}^{13/8} u^{11/8} g_s^{5/6}}{33
   r_h^3}\right)}{33 33^{5/11} r_h^{48/11}}.
\end{eqnarray}
Substituting (\ref{Htyq0u0solution}), the $\phi(u)$ EOM near $u=0$ can be approximated by:
\begin{eqnarray}
\label{phi EOM i}
\hskip -0.6in -4 c {g_s} {r_h}^4 u {H_{ty}}'(u)+8 c {g_s} {r_h}^4 {H_{ty}}(u)+16 \pi  {g_s} i N {r_h}^{9/4} u^{11/4} \left(q^2-w^2\right) \phi
   (u)+\frac{17 {r_h}^{25/4} \phi '(u)}{u^{9/4}}-\frac{4 {r_h}^{25/4} \phi ''(u)}{u^{5/4}}=0,\nonumber\\
   & &
\end{eqnarray}
or
\begin{eqnarray}
\label{phi EOM ii}
& & \hskip -0.6in\frac{{r_h}^{7/4} \left(c {g_s} u^{9/4} \left(-8 \gamma  {k_2} u^4+8 {k_1}-3 {k_2} u^{11/4}\right)-4 {r_h}^{9/4} u \phi ''(u)+17
   {r_h}^{9/4} \phi '(u)\right)+16 \pi  {g_s} i N u^5 \left(q^2-w^2\right) \phi (u)}{u^{9/4}}=0,\nonumber\\
   & &
   \end{eqnarray}
whose solution is given by:{\footnotesize
\begin{eqnarray}
\label{phiu0solution_i}
& &\hskip -1.4in \phi(u) = -\frac{1}{3024 \pi
   ^{9/16} {r_h}^{9/4} \left(-w^2\right)^{23/16} \left(\frac{u^3 \sqrt{-{g_s} i N w^2}}{{r_h}^2}\right)^{7/8} \Gamma \left(\frac{2}{3}\right) \Gamma
   \left(\frac{4}{3}\right) \Gamma \left(\frac{37}{24}\right) \Gamma \left(\frac{53}{24}\right)}\nonumber\\
   & & \hskip -1.4in \times\Biggl\{{g_s}^{7/16} u^{13/4} \Biggl(\frac{1}{{r_h}^4}
   \Biggl\{u^{19/8} \Gamma \left(\frac{2}{3}\right) \Biggl[192 3^{7/8} c \gamma  {g_s}^{9/16} {k_2} \sqrt[8]{\pi }
   {r_h}^4 u^{13/8} \left(-w^2\right)^{23/16} I_{\frac{7}{8}}\left(\frac{2 \sqrt{\pi } u^3 \sqrt{-{g_s} i N w^2}}{3 {r_h}^2}\right) \Gamma
   \left(\frac{1}{3}\right) \Gamma \left(\frac{37}{24}\right) \Gamma \left(\frac{15}{8}\right) \Gamma \left(\frac{53}{24}\right)\nonumber\\
   & &\hskip -1.4in   \,
   _1F_2\left(\frac{1}{3};\frac{1}{8},\frac{4}{3};-\frac{{g_s} i N \pi  u^6 w^2}{9 {r_h}^4}\right)-\frac{1}{\sqrt[4]{\frac{u^3 \sqrt{-{g_s} i N w^2}}{{r_h}^2}}}
   \Biggl\{w^2 \Gamma \left(\frac{4}{3}\right)
   \Biggl(64
   \sqrt[8]{3} c {g_s}^{25/16} i {k_1} N \pi  \left(-w^2\right)^{23/16} I_{-\frac{7}{8}}\left(\frac{2 \sqrt{\pi } u^3 \sqrt{-{g_s} i N w^2}}{3
   {r_h}^2}\right) \Gamma \left(\frac{1}{8}\right) \Gamma \left(\frac{13}{24}\right) \Gamma \left(\frac{53}{24}\right)\nonumber\\
   & & \hskip -1.4in\times \,
   _1F_2\left(\frac{13}{24};\frac{37}{24},\frac{15}{8};-\frac{{g_s} i N \pi  u^6 w^2}{9 {r_h}^4}\right) u^{29/8}\nonumber\\
   & &\hskip -1.4in +\Gamma \left(\frac{37}{24}\right)
   \Biggl[9 {r_h}^{5/2} \Gamma \left(\frac{53}{24}\right) \Biggl(\frac{1}{u^3}\Biggl\{\left(-w^2\right)^{7/16} I_{-\frac{7}{8}}\left(\frac{2 \sqrt{\pi } u^3 \sqrt{-{g_s}
   i N w^2}}{3 {r_h}^2}\right) \Gamma \left(\frac{1}{8}\right)\nonumber\\
    & & \hskip -1.2in\times\Biggl[21 \sqrt[8]{3} c {g_s}^{9/16} {k_2} u^{27/8}-24 c {g_s}^{9/16} {k_2}
   \sqrt[16]{\pi } \sqrt[8]{\frac{\sqrt{{g_s}} \sqrt{i} \sqrt{N} u^3 \sqrt{-w^2}}{{r_h}^2}} I_{-\frac{1}{8}}\left(\frac{2 \sqrt{{g_s}} \sqrt{i}
   \sqrt{N} \sqrt{\pi } u^3 \sqrt{-w^2}}{3 {r_h}^2}\right)\nonumber\\
   & & \hskip -1.4in \Gamma \left(\frac{15}{8}\right) u^{27/8}-112 \sqrt[8]{3} i^{7/16} N^{7/16} \pi  \sqrt{{r_h}}
   \left(-w^2\right)^{7/16} \left(\frac{u^3 \sqrt{-{g_s} i N w^2}}{{r_h}^2}\right)^{9/8} c_1\Biggr] {r_h}^{3/2}\Biggr\}
    +8 \sqrt[8]{{g_s}}
   \sqrt[8]{3 \pi } \sqrt[8]{\frac{u^3 \sqrt{-{g_s} i N w^2}}{{r_h}^2}} I_{\frac{7}{8}}\left(\frac{2 \sqrt{\pi } u^3 \sqrt{-{g_s} i N w^2}}{3
   {r_h}^2}\right) \Gamma \left(\frac{15}{8}\right)\nonumber\\
   & & \hskip -1.4in \left(14 i^{9/16} N^{9/16} (-\pi )^{7/8} \left(-{g_s} i N w^2\right)^{3/8} c_2 w^2+3^{3/4} c
   {g_s}^{7/16} {k_2} {r_h}^{3/2} u^{3/8} \left(-w^2\right)^{7/16} \sqrt[8]{\frac{u^3 \sqrt{-{g_s} i N w^2}}{{r_h}^2}} \Gamma
   \left(\frac{1}{8}\right) \left(;\frac{9}{8};-\frac{{g_s} i N \pi  u^6 w^2}{9 {r_h}^4}\right)\right)\Biggr)\nonumber\\
   & & \hskip -1.2in-64 \sqrt[8]{3} c \gamma  {g_s}^{25/16}
   i {k_2} N \pi  u^{61/8} \left(-w^2\right)^{23/16} I_{-\frac{7}{8}}\left(\frac{2 \sqrt{\pi } u^3 \sqrt{-{g_s} i N w^2}}{3 {r_h}^2}\right) \Gamma
   \left(\frac{1}{8}\right) \Gamma \left(\frac{29}{24}\right) \, _1F_2\left(\frac{29}{24};\frac{15}{8},\frac{53}{24};-\frac{{g_s} i N \pi  u^6 w^2}{9
   {r_h}^4}\right)\Biggr]
   \Biggr)\Biggr\}\Biggr]\Biggr\}\nonumber\\
& &  \hskip -1.4in  -192 3^{7/8} c {g_s}^{9/16} {k_1}
   \sqrt[8]{\pi } \left(-w^2\right)^{23/16} I_{\frac{7}{8}}\left(\frac{2 \sqrt{\pi } u^3 \sqrt{-{g_s} i N w^2}}{3 {r_h}^2}\right) \Gamma
   \left(-\frac{1}{3}\right) \Gamma \left(\frac{4}{3}\right) \Gamma \left(\frac{37}{24}\right) \Gamma \left(\frac{15}{8}\right) \Gamma
   \left(\frac{53}{24}\right) \, _1F_2\left(-\frac{1}{3};\frac{1}{8},\frac{2}{3};-\frac{{g_s} i N \pi  u^6 w^2}{9 {r_h}^4}\right)\Biggr)\Biggr\}.
       \end{eqnarray}}

\section{Frobenius Solution  of  EOM of Gauge-Invariant $Z_s(u)$ for Scalar Modes of Metric Fluctuations for
($\alpha^\prime=1$) $r:\log r<\log N$}
\setcounter{equation}{0} \seceqcc

The $Z_s(u)$ EOM can be rewritten as:
\begin{equation}
(u-1)^2Z_s^{\prime\prime}(u) + (u-1)P(u-1) Z_s^\prime(u) + Q(u-1) Z_s(u) = 0,
\end{equation}
in which $P(u-1) = \sum_{n=0}^\infty p_n(u-1)^n$ and $Q(u-1) = \sum_{m=0}^\infty q_n (u-1)^n$  wherein, up to ${\cal O}\left(\frac{1}{N}\right)$:
{\footnotesize
\begin{eqnarray}
\label{pn+qn_up_to_2nd_order}
& & p_0 = 1,\nonumber\\
& & p_1 = \frac{3 {g_s}^2 M^2 {N_f} \log (N) \left(28 {q_3}^4+36 {q_3}^2 {\omega_3}^2-81 {\omega_3}^4\right)}{64 \pi ^2 N \left(2 {q_3}^2-3
   {\omega_3}^2\right)^2}+\frac{10 {q_3}^2+9 {\omega_3}^2}{4 {q_3}^2-6 {\omega_3}^2},\nonumber\\
   & & p_2 = \frac{3 {g_s}^2 M^2 {N_f} \log (N) \left(712 {q_3}^6-948 {q_3}^4 {\omega_3}^2-162 {q_3}^2 {\omega_3}^4+405 {\omega_3}^6\right)}{64 \pi ^2 N \left(2
   {q_3}^2-3 {\omega_3}^2\right)^3}+\frac{364 {q_3}^4-420 {q_3}^2 {\omega_3}^2+99 {\omega_3}^4}{4 \left(2 {q_3}^2-3 {\omega_3}^2\right)^2};\nonumber\\
   & & q_0 = \frac{3 {g_s}^2 M^2 {N_f} \log (N) \left(\left({\omega_3}^2+4\right) \left(27 {\omega_3}^2-10 {q_3}^2\right)-8 {q_3}^2 {\omega_3}^2 \log
   (N)\right)}{4096 \pi ^2 N {q_3}^2}+\frac{{\omega_3}^2}{16},\nonumber\\
   & & q_1 = \frac{1}{4096 \pi ^2 N {q_3}^2 \left(2 {q_3}^2-3 {\omega_3}^2\right)^2}\nonumber\\
   & & \times\Biggl\{3 {g_s}^2 M^2 {N_f} \log (N) \biggl(-8 \log (N) \left(4 {q_3}^2-3 {\omega_3}^2\right) \left(2 {q_3}^3-3 {q_3} {\omega_3}^2\right)^2 \nonumber\\
   & & -96
   {q_3}^8+8 {q_3}^6 \left(51 {\omega_3}^2-52\right)-36 {q_3}^4 {\omega_3}^2 \left(9 {\omega_3}^2+52\right)-54 {q_3}^2 {\omega_3}^4 \left(9
   {\omega_3}^2-28\right)+81 {\omega_3}^6 \left(7 {\omega_3}^2+20\right)\biggr)\Biggr\}\nonumber\\
   & & +\frac{8
   {q_3}^4-2 {q_3}^2 \left(9 {\omega_3}^2+32\right)+9 {\omega_3}^4}{32 {q_3}^2-48 {\omega_3}^2},\nonumber\\
   & & q_2 = \frac{-96 {q_3}^6+52 {q_3}^4 \left(7 {\omega_3}^2-64\right)+{q_3}^2 \left(3456 {\omega_3}^2-444 {\omega_3}^4\right)+171 {\omega_3}^6}{64 \left(2
   {q_3}^2-3 {\omega_3}^2\right)^2}\nonumber\\
   & & -\frac{3 {g_s}^2 M^2 {N_f} \log (N) }{16384 \pi ^2 N {q_3}^2 \left(2 {q_3}^2-3 {\omega_3}^2\right)^3}\nonumber\\
   & & \times \Biggl[-8 {q_3}^2 \log (N) \left(24 {q_3}^2-19 {\omega_3}^2\right) \left(2
   {q_3}^2-3 {\omega_3}^2\right)^3+1920 {q_3}^{10}-16 {q_3}^8 \left(157 {\omega_3}^2-6268\right)\nonumber\\
   & & -96 {q_3}^6 {\omega_3}^2 \left(351
   {\omega_3}^2+1444\right)+216 {q_3}^4 {\omega_3}^4 \left(559 {\omega_3}^2+76\right)-864 {q_3}^2 {\omega_3}^6 \left(173 {\omega_3}^2+53\right)+243 {\omega_3}^8
   \left(265 {\omega_3}^2+308\right)\Biggr].
\end{eqnarray}}
The Frobenius method then dictates that the solution is given by:
\begin{equation}
\label{solution-i}
Z_s(u) = \left(1 - u \right)^{\frac{3 {g_s}^2 M^2 {N_f} {q_3} {\omega_3}^2 \log (N) \left(8 {q_3}^2 {\omega_3}^2 \log (N)+\left({\omega_3}^2+4\right) \left(10 {q_3}^2-27
   {\omega_3}^2\right)\right)}{2048 \pi ^2 N \left(-{q_3}^2 {\omega_3}^2\right)^{3/2}}-\frac{i {\omega_3}}{4}}\left(1 + \sum_{m=1}a_m (u - 1)^m\right),
\end{equation}
where
{\footnotesize
\begin{eqnarray}
\label{a1a2-sound}
& & a_1 = \frac{8 i {q_3}^4+2 {q_3}^2 \left(-9 i {\omega_3}^2+10 {\omega_3}-32 i\right)+9 (2+i {\omega_3}) {\omega_3}^3}{8 ({\omega_3}+2 i) \left(3 {\omega_3}^2-2
   {q_3}^2\right)}\nonumber\\
   & & -\frac{1}{4096 \pi ^2 N
   {q_3}^2 {\omega_3} ({\omega_3}+2 i)^2 \left(2 {q_3}^2-3 {\omega_3}^2\right)^2}\nonumber\\
   & & \times\Biggl\{3 i {g_s}^2 M^2 {N_f} \log (N) \biggl(({\omega_3}+2 i) \biggl(-32 {q_3}^8 ({\omega_3}+10 i)-8 {q_3}^6 \left(27
   {\omega_3}^3-146 i {\omega_3}^2+364 {\omega_3}-520 i\right)\nonumber\\
   & & +12 {q_3}^4 {\omega_3}^2 \left(141 {\omega_3}^3-486 i {\omega_3}^2+356 {\omega_3}-1336 i\right)-54
   {q_3}^2 {\omega_3}^4 \left(59 {\omega_3}^3-130 i {\omega_3}^2+44 {\omega_3}-200 i\right)\nonumber\\
   & & +81 {\omega_3}^6 \left(23 {\omega_3}^3-18 i {\omega_3}^2+4 {\omega_3}+72
   i\right)\biggr)-8 {q_3}^2 {\omega_3} \log (N) \biggl[16 {q_3}^6 ({\omega_3}+4 i)-4 {q_3}^4 {\omega_3} \left(15 {\omega_3}^2+60 i {\omega_3}-52\right)\nonumber\\
   & & +24
   {q_3}^2 {\omega_3}^3 \left(3 {\omega_3}^2+12 i {\omega_3}-10\right)-27 {\omega_3}^5 \left({\omega_3}^2+4 i {\omega_3}+4\right)\biggr]\biggr)\Biggr\},\nonumber\\
   & & a_2 = \frac{1}{128 ({\omega_3}+2 i) ({\omega_3}+4 i) \left(3 {\omega_3}^2-2 {q_3}^2\right)}\nonumber\\
   & & \times{32 {q_3}^6-32 {q_3}^4 {\omega_3} (3 {\omega_3}+8 i)+2 {q_3}^2 \left(45 {\omega_3}^4+98 i {\omega_3}^3+624 {\omega_3}^2+32 i {\omega_3}+3072\right)-3
   {\omega_3}^3 \left(9 {\omega_3}^3+2 i {\omega_3}^2+48 {\omega_3}+32 i\right)}\nonumber\\
   & & -\frac{1}{32768 \pi ^2 N {q_3}^2 {\omega_3} ({\omega_3}+2 i)^2
   ({\omega_3}+4 i)^2 \left(2 {q_3}^2-3 {\omega_3}^2\right)^2}\nonumber\\
   & & \times\Biggl\{3
   {g_s}^2 M^2 {N_f} \log (N) \Biggl[(-{\omega_3}-2 i) \biggl(128 {q_3}^{10} \left({\omega_3}^2+19 i {\omega_3}-30\right)+256 {q_3}^8 \left(3
   {\omega_3}^4-18 i {\omega_3}^3+33 {\omega_3}^2-76 i {\omega_3}+160\right)\nonumber\\
   & & -8 {q_3}^6 \left(927 {\omega_3}^6-104 i {\omega_3}^5+17784 {\omega_3}^4+11808 i
   {\omega_3}^3+70256 {\omega_3}^2+42368 i {\omega_3}+93440\right)\nonumber\\
   & & +12 {q_3}^4 {\omega_3}^2 \left(1485 {\omega_3}^6+128 i {\omega_3}^5+31912 {\omega_3}^4+14176 i
   {\omega_3}^3+96656 {\omega_3}^2-68992 i {\omega_3}+262912\right)\nonumber\\
   & & -18 {q_3}^2 {\omega_3}^4 \left(945 {\omega_3}^6-1388 i {\omega_3}^5+27920 {\omega_3}^4+13664 i
   {\omega_3}^3+54736 {\omega_3}^2-80768 i {\omega_3}+171776\right)\nonumber\\
   & & +81 {\omega_3}^6 \left(69 {\omega_3}^6-416 i {\omega_3}^5+3960 {\omega_3}^4+4128 i {\omega_3}^3+5520
   {\omega_3}^2+4736 i {\omega_3}+768\right)\biggr)-8 {q_3}^2 {\omega_3} \log (N)\nonumber\\
    & & \times\biggl(64 {q_3}^8 \left({\omega_3}^2+9 i {\omega_3}-16\right)-32 {q_3}^6
   {\omega_3} \left(9 {\omega_3}^3+89 i {\omega_3}^2-240 {\omega_3}-192 i\right)\nonumber\\
   & & +4 {q_3}^4 {\omega_3} \left(117 {\omega_3}^5+1198 i {\omega_3}^4-3612 {\omega_3}^3-1624
   i {\omega_3}^2-8064 {\omega_3}-9344 i\right)\nonumber\\
   & & -12 {q_3}^2 {\omega_3}^3 ({\omega_3}+4 i)^2 \left(27 {\omega_3}^3+52 i {\omega_3}^2+116 {\omega_3}+296 i\right)\nonumber\\
   & & +9
   {\omega_3}^5 ({\omega_3}+2 i)^2 \left(9 {\omega_3}^3+46 i {\omega_3}^2+64 {\omega_3}+32 i\right)\biggr)\Biggr]\Biggr\}.
   \end{eqnarray}}
   As stated in {\bf 5.3.1}, imposing Dirichlet boundary condition $Z_s(u=0)=0$ and going up to second order in powers of $(u-1)$ in (\ref{solution-i}) and considering in the hydrodynamical limit $\omega_3^nq_3^m:m+n=2$ one obtains:
\begin{equation}
\label{vs+Gammas-second_order}
\omega_3 = -\frac{2 {q_3}}{\sqrt{3}}-\frac{9 i {q_3}^2}{32},
\end{equation}
which yields a result for the speed of sound similar to (\ref{v_s 1}) for $n=0,1$.

To get the LO or conformal result for the speed of sound $v_s = \frac{1}{\sqrt{3}}$, let us go to the fourth order in (\ref{solution-i}). For this, up to ${\cal O}\left(\frac{1}{N}\right)$, we will need:
{\footnotesize
\begin{eqnarray}
\label{pn+qn_up_to_fourth_order}
& & \hskip -0.5in p_3 = \frac{3 {g_s}^2 M^2 {N_f} \log (N) \left(6256 {q_3}^8-9600 {q_3}^6 {\omega_3}^2-4104 {q_3}^4 {\omega_3}^4+12960 {q_3}^2 {\omega_3}^6-5589
   {\omega_3}^8\right)}{64 \pi ^2 N \left(2 {q_3}^2-3 {\omega_3}^2\right)^4}\nonumber\\
   & & \hskip -0.5in+\frac{3880 {q_3}^6-4788 {q_3}^4 {\omega_3}^2+270 {q_3}^2 {\omega_3}^4+729
   {\omega_3}^6}{8 \left(2 {q_3}^2-3 {\omega_3}^2\right)^3},\nonumber\\
   & &\hskip -0.5in p_4 = \frac{3 {g_s}^2 M^2 {N_f} \log (N) }{64 \pi ^2 N \left(2 {q_3}^2-3 {\omega_3}^2\right)^5}\nonumber\\
   & &\hskip -0.5in \times \left(53536 {q_3}^{10}-110256 {q_3}^8 {\omega_3}^2-2736 {q_3}^6 {\omega_3}^4+168264 {q_3}^4
   {\omega_3}^6-156006 {q_3}^2 {\omega_3}^8+47385 {\omega_3}^{10}\right)\nonumber\\
   & &\hskip -0.5in +\frac{3 \left(17360
   {q_3}^8-32992 {q_3}^6 {\omega_3}^2+19320 {q_3}^4 {\omega_3}^4-5112 {q_3}^2 {\omega_3}^6+1485 {\omega_3}^8\right)}{16 \left(2 {q_3}^2-3
   {\omega_3}^2\right)^4};\nonumber\\
   & &\hskip -0.5in q_3 = \frac{3 {g_s}^2 M^2 {N_f} \log (N) }{4096 \pi ^2 N {q_3}^2 \left(2 {q_3}^2-3
   {\omega_3}^2\right)^4}\nonumber\\
   & &\hskip -0.5in \times \biggl[-40 {q_3}^2 \log (N) \left(2 {q_3}^2-3 {\omega_3}^2\right)^4 \left({q_3}^2-{\omega_3}^2\right)-3552
   {q_3}^{12}+416 {q_3}^{10} \left(31 {\omega_3}^2-648\right)+96 {q_3}^8 {\omega_3}^2 \left(466 {\omega_3}^2+5695\right)
   \nonumber\\
& & \hskip -0.5in   -288 {q_3}^6 {\omega_3}^4
   \left(1107 {\omega_3}^2+818\right)+54 {q_3}^4 {\omega_3}^6 \left(12319 {\omega_3}^2+840\right)-162 {q_3}^2 {\omega_3}^8 \left(3779
   {\omega_3}^2+1632\right)\nonumber\\
   & & \hskip -0.5in+729 {\omega_3}^{10} \left(293 {\omega_3}^2+250\right)\biggr]\nonumber\\
   & &\hskip -0.5in+\frac{40 {q_3}^8-4 {q_3}^6 \left(55 {\omega_3}^2+1488\right)+6 {q_3}^4 {\omega_3}^2 \left(75 {\omega_3}^2+1696\right)-9
   {q_3}^2 {\omega_3}^4 \left(45 {\omega_3}^2+464\right)+135 {\omega_3}^8}{16 \left(2 {q_3}^2-3 {\omega_3}^2\right)^3},\nonumber\\
   & &\hskip -0.5in q_4 = \nonumber\\
   & & \hskip -0.5in\frac{-640 {q_3}^{10}+48 {q_3}^8 \left(99 {\omega_3}^2-12352\right)-288 {q_3}^6 {\omega_3}^2 \left(49 {\omega_3}^2-4400\right)+216 {q_3}^4 {\omega_3}^4
   \left(97 {\omega_3}^2-3680\right)-1728 {q_3}^2 {\omega_3}^6 \left(9 {\omega_3}^2-70\right)+4617 {\omega_3}^{10}}{256 \left(2 {q_3}^2-3
   {\omega_3}^2\right)^4}\nonumber\\
   & &\hskip -0.5in -\frac{1}{65536 \pi ^2 N {q_3}^2 \left(2 {q_3}^2-3 {\omega_3}^2\right)^5}\nonumber\\
   & & \hskip -0.5in\times\Biggl\{3 {g_s}^2 M^2 {N_f} \log (N) \biggl(3 \biggl[112128 {q_3}^{14}-64 {q_3}^{12} \left(6133 {\omega_3}^2-190428\right)-192
   {q_3}^{10} {\omega_3}^2 \left(15391 {\omega_3}^2+151076\right)\nonumber\\
   & & \hskip -0.5in+720 {q_3}^8 {\omega_3}^4 \left(30785 {\omega_3}^2+17332\right)-28800 {q_3}^6 {\omega_3}^6
   \left(2079 {\omega_3}^2-443\right)+324 {q_3}^4 {\omega_3}^8 \left(252677 {\omega_3}^2+420\right)-972 {q_3}^2 {\omega_3}^{10} \left(58571
   {\omega_3}^2+17828\right)\nonumber\\
   & & \hskip -0.5in+729 {\omega_3}^{12} \left(22027 {\omega_3}^2+12156\right)\biggr]-8 {q_3}^2 \log (N) \left(40 {q_3}^2-57 {\omega_3}^2\right) \left(2
   {q_3}^2-3 {\omega_3}^2\right)^5\biggr)\Biggr\}.
\end{eqnarray}}
We will not quote the expressions for $a_3$ and $a_4$ because they are too cumbersome. Substituting the expressions for $a_{1,2,3,4}$ into $Z(u)$ and implementing the Dirichlet boundary condition: $Z_s(u=0)=0$, in the hydrodynamical limit, going up to ${\cal O}(\omega_3^4)$ one sees that one can write the Dirichlet boundary condition as a quartic: $a \omega_3^4 + b \omega_3^3 + c \omega_3^2 + f \omega_3 + g = 0$ where:
\begin{eqnarray}
\label{a+b+c+f+g}
& & a = -\frac{(17978967982080+432082299912192 i) {g_s}^2 M^2 {N_f} \log (N)}{N},\nonumber\\
& & b = -16384 {q_3}^2 \left(55717134336 \pi ^2-\frac{(8065585152-2189804544 i) {g_s}^2 M^2 {N_f} \log (N)}{N}\right),\nonumber\\
& & c = \frac{(6351753314304+163465918414848 i) {g_s}^2 M^2 {N_f} {q_3}^2 \log (N)}{N},\nonumber\\
& & f = 196608 {q_3}^4 \left(981467136 \pi ^2-\frac{(25958400-36690432 i) {g_s}^2 M^2 {N_f} \log (N)}{N}\right),\nonumber\\
& & g = -\frac{(842551787520+22613002813440 i) {g_s}^2 M^2 {N_f} {q_3}^4 \log (N)}{N}.
\end{eqnarray}

\section{Frobenius Solution of  EOM of Gauge-Invariant $Z_s(u)$  for Scalar Modes of Metric Fluctuations for $ (\alpha^\prime=1)\ r: \log r\sim\log N$}
\setcounter{equation}{0} \seceqdd

Constructing a $Z_s(u)$ which is invariant under diffeomorphisms: $h_{\mu\nu}\rightarrow h_{\mu\nu} - \nabla_{(\mu}\xi_{\nu)}$, one sees one obtain the following equation of motion for $Z_s(u)$:
\begin{eqnarray}
\label{Z-EOM}
& & Z_s^{\prime\prime}(u) = \Biggl[\frac{q_3^2 \left(7 u^8-8 u^4+9\right)-3 \left(u^4+3\right) \omega_3^2}{u \left(u^4-1\right) \left(q_3^2 \left(u^4-3\right)+3
   \omega_3^2\right)}\nonumber\\
   & &   -\frac{1}{64 \pi ^2 N u \left(u^8-4 u^4+3\right)
   \left(q_3^2 \left(u^4-3\right)+3 \omega_3^2\right)^2}\Biggl\{-3 {g_s}^2 M^2 {N_f} \log N\nonumber\\
   & &  \times \Biggl(q_3^4 \left(5 u^{16}-98 u^{12}+372 u^8-414 u^4+135\right)+2
   q_3^2 \left(32 u^{12}-183 u^8+306 u^4-135\right) \omega_3^2\nonumber\\
   & & +3 \left(u^8-66 u^4+45\right) \omega_3^4\Biggr)\Biggr\}\Biggr]Z_s^\prime(u)\nonumber\\
& & + \Biggl[\frac{1}{128 \pi ^2 N
   q_3^2 \left(u^4-3\right) \left(u^4-1\right)^3 \left(q_3^2 \left(u^4-3\right)+3 \omega_3^2\right)^2}\Biggl\{-3 {g_s}^2 M^2 {N_f} \log N\nonumber\\
   & & \times \Biggl(30 q_3^6 u^{22}-542 q_3^6 u^{18}-7 q_3^6 u^{16} \omega_3^2+2540
   q_3^6 u^{14}+46 q_3^6 u^{12} \omega_3^2-4764 q_3^6 u^{10}-84 q_3^6 u^8 \omega_3^2\nonumber\\
   & & +4086 q_3^6 u^6+18 q_3^6 u^4
   \omega_3^2-1350 q_3^6 u^2+27 q_3^6 \omega_3^2+318 q_3^4 u^{18} \omega_3^2-2464 q_3^4 u^{14} \omega_3^2\nonumber\\
   & & -49 q_3^4 u^{12}
   \omega_3^4+6972 q_3^4 u^{10} \omega_3^2+189 q_3^4 u^8 \omega_3^4-8496 q_3^4 u^6 \omega_3^2-99 q_3^4 u^4 \omega_3^4+3510
   q_3^4 u^2 \omega_3^2-81 q_3^4 \omega_3^4\nonumber\\
   & & +114 q_3^2 u^{14} \omega_3^4-2262 q_3^2 u^{10} \omega_3^4-105 q_3^2 u^8
   \omega_3^6+5598 q_3^2 u^6 \omega_3^4+144 q_3^2 u^4 \omega_3^6-2970 q_3^2 u^2 \omega_3^4+81 q_3^2 \omega_3^6\nonumber\\
   & & -8 \left(u^8-4
   u^4+3\right) \left(q_3^2 \left(u^4-1\right)+\omega_3^2\right) \left(q_3^3 \left(u^4-3\right)+3 q_3 \omega_3^2\right)^2 \log
   \left(\frac{{r_h}}{u}\right)\nonumber\\
   & & +18 u^{10} \omega_3^6-1188 u^6 \omega_3^6-63 u^4 \omega_3^8+810 u^2 \omega_3^6-27 \omega_3^8\Biggr)\Biggr\}\nonumber\\
   & & -\frac{q_3^4 \left(u^8-4 u^4+3\right)+2
   q_3^2 \left(8 u^{10}-8 u^6+2 u^4 \omega_3^2-3 \omega_3^2\right)+3 \omega_3^4}{\left(u^4-1\right)^2 \left(q_3^2 \left(u^4-3\right)+3
   \omega_3^2\right)}\Biggr]Z_s(u).
\end{eqnarray}
The horizon $u=1$ due to inclusion of the non-conformal corrections to the metric, ceases to be an irregular singular point. One then tries the ansatz: $Z_s(u) = e^{S(u)}$ near $u=1$. Assuming that $\left(S^{\prime}\right)^2\gg S^{\prime\prime}(u)$ near $u=1$ the differential equation (\ref{Z-EOM}), which could written as $Z_s^{\prime\prime}(u) = m(u)Z_s^\prime + l(u) Z_s(u)$
can be approximated by:
\begin{equation}
\label{S_EOM}
\left(S^\prime\right)^2 - m(u) S^\prime(u) - l(u) \approx 0.
\end{equation}
A solution to (\ref{S_EOM}) is:
\begin{eqnarray}
\label{S-solution}
& & S(u) = \frac{1}{2}\left(m(u) - \sqrt{m^2(u) + 4 l(u)}\right)\nonumber\\
& & = -\frac{\sqrt{\frac{15}{2}} \sqrt{\frac{{g_s}^2 M^2 {N_f} {\omega_3}^2 \left({\omega_3}^2+4\right) \log
   \left(\frac{1}{N}\right)}{N {q_3}^2}}}{64 \pi  (u-1)^{3/2}} + \frac{\frac{15 {g_s}^2 M^2 {N_f} {\omega_3}^2 \log \left(\frac{1}{N}\right)}{256 \pi ^2 N \left(2 {q_3}^2-3
   {\omega_3}^2\right)}-\frac{1}{2}}{u-1}\nonumber\\
   & & +\frac{3 {g_s}^2 M^2 {N_f} \log \left(\frac{1}{N}\right) \left(112 {q_3}^4+214
   {q_3}^2 {\omega_3}^2-369 {\omega_3}^4\right)+128 \pi ^2 N \left(-20 {q_3}^4+12 {q_3}^2 {\omega_3}^2+27 {\omega_3}^4\right)}{512 \pi ^2 N \left(2
   {q_3}^2-3 {\omega_3}^2\right)^2}\nonumber\\
   & & +\frac{1}{1024 \sqrt{30} \pi ^3
   N^2 {q_3}^2 \sqrt{u-1} \left(2 {q_3}^2-3 {\omega_3}^2\right)^2 \sqrt{\frac{{g_s}^2 M^2 {N_f} {\omega_3}^2 \left({\omega_3}^2+4\right) \log
   \left(\frac{1}{N}\right)}{N {q_3}^2}}}\nonumber\\
   & & \times\Biggl\{-225 {g_s}^4 M^4 {N_f}^2 {q_3}^2 {\omega_3}^4 \log ^2\left(\frac{1}{N}\right)+12 \pi ^2 {g_s}^2 M^2
   N {N_f} \log \left(\frac{1}{N}\right) \left(2 {q_3}^2-3 {\omega_3}^2\right)\nonumber\\
   & & \times \left(-80 {q_3}^4 \left({\omega_3}^2-4\right)+2 {q_3}^2 {\omega_3}^2
   \left(57 {\omega_3}^2-412\right)+64 \left(2 {q_3}^4 {\omega_3}^2-3 {q_3}^2 {\omega_3}^4\right) \log ({r_h})+9 {\omega_3}^4
   \left({\omega_3}^2+164\right)\right)\nonumber\\
   & & +4096 \pi ^4 N^2 {q_3}^2 \left({\omega_3}^2-4\right) \left(2 {q_3}^2-3 {\omega_3}^2\right)^2\Biggr\}  + {\cal O}\left(\sqrt{u-1}\right).
\end{eqnarray}
Taking first the MQGP limit, the first term in the RHS of (\ref{S-solution}) can be dropped. After integrating with respect to $u$, the solution (\ref{S-solution}) to equation (\ref{S_EOM}) will reflect the singular nature of $Z(u)$'s equation of motion (\ref{S-solution}) via
\begin{equation}
\label{pole-soln-Z}
Z_s(u)\sim \left(1 - u \right)^{-\frac{1}{2} + \frac{15 {g_s}^2 M^2 {N_f} {\omega_3}^2 \log \left(\frac{1}{N}\right)}{256 \pi ^2 N \left(2 {q_3}^2-3{\omega_3}^2\right)}}F(u),
\end{equation}
 where $F(u)$ is regular in $u$ and its equation of motion, around $u=0$, is given by:
\begin{eqnarray}
\label{F-EOM}
& & 256 F''(u)+\frac{F'(u) \left(\frac{60 {g_s}^2 M^2 {N_f} \log \left(\frac{1}{N}\right)}{\pi ^2 N}-768\right)}{u}\nonumber\\
& & -\frac{3 F(u) \left(64 \pi ^2 N-5
   {g_s}^2 M^2 {N_f} \log \left(\frac{1}{N}\right)\right) \left(15 {g_s}^2 M^2 {N_f} {\omega_3}^2 \log (N)+128 \pi ^2 N \left(2 {q_3}^2-3
   {\omega_3}^2\right)\right)}{64 \pi ^4 N^2 u \left(2 {q_3}^2-3 {\omega_3}^2\right)} = 0.\nonumber\\
   & &
\end{eqnarray}
The solution to (\ref{F-EOM}) is given by:
{\footnotesize
\begin{eqnarray}
\label{F-eom-solution}
& & F(u) = 2^{-\frac{105 {g_s}^2 M^2 {N_f} \log (N)}{64 \pi ^2 N}-28} 3^{\frac{15 {g_s}^2 M^2 {N_f} \log (N)}{128 \pi ^2 N}+2} N^{-\frac{15 {g_s}^2 M^2
   {N_f} \log (N)}{64 \pi ^2 N}-4} \pi^{-\frac{15 {g_s}^2 M^2 {N_f} \log (N)}{32 \pi ^2 N}-8}\nonumber\\
  & & \times  \left(5 {g_s}^2 M^2 {N_f} \log (N)+64 \pi ^2
   N\right)^{2-\frac{15 {g_s}^2 M^2 {N_f} \log (N)}{128 \pi ^2 N}} u^{\frac{15 {g_s}^2 M^2 {N_f} \log (N)}{128 \pi ^2 N}+2} \left(2 {q_3}^2-3
   {\omega_3}^2\right)^{-\frac{15 {g_s}^2 M^2 {N_f} \log (N)}{128 \pi ^2 N}-2}\nonumber\\
   & & \times \left(15 {g_s}^2 M^2 {N_f} {\omega_3}^2 \log (N)+128 \pi ^2 N \left(2
   {q_3}^2-3 {\omega_3}^2\right)\right)^{2-\frac{15 {g_s}^2 M^2 {N_f} \log (N)}{128 \pi ^2 N}}\nonumber\\
 & & \times   \Biggl(c_1 \left(5 {g_s}^2 M^2 {N_f} \log (N)+64
   \pi ^2 N\right)^{\frac{15 {g_s}^2 M^2 {N_f} \log (N)}{64 \pi ^2 N}} \Gamma \left(-\frac{15 {g_s}^2 {N_f} \log (N) M^2}{64 N \pi ^2}-3\right)\nonumber\\
& & \times   \left(15 {g_s}^2 M^2 {N_f} {\omega_3}^2 \log (N)+128 \pi ^2 N \left(2 {q_3}^2-3 {\omega_3}^2\right)\right)^{\frac{15 {g_s}^2 M^2 {N_f} \log
   (N)}{64 \pi ^2 N}}\nonumber\\
   & & \times I_{-\frac{15 {g_s}^2 {N_f} \log (N) M^2}{64 N \pi ^2}-4}\left(\frac{\sqrt{3} \sqrt{u} \sqrt{\left(5 {g_s}^2 {N_f} \log (N)
   M^2+64 N \pi ^2\right) \left(15 {g_s}^2 M^2 {N_f} \log (N) {\omega_3}^2+128 N \pi ^2 \left(2 {q_3}^2-3 {\omega_3}^2\right)\right)}}{64 \pi ^2
   \sqrt{N^2 \left(2 {q_3}^2-3 {\omega_3}^2\right)}}\right)\nonumber\\
   & & +c_2 N^{\frac{15 {g_s}^2 M^2 {N_f} (2 \log (N)+i \pi )}{64 \pi ^2 N}} \Gamma \left(\frac{15
   {g_s}^2 {N_f} \log (N) M^2}{64 N \pi ^2}+5\right) \left(2 {q_3}^2-3 {\omega_3}^2\right)^{\frac{15 {g_s}^2 M^2 {N_f} \log (N)}{64 \pi ^2 N}}\nonumber\\
& & \times   \left(\left(5 {g_s}^2 M^2 {N_f} \log (N)+64 \pi ^2 N\right) \left(15 {g_s}^2 M^2 {N_f} {\omega_3}^2 \log (N)+128 \pi ^2 N \left(2 {q_3}^2-3
   {\omega_3}^2\right)\right)\right)^{\frac{15 {g_s}^2 M^2 {N_f} \log (N)}{64 \pi ^2 N}}\nonumber\\
   & & \times \left(N^2 \left(2 {q_3}^2-3
   {\omega_3}^2\right)\right)^{-\frac{15 {g_s}^2 M^2 {N_f} \log (N)}{64 \pi ^2 N}}\nonumber\\
   & & \times I_{\frac{15 {g_s}^2 {N_f} \log (N) M^2}{64 N \pi
   ^2}+4}\left(\frac{\sqrt{3} \sqrt{u} \sqrt{\left(5 {g_s}^2 {N_f} \log (N) M^2+64 N \pi ^2\right) \left(15 {g_s}^2 M^2 {N_f} \log (N)
   {\omega_3}^2+128 N \pi ^2 \left(2 {q_3}^2-3 {\omega_3}^2\right)\right)}}{64 \pi ^2 \sqrt{N^2 \left(2 {q_3}^2-3 {\omega_3}^2\right)}}\right)\Biggr).\nonumber\\
   & &
\end{eqnarray}}
One notes from (\ref{F-eom-solution}) that $F(u\sim0) = c_1$. This needs to be improved upon by including the sub-leading terms in $u$ in $F'(u)$ in (\ref{F-EOM}), implying that we should look at:
\begin{eqnarray}
\label{improved_F_EOM}
& & \hskip -1.2in 256 F''(u)\nonumber\\
& & \hskip -1.2in+F'(u) \left(\frac{120 {g_s}^2 M^2 {N_f} \log N\left(2 {q_3}^2-3 {\omega_3}^2\right)+30 {g_s}^2 M^2
   {N_f} {\omega_3}^2 \log (N)+1792 \pi ^2 N \left(2 {q_3}^2-3 {\omega_3}^2\right)}{\pi ^2 N \left(2 {q_3}^2-3
   {\omega_3}^2\right)}+\frac{\frac{-60 {g_s}^2 M^2 {N_f} \log N}{\pi ^2 N}-768}{u}\right)\nonumber\\
   & &\hskip -1.2in -\frac{3 F(u) \left(64 \pi ^2 N+5
   {g_s}^2 M^2 {N_f} \log N\right) \left(15 {g_s}^2 M^2 {N_f} {\omega_3}^2 \log (N)+128 \pi ^2 N \left(2
   {q_3}^2-3 {\omega_3}^2\right)\right)}{64 \pi ^4 N^2 u \left(2 {q_3}^2-3 {\omega_3}^2\right)} = 0.
   \end{eqnarray}
   The solution to (\ref{improved_F_EOM}) near $u=0$, is given as under:
   \begin{eqnarray}
   \label{solution-improved-F-EOM}
   & & N^{\frac{15 {g_s}^2 M^2 {N_f} u \left(11 {\omega_3}^2-8 {q_3}^2\right)}{128 \pi ^2 N \left(2 {q_3}^2-3 {\omega_3}^2\right)}}
   \nonumber\\
   & & \times\Biggl[u^4
   \Biggl(\frac{c_1 \Gamma \left(-\frac{15 {g_s}^2 {N_f} \log (N) M^2}{64 N \pi ^2}-4\right)}{\Gamma \left(-\frac{3 \left(5 {g_s}^2 {N_f}
   \log (N) M^2+64 N \pi ^2\right) \left(15 {g_s}^2 {N_f} \left(16 {q_3}^2-23 {\omega_3}^2\right) \log (N) M^2+1664 N \pi ^2 \left(2
   {q_3}^2-3 {\omega_3}^2\right)\right)}{128 N \pi ^2 \left(15 {g_s}^2 {N_f} \left(8 {q_3}^2-11 {\omega_3}^2\right) \log (N) M^2+896 N \pi
   ^2 \left(2 {q_3}^2-3 {\omega_3}^2\right)\right)}\right)}\nonumber\\
   & & +c_2 L_{-\frac{225 {g_s}^4 M^4 {N_f}^2 {\omega_3}^2 \log ^2(N)+4800 \pi ^2
   {g_s}^2 M^2 N {N_f} \log (N) \left(4 {q_3}^2-5 {\omega_3}^2\right)+139264 \pi ^4 N^2 \left(2 {q_3}^2-3 {\omega_3}^2\right)}{128 \pi ^2 N
   \left(15 {g_s}^2 M^2 {N_f} \log (N) \left(8 {q_3}^2-11 {\omega_3}^2\right)+896 \pi ^2 N \left(2 {q_3}^2-3
   {\omega_3}^2\right)\right)}}^{\frac{15 {g_s}^2 M^2 {N_f} \log (N)}{64 \pi ^2 N}+4}(0)\Biggr)\nonumber\\
   & & +\frac{1}{\Gamma \left(\frac{225 {g_s}^4 {N_f}^2 {\omega_3}^2 \log ^2(N) M^4+4800 {g_s}^2 N
   {N_f} \pi ^2 \left(4 {q_3}^2-5 {\omega_3}^2\right) \log (N) M^2+139264 N^2 \pi ^4 \left(2 {q_3}^2-3 {\omega_3}^2\right)}{128 N \pi ^2
   \left(15 {g_s}^2 {N_f} \left(8 {q_3}^2-11 {\omega_3}^2\right) \log (N) M^2+896 N \pi ^2 \left(2 {q_3}^2-3
   {\omega_3}^2\right)\right)}\right)}\nonumber\\
   & & \times\Biggl\{c_1 2^{\frac{105 {g_s}^2 M^2 {N_f}
   \log (N)}{64 \pi ^2 N}+28} \pi ^{\frac{15 {g_s}^2 M^2 {N_f} \log (N)}{32 \pi ^2 N}+8} u^{-\frac{15 {g_s}^2 M^2 {N_f} \log (N)}{64 \pi
   ^2 N}} \Gamma \left(\frac{15 {g_s}^2 {N_f} \log (N) M^2}{64 N \pi ^2}+4\right)\nonumber\\
   & &  \left(\frac{15 {g_s}^2 M^2 {N_f} \log (N) \left(8
   {q_3}^2-11 {\omega_3}^2\right)+896 \pi ^2 N \left(2 {q_3}^2-3 {\omega_3}^2\right)}{N \left(2 {q_3}^2-3 {\omega_3}^2\right)}\right)^{-\frac{15
   {g_s}^2 M^2 {N_f} \log (N)}{64 \pi ^2 N}-4}\Biggr\}\Biggr].\nonumber\\
   & &
   \end{eqnarray}

\section{Gauge Transformations Preserving $h_{m\ \mu}=0$,  Pole Structure of $\Omega(\omega_3,q_3)$ and Solutions to ${\cal H}_{ab}(u)$}
\setcounter{equation}{0} \seceqee

\subsection{Gauge Transformations Preserving $h_{m\ \mu}=0$}

There are three gauge transformations that preserve $h_{\mu u}=0$, for the black $M3$-brane metric having integrated out the $M_6$ in the (asymptotic) $AdS_5\times M_6$ in the MQGP limit. They are given below:

{\bf Set I}: The Gauge transformations are generated by
\begin{eqnarray}
&& \xi_{{x}}= \frac{C_{{x}}(t,{x})}{u^2}+\xi^{(1)}_{{x}}(u,t,{x})\nonumber\\
&& \xi_{t}=\xi^{(1)}_{t}(u,t,{x})
\end{eqnarray}
The Gauge Solutions for the above kind of  transformations are given as:
\begin{eqnarray}
\label{GTI}
&& H^{\rm Gauge(I)}_{tt}=\frac{g^{\frac{2}{3}}_{s}}{i L^2}\left[\frac{2 \omega_3 u^2}{g_1}\tilde{\xi}^{(1)}_{t}-q_3 \tilde{C}_{{x}} H_{t t}-\frac{ 2 \omega_3 \tilde{C}_{{x}}}{g_1} H_{t {x}}\right]\nonumber\\
&& H^{\rm Gauge(I)}_{{x} {x}}=\frac{g^{\frac{2}{3}}_{s}}{i L^2}\left[ q_3 \tilde{C}_{{x}} H_{{x} {x}}-2 q_3 u^2 \tilde{\xi}^{(1)}_{{x}}- { 2 q_3 \tilde{C}_{{x}}} \right]\nonumber\\
&&H^{\rm Gauge(I)}_{t {x}}=\frac{g^{\frac{2}{3}}_{s}}{i L^2}\left[\omega_3 \tilde{C}_{{x}}+ u^2 \omega_3 \tilde{\xi}^{(1)}_{x} - u^2 q_3 \tilde{\xi}^{(1)}_{t}-\omega_3 \tilde{C}_{{x}}H_{{x} {x}}\right]\nonumber\\
&&  H^{\rm gauge (I)}_{a a}= \frac{g^{\frac{2}{3}}_{s}}{i L^2}\left[ -q_3 \tilde{C}_{{x}} H_{aa}\right]
\end{eqnarray}
where $H_{aa}=H_{{y} {y}}+H_{{z} {z}}, \tilde{C}_{{x}}\equiv\frac{C_{{x}}}{\pi T}, \tilde{\xi}^{(1)}_t\equiv\frac{\xi^{(1)}_t}{\pi T}$.

{\bf Set II}: The Gauge transformations are generated by
\begin{eqnarray}
&& \xi_{t}=- \frac{ g_1 C_t(t,{x})}{u^2}+\xi^{(1)}_{t}(u,t,{x})\nonumber\\
&& \xi_{{x}}=\xi^{(1)}_{{x}}(u,t,{x})
\end{eqnarray}
The Gauge Solutions for the above kind of  transformations are given as:
\begin{eqnarray}
\label{GTII}
&& H^{\rm Gauge(II)}_{tt}=\frac{g^{\frac{2}{3}}_{s}}{i L^2}\left[-2 \omega_3 \tilde{C}_{t}+\frac{2 \omega_3 u^2}{g_1}\tilde{\xi}^{(1)}_{t}-\omega_3 \tilde{C}_{t} H_{tt} \right]\nonumber\\
&& H^{\rm Gauge(II)}_{{x} {x}}=\frac{g^{\frac{2}{3}}_{s}}{i L^2}\left[ -2 q_3 u^2 \tilde{\xi}^{(1)}_{{x}}+ 2 q_3 \tilde{C}_{t} H_{t {x}}+ \omega_3   \tilde{C}_t H_{{x} {x}} \right]\nonumber\\
&&H^{\rm Gauge(II)}_{t {x}}=\frac{g^{\frac{2}{3}}_{s}}{i L^2}\left[q_3 g_1 \tilde{C}_{t}- u^2 q_3 \tilde{\xi}^{(1)}_{t}+   u^2 \omega_3 \tilde{\xi}^{(1)}_{{x}}+ q_3 g_1 \tilde{C}_{t}H_{tt}\right]\nonumber\\
&&  H^{\rm gauge (II)}_{a a}= \frac{g^{\frac{2}{3}}_{s}}{i L^2}\left[ \omega_3 \tilde{C}_{t} H_{aa}\right]
\end{eqnarray}
where $H_{aa}=H_{{y} {y}}+H_{{z} {z}}, \tilde{\xi}^{(1)}_{{x}}\equiv\frac{\xi^{(1)}_{{x}}}{\pi T}$.

{\bf Set III}: Writing $\xi_u^{(0)}=\frac{C_u(t,{x})}{u\sqrt{g}}, \xi^{(0)}_t = - \partial_t C_u(t,{x})\psi(u), \xi^{(0)}_{{x}} = - \partial_{{x}}C_u(t,{x})\chi(u)$, and demanding the solutions to be well behave at $u=0$, one obtains:
\begin{eqnarray}
\label{GT-III}
& & \xi_u^{(0)} = \frac{C_u(t,{x})}{u\sqrt{g}};\nonumber\\
& & \xi^{(0)}_t = - \left(\frac{1}{2} - \frac{u^4}{3}\right)\sqrt{g} \partial_t C_u(t,{x});
\nonumber\\
& & \xi^{(0)}_{{x}} = - \partial_{{x}}C_u(t,{x}) \frac{F(\sin^{-1}u|1)}{u}  = - \partial_{{x}}C_u(t,{x})\left(1 + \frac{u^4}{10} + {\cal O}(u^8)\right).
\end{eqnarray}

This yields the following:
\begin{eqnarray}
\label{GTIII}
& & H^{\rm Gauge(III)}_{tt}=2\frac{g_s^{\frac{2}{3}}u^2\omega_3^2C_u(t,{x})}{L^2 \sqrt{1-u^4}}\left(\frac{1}{2} - \frac{u^4}{3}\right) - 2\frac{g_s^{\frac{2}{3}}}{L^2\sqrt{1-u^4}}C_u(t,{x})(1+u^4) \nonumber\\
& & + \frac{\omega_3q_3g_s^{\frac{2}{3}}}{L^2} H_{tt} C_u F\left(\sin^{-1}u|1\right)u - \omega_3^2g_s^{\frac{2}{3}}H_{tt}\frac{\left(\frac{1}{2} - \frac{u^4}{3}\right)}{\sqrt{1-u^4}} + \frac{g_s^{\frac{2}{3}}C_u\left(u H_{t{x}}' - 2 H_{t{x}}\right)}{2L^2\sqrt{1-u^4}};\nonumber\\
& & H^{\rm Gauge(III)}_{{x}t}=-\frac{\omega_3q_3g_s^{\frac{2}{3}}}{L^2}\sqrt{1-u^4}C_u u^2\left(\frac{1}{2} - \frac{7u^4}{30}\right);\nonumber\\
& & H^{\rm Gauge(III)}_{{x} {x}}=-2\frac{q_3^2g_s^{\frac{2}{3}}}{L^2}
C_u\left(1+\frac{u^4}{10}\right) + 2\frac{\sqrt{g}g_s^{\frac{2}{3}}}{L^2}C_u(1+u^4);
\nonumber\\
& & H^{\rm Gauge(III)}_{{y}{y}}=\frac{g_s^{\frac{2}{3}}\sqrt{1-u^4}}{L^2}\left(-H_{{x}{x}} + u H_{{x}{x}}'\right).
\end{eqnarray}

\subsection{Pole Structure of $\Omega(\omega_3,q_3)$}

The equation (\ref{pole-speed_s}) can be solved for $\omega_3$ and the solution is given by:
\begin{eqnarray}
\label{pole-speed_s_ii}
& & \omega_3 =  -\frac{2 \left({\alpha_{yy}^{(1,0)}}+\sqrt{{\alpha_{yy}^{(1,0)}}^2-{\alpha_{yy}^{(0,0)}} (4 {C_{1yy}^{(0,2)}}+4
   {C_{2yy}^{(0,2)}}+i {\Sigma_{2yy}^{(0,1)}})}\right)}{4 {C_{1yy}^{(0,2)}}+4 {C_{2yy}^{(0,2)}}+i {\Sigma_{2\ yy}^{(0,1)}}} \nonumber\\
   & & + \frac{{q_3} \left(\frac{{\alpha_{yy}^{(1,0)}} \left(72 {C_{1yy}^{(0,2)}}+8 e^3 {C_{1yy}^{(1,1)}}+72 {C_{2yy}^{(0,2)}}-36 {C_{2yy}^{(1,1)}}+18 i {\Sigma_{2\ yy}^{(0,1)}}+9
   i\right)}{\sqrt{{\alpha_{yy}^{(1,0)}}^2-{\alpha_{yy}^{(0,0)}} (4 {C_{1yy}^{(0,2)}}+4 {C_{2yy}^{02}}+i
   {\Sigma_{2\ yy}^{(0,1)}})}}+8 e^3 {C_{1yy}^{(1,1)}}-36 {C_{2yy}^{(1,1)}}+9 i\right)}{18
   (4 {C_{1yy}^{(0,2)}}+4 {C_{2yy}^{(0,2)}}+i {\Sigma_{2\ yy}^{(0,1)}})}
   \nonumber\\
   & & + {\cal O}(q_3^2);\nonumber\\
   & & -\frac{2 \left({\alpha_{yy}^{(1,0)}}-\sqrt{{\alpha_{yy}^{(1,0)}}^2-{\alpha_{yy}^{(0,0)}} (4 {C_{1yy}^{(0,2)}}+4
   {C_{2yy}^{(0,2)}}+i {\Sigma_{2yy}^{(0,1)}})}\right)}{4 {C_{1yy}^{(0,2)}}+4 {C_{2yy}^{(0,2)}}+i {\Sigma_{2\ yy}^{(0,1)}}} \nonumber\\
   & & + \frac{{q_3} \left(-\frac{{\alpha_{yy}^{(1,0)}} \left(72 {C_{1yy}^{(0,2)}}+8 e^3 {C_{1yy}^{(1,1)}}+72 {C_{2yy}^{(0,2)}}-36 {C_{2yy}^{(1,1)}}+18 i {\Sigma_{2\ yy}^{(0,1)}}+9
   i\right)}{\sqrt{{\alpha_{yy}^{(1,0)}}^2-{\alpha_{yy}^{(0,0)}} (4 {C_{1yy}^{(0,2)}}+4 {C_{2yy}^{02}}+i
   {\Sigma_{2\ yy}^{(0,1)}})}}+8 e^3 {C_{1yy}^{(1,1)}}-36 {C_{2yy}^{(1,1)}}+9 i\right)}{18
   (4 {C_{1yy}^{(0,2)}}+4 {C_{2yy}^{(0,2)}}+i {\Sigma_{2\ yy}^{(0,1)}})}
   \nonumber\\
   & & + {\cal O}(q_3^2)
   \end{eqnarray}
 Assuming $\alpha_{yy}^{(0,0)}\ll1, |\Sigma_{2\ yy}^{(0,1)}|\gg1(i \Sigma_{2\ yy}^{(0,1)}\in\mathbb{R}): \alpha_{yy}^{(0,0)}\Sigma_{2\ yy}^{(0,1)}<1; \alpha_{yy}^{(1,0)} = - |\alpha_{yy}^{(1,0)}|$, consistent with the constraints (\ref{constraints_I},\ref{costrainst_ii}),  (\ref{pole-speed_s_ii}) implies the following.

 \noindent\underline{Root 1}:
 {
 \begin{eqnarray}
 \label{root1-i}
 & & \hskip -0.6in \omega_3 = \left(-\frac{2 \left(\sqrt{{\alpha_{yy}^{(1,0)}}^2}+{\alpha_{yy}^{(1,0)}}\right)}{4 {C_{1yy}^{(0,2)}}+4 {C_{2yy}^{(0,2)}}+i {\Sigma_{2yy}^{(0,1)}}}-\frac{{\alpha_{yy}^{(0,0)}}
   \sqrt{{\alpha_{yy}^{(1,0)}}^2} (-4 {C_{1yy}^{(0,2)}}-4 {C_{2yy}^{(0,2)}}-i {\Sigma_{2yy}^{(0,1)}})}{{\alpha_{yy}^{(1,0)}}^2 (4 {C_{1yy}^{(0,2)}}+4 {C_{2yy}^{(0,2)}}+i
   {\Sigma_{2yy}^{(0,1)}})}+O\left({\alpha_{yy}^{(0,0)}}^2\right)\right)+\nonumber\\
   & & \hskip -1in q_3 \Biggl[\frac{\sqrt{{\alpha_{yy}^{(1,0)}}^2} \left(8 e^3 {C_{1yy}^{(1,1)}}-36
   {C_{2yy}^{(1,1)}}+9 i\right)+{\alpha_{yy}^{(1,0)}} \left(72 {C_{1yy}^{(0,2)}}+72 {C_{2yy}^{(0,2)}}-36 {C_{2yy}^{(1,1)}}+18 i {\Sigma_{2yy}^{(0,1)}}+8 {C_{1yy}^{(1,1)}} e^3+9
   i\right)}{18 \sqrt{{\alpha_{yy}^{(1,0)}}^2} (4 {C_{1yy}^{(0,2)}}+4 {C_{2yy}^{(0,2)}}+i {\Sigma_{2yy}^{(0,1)}})}+\nonumber\\
   & & \hskip -1in\frac{1}{18 (4 {C_{1yy}^{(0,2)}}+4 {C_{2yy}^{(0,2)}}+i
   {\Sigma_{2yy}^{(0,1)}})} \Biggl(\frac{\left(8 e^3 {C_{1yy}^{(1,1)}}-36
   {C_{2yy}^{(1,1)}}+9 i\right) (-4 {C_{1yy}^{(0,2)}}-4 {C_{2yy}^{(0,2)}}-i {\Sigma_{2yy}^{(0,1)}})}{2 {\alpha_{yy}^{(1,0)}}^2}\nonumber\\
   & & \hskip -1.2in -\frac{\left(\sqrt{{\alpha_{yy}^{(1,0)}}^2}
   \left(8 e^3 {C_{1yy}^{(1,1)}}-36 {C_{2yy}^{(1,1)}}+9 i\right)+{\alpha_{yy}^{(1,0)}} \left(72 {C_{1yy}^{(0,2)}}+72 {C_{2yy}^{(0,2)}}-36 {C_{2yy}^{(1,1)}}+18 i
   {\Sigma_{2yy}^{(0,1)}}+8 {C_{1yy}^{(1,1)}} e^3+9 i\right)\right)}{2 {\alpha_{yy}^{(1,0)}}^2
   \sqrt{{\alpha_{yy}^{(1,0)}}^2}}\Biggr) \nonumber\\
   & & \hskip -1in\times  (-4 {C_{1yy}^{(0,2)}}-4 {C_{2yy}^{(0,2)}}-i {\Sigma_{2yy}^{(0,1)}}){\alpha_{yy}^{(0,0)}} +O\left({\alpha_{yy}^{(0,0)}}^2\right)\Biggr]+O\left(q_3^2\right).
 \end{eqnarray}}
 The expansion (\ref{root1-i}) implies:
 {
\begin{eqnarray}
\label{root1-ii}
& &  \hskip -0.4in \omega_3= -\frac{{\alpha_{yy}^{(0,0)}}}{{\alpha_{yy}^{(1,0)}}} + \nonumber\\
& & \hskip -0.4in q_3\Biggl[ -\frac{72 {\alpha_{yy}^{(0,0)}} {C_{1yy}^{(0,2)}}+8 e^3 {\alpha_{yy}^{(0,0)}} {C_{1yy}^{(1,1)}}+72 {\alpha_{yy}^{(0,0)}} {C_{2yy}^{(0,2)}}-36 {\alpha_{yy}^{(0,0)}} {C_{2yy}^{(1,1)}}+18
   i {\alpha_{yy}^{(0,0)}} {\Sigma_{2yy}^{(0,1)}}+9 i {\alpha_{yy}^{(0,0)}}+36 {\alpha_{yy}^{(1,0)}}^2}{36 {\alpha_{yy}^{(1,0)}}^2}\Biggr]\nonumber\\
   & & \hskip -0.4in \approx - q_3\left(1 + i \frac{\alpha_{yy}^{(00)}\Sigma_{2\ yy}^{(0,1)}}{2\left(\alpha_{yy}^{(1,0)}\right)^2}\right).
\end{eqnarray} }

 \noindent\underline{Root 2}:
 {
 \begin{eqnarray}
 \label{root2-i}
  && \hskip -0.6in \omega_3 = \left(\frac{2 \left(\sqrt{{\alpha_{yy}^{(1,0)}}^2}-{\alpha_{yy}^{(1,0)}}\right)}{4 {C_{1yy}^{(0,2)}}+4 {C_{2yy}^{(0,2)}}+i {\Sigma_{2yy}^{(0,1)}}}-\frac{{\alpha_{yy}^{(0,0)}}
   \sqrt{{\alpha_{yy}^{(1,0)}}^2}}{{\alpha_{yy}^{(1,0)}}^2}+O\left({\alpha_{yy}^{(0,0)}}^2\right)\right)
   + \nonumber\\
& & \hskip -0.6in   q_3 \left(\frac{\sqrt{{\alpha_{yy}^{(1,0)}}^2} \left(8 e^3
   {C_{1yy}^{(1,1)}}-36 {C_{2yy}^{(1,1)}}+9 i\right)-{\alpha_{yy}^{(1,0)}} \left(72 {C_{1yy}^{(0,2)}}+72 {C_{2yy}^{(0,2)}}-36 {C_{2yy}^{(1,1)}}+18 i {\Sigma_{2yy}^{(0,1)}}+8
   {C_{1yy}^{(1,1)}} e^3+9 i\right)}{18 \sqrt{{\alpha_{yy}^{(1,0)}}^2} (4 {C_{1yy}^{(0,2)}}+4 {C_{2yy}^{(0,2)}}+i
   {\Sigma_{2yy}^{(0,1)}})}\right.\nonumber\\
   & & \hskip -0.6in \left.-\frac{\left(\sqrt{{\alpha_{yy}^{(1,0)}}^2} \left(72 {C_{1yy}^{(0,2)}}+72 {C_{2yy}^{(0,2)}}-36 {C_{2yy}^{(1,1)}}+18 i {\Sigma_{2yy}^{(0,1)}}+8
   {C_{1yy}^{(1,1)}} e^3+9 i\right)\right) {\alpha_{yy}^{(0,0)}}}{36 {\alpha_{yy}^{(1,0)}}^3}\right.\nonumber\\
   & &\hskip -0.6in \left.+O\left({\alpha_{yy}^{(0,0)}}^2\right)\right)+O\left(q_3^2\right).
 \end{eqnarray}}
 The expansion (\ref{root2-i}) implies:
 {
 \begin{eqnarray}
 \label{root2-ii}
 & &  \omega_3 =  \sqrt{{\alpha_{yy}^{(1,0)}}^2} \left(-\frac{{\alpha_{yy}^{(0,0)}}}{{\alpha_{yy}^{(1,0)}}^2}+\frac{4}{4 {C_{1yy}^{(0,2)}}+4 {C_{2yy}^{(0,2)}}+i {\Sigma_{2yy}^{(0,1)}}}\right)\nonumber\\
 & & + q_3\Biggl[\frac{ \left(36 {C_{1yy}^{(0,2)}}+8 e^3 {C_{1yy}^{(1,1)}}+36 {C_{2yy}^{(0,2)}}-36 {C_{2yy}^{(1,1)}}+9 i {\Sigma_{2yy}^{(0,1)}}+9 i\right)}{9 (4 {C_{1yy}^{(0,2)}}+4
   {C_{2yy}^{(0,2)}}+i {\Sigma_{2yy}^{(0,1)}})} \nonumber\\
   & &  + \frac{\alpha_{yy}^{(0,0)} \left(72 {C_{1yy}^{(0,2)}}+8 e^3 {C_{1yy}^{(1,1)}}+72 {C_{2yy}^{(0,2)}}-36 {C_{2yy}^{(1,1)}}+18 i {\Sigma_{2yy}^{(0,1)}}+9 i\right)}{36
   {\alpha_{yy}^{(1,0)}}^2} \Biggr]\nonumber\\
   & &  \approx q_3\left(1 + i \frac{\alpha_{yy}^{(00)}\Sigma_{2\ yy}^{(0,1)}}{2\left(\alpha_{yy}^{(1,0)}\right)^2}\right).
 \end{eqnarray}}

\subsection{Solutions to ${\cal H}_{ab}(u)$}

Making double perturbative ansatze:
${\cal H}_{ab}(u) = \sum_{m=0}^\infty\sum_{n=0}^\infty {\cal H}_{ab}^{(m,n)}(u)q_3^m\omega_3^n$, one obtains near $u=0$ the solutions to the scalar modes' EOMs (\ref{7scalar_EOMs}):
\begin{eqnarray*}
& & {\cal H}_{yy}^{(0,0)}(u) = \alpha_{yy}^{(0,0)} + \beta_{yy}^{(0,0)} u^4,\nonumber\\
& & {\cal H}_{yy}^{(1,0)}(u) = \alpha_{yy}^{(1,0)} + \beta_{yy}^{(1,0)} u^4,\nonumber\\
& & {\cal H}_{yy}^{(0,1)}(u) = \Sigma_{2\ yy}^{(0,1)} + \frac{i}{16} \alpha_{yy}^{(0,0)}u^2\nonumber\\
& & {\rm where:}\Sigma_{2\ yy}^{(0,1)}\equiv \frac{1}{8}\left(- 195 i - 60 \pi - 35\beta_{yy}^{(0,0)}(13 i + 4\pi) + 8 C_{2yy}^{(0,1)}\right),\nonumber\\
& & {\cal H}_{yy}^{(1,1)} = -\frac{i}{4} - \frac{2}{9} e^3 C_{1yy}^{(1,1)} + C_{2yy}^{(1,1)} +
i \frac{u}{4},\nonumber\\
& & {\cal H}_{yy}^{(2,0)}(u) = C_{1yy}^{(2,0)} + C_{2yy}^{(2,0)} u^4,\nonumber\\
& & {\cal H}_{yy}^{(0,2)}(u) = i \frac{\Sigma_{2\ yy}^{(0,1)}}{4} + C_{1yy}^{(0,2)} + C_{2yy}^{(0,2)} - \frac{i}{4} \Sigma_{2\ yy}^{(0,1)} u;\nonumber\\
& & {\cal H}_{xt}^{(0,0)}(u) = \alpha_{xt}{(0,0)},\nonumber\\
& & {\cal H}_{xt}^{(1,0)}(u) = \alpha_{xt}^{(1,0)} + \beta_{xt}^{(1,0)} u^4,\nonumber\\
& & {\cal H}_{xt}^{(0,1)}(u) = C_{2xt}^{(0,1)} + \frac{i}{4} \alpha_{xt}^{(0,0)} u,\nonumber\\
& & {\cal H}_{xt}^{(1,1)}(u) = C_{2xt}^{(1,1)} + \frac{1}{96} \left[12 i \left(\alpha_{xt}^{(1,1)} + 5 \beta_{xt}^{(1,1)}\right) u +
     6 \left(3 i \alpha_{xt}^{(1,1)} + 4 \alpha_{yy} + 5 i \beta_{xt}^{(1,1)} - 4 \beta_{yy}\right) u^2\right],\nonumber\\
& & {\cal H}_{xt}^{(2,0)}(u) = \alpha_{xt}^{(2,0)} + \beta_{xt}^{(2,0)} u^4,\nonumber\\
& & {\cal H}_{xt}^{(0,2)}(u) = \frac{i}{4} C_{2xt}^{(0,1)} u + \frac{1}{4} u^4 C_{1xt}^{(0,2)} + C_{2xt}^{(0,2)};\nonumber\\
& & {\cal H}_{tt}^{(0,0)}(u) = \left(\frac{4}{3} \beta_{yy}^{(0,0)} -
    i C_{1tt}^{(0,0)}\right) + \left(-4 \beta_{yy}^{(0,0)}/3 - \frac{i}{2} C_{1tt}^{(0,0)}\right) u^4 - \frac{3i}{8}C_{1tt}^{(0,0)}u^8,\nonumber\\
& & {\cal H}_{tt}^{(0,1)}(u) =
 \alpha_{tt}^{(0,1)} + \frac{i}{12} \left(6 \alpha_{yy}^{(0,0)} + 4 \beta_{yy}^{(0,0)} - 3 i C_{1tt}^{(0,0)}\right) u +
  \frac{i}{24} \left(3 \alpha + 4 \beta - 3 i C_{tt}^{(0,0)}\right) u^2,\nonumber\\
& & {\cal H}_{tt}^{(1,0)}(u) = \left(\frac{4}{3} \beta_{yy}^{(1,0)} -
    i C_{1tt}^{(1,0)}\right) + \left(-\frac{4}{3} \beta_{yy}^{(1,0)} - (i/2) C_{1yy}^{(1,0)}\right) u^4,\nonumber\\
    & & {\cal H}_{tt}^{(1,1)}(u) =
 C_{1tt}^{(1,1)} + 1/12 i \left(-6 + 6 \alpha_{yy}^{(1,0)} + 4 \beta_{yy}^{(1,0)} - 3 i C1tt^{(1,0)}\right) u +
  1/24 i \left(6 \alpha_{yy}^{(1,0)} + 4 \beta_{yy}^{(1,0)} - 3 i C_{1tt}^{(1,0)}\right) u^2,\nonumber\\
& & {\cal H}_{tt}^{(0,2)}(u) =
 \frac{i}{192} \left(12 \alpha_{tt}^{(0,1)} \pi + 12 \alpha_{xt}^{(1,0)} \pi +
     6 i \alpha_{yy}^{(0,0)} \pi + 4 i \beta_{yy}^{(0,0)} \pi  + 3 C_{1tt}^{(0,0)} \pi -
     24 \pi \Sigma_{2\ yy}^{(0,1)} - 192 C_{1tt}^{(0,2)}\right)\nonumber\\
     & &  -
  \frac{i}{8} \left(2 \alpha_{tt}^{(0,1)} + 2 \alpha_{xt}^{(1,0)} - \alpha_{xt}^{(1,1)} - 5 \beta_{xt}^{(1,1)}-
     12 \Sigma_{2\ yy}^{(0,1)}\right) u \nonumber\\
  & &
  -\frac{i}{96} \left(12 \alpha_{tt}^{(0,1)} + 12 \alpha_{xt}^{(1,0)} + 6 i \alpha_{yy}^{(0,0)} +
     4 i \beta_{yy}^{(0,0)} + 3 C_{1tt}^{(0,0)} - 24 \Sigma_{2\ yy}^{(0,1)}\right) u^2,\nonumber\\
& & {\cal H}_{tt}^{(2,0)}(u) = \left(\frac{4}{3} C_{2yy}^{(2,0)} -
    i C_{1tt}^{(2,0)}\right) + \left(-\frac{4}{3} C_{2yy}^{(2,0)} - 1/2 i C_{1tt}^{(2,0)}\right) u^4;\nonumber\\
    \end{eqnarray*}

 \begin{eqnarray}
 \label{hab}
    && {\cal H}_s^{(0,0)}(u) = \frac{C_{1s}^{(0,0)}}{2}u^2  + C_{2s}^{(0,0)},\nonumber\\
& & {\cal H}_s^{(1,0)}(u) = \frac{C_{1s}^{(1,0)}}{2} u^2 + C_{2s}^{(1,0)},\nonumber\\
& & {\cal H}_s^{(0,1)}(u) = -\frac{(2 + 2 i) C_{1s}^{(0,1)}}{\pi},\nonumber\\
& & {\cal H}_s^{(2,0)}(u) = \frac{C_{1s}^{(2,0)}}{2} u^2 + C_{2s}^{(2,0)},\nonumber\\
& & {\cal H}_s^{(0,2)}(u) = \Sigma_{s}^{(0,2)} u + \frac{C_{1s}^{(0,2)}}{2} u^2 + C_{2s}^{(0,2)}.
\end{eqnarray}
such that:
\begin{eqnarray}
\label{constraints_I}
& & 171 i + 2 i \alpha_{yy}^{(0,0)} + 319 i  \beta_{yy}^{(0,0)} + 24 C_{1yy}^{(0,0)} = 0;\nonumber\\
& & 3 \alpha_{yy}^{(0,0)} + 4 \beta_{yy}^{(0,0)} - 3 i C_{1tt}^{(0,0)} - 3 C_{2s}^{(0,0)} = 0
\end{eqnarray}
For consistency checks, we have ensured that (\ref{hab}) obtained from the fourth, fifth and the sixth equations of (\ref{7scalar_EOMs}), also solve the first, second, third and seventh equations near $u=0$ and up to ${\cal O}(q_3^m\omega_3^n):m+n=2$ by imposing suitable additional constraints on the constants appearing in (\ref{hab}).

\section{Frobenius Solution to EOM of Gauge-Invariant $Z_v(u)$ for Vector Modes of Metric Fluctuations}
\setcounter{equation}{0} \seceqff

The equations of motion for the vector perturbation modes up next-to-leading order in $N$, can be reduced to the following single equation of motion in terms of a gauge-invariant variable $Z_v(u)$:
\begin{equation}
\label{vector-modes-ZEOM}
Z_v^{\prime\prime}(u) - m(u) Z_v^\prime(u) - l(u) Z_v(u) = 0,
\end{equation}
where
\begin{eqnarray}
\label{m+l-vec-definitions}
& & m(u)\equiv\frac{15 {g_s}^2 M^2 {N_f} \left(u^4-1\right) \log (N) \left({q_3}^2 \left(u^4-1\right)+{\omega_3}^2\right)+64 \pi ^2 N \left(3 {q_3}^2
   \left(u^4-1\right)^2-\left(u^4+3\right) {\omega_3}^2\right)}{64 \pi ^2 N u \left(u^4-1\right) \left({q_3}^2 \left(u^4-1\right)+{\omega_3}^2\right)},\nonumber\\
   & & l(u)\equiv -\frac{\left({q_3}^2 \left(u^4-1\right)+{\omega_3}^2\right) \left(32 \pi ^2 N-3 {g_s}^2 M^2 {N_f} \log ^2(N)\right)}{32 \pi ^2 N \left(u^4-1\right)^2}.
\end{eqnarray}
The horizon $u=1$ is a regular singular point of (\ref{vector-modes-ZEOM}) and the root of the indicial equation corresponding to the incoming-wave solution is given by:
\begin{equation}
\label{root-solution-incoming-wave}
-\frac{i {\omega_3}}{4} + \frac{3 i {g_s}^2 M^2 {N_f} {\omega_3} \log ^2(N)}{256 \pi ^2 N}.
\end{equation}
(a) Using the Frobenius method, taking the solution about $u=1$ to be:
\begin{equation}
\label{solution1}
Z_v(u) = (1 - u)^{-\frac{i {\omega_3}}{4} + \frac{3 i {g_s}^2 M^2 {N_f} {\omega_3} \log ^2(N)}{256 \pi ^2 N}}\left(1 + \sum_{n=1}^\infty a_n (u - 1)^n\right),
\end{equation}
by truncating the infinite series in (\ref{solution1}) to ${\cal O}((u-1)^2)$ one obtains:
{\footnotesize
\begin{eqnarray}
\label{a1a2}
& & \hskip -0.8in a_1 = \frac{1}{512 \pi ^2 N {\omega_3} ({\omega_3}+2 i)^2}\Biggl\{3 i {g_s}^2 M^2 {N_f} \log (N)\left(20 (2-i {\omega_3}) {\omega_3}^2-\log (N) \left(3 {\omega_3}^2 \left({\omega_3}^2+4 i {\omega_3}+4\right)-4
   {q_3}^2 \left({\omega_3}^2+4 i {\omega_3}-8\right)\right)\right)\Biggr\}\nonumber\\
   & & \hskip -0.8in +\frac{4 {q_3}^2 (4-i {\omega_3})+3 (2+i
   {\omega_3}) {\omega_3}^2}{8 {\omega_3} ({\omega_3}+2 i)} + {\cal O}\left(\frac{1}{N^2}\right),
   \nonumber\\
& & \hskip -0.8in  a_2 = -\frac{1}{\Sigma}\Biggl\{4 \Biggl(405 i {g_s}^8 M^8 {N_f}^4 {\omega_3}^5 \log ^7(N)+8640 i \pi ^2 {g_s}^6 M^6 N {N_f}^3 {\omega_3}^2 \log ^5(N) \left(16 {q_3}^2
   ({\omega_3}+2 i)+(-13 {\omega_3}+4 i) {\omega_3}^2\right)\nonumber\\
   & &\hskip -0.8in +368640 \pi ^4 {g_s}^4 M^4 N^2 {N_f}^2 {\omega_3} \log ^3(N) \left(4 {q_3}^2 \left(-3 i
   {\omega_3}^2+12 {\omega_3}+16 i\right)+i {\omega_3}^2 \left(9 {\omega_3}^2+8 i {\omega_3}+16\right)\right)\nonumber\\
   & &\hskip -0.5in +7864320 \pi ^6 {g_s}^2 M^2 N^3 {N_f} {\omega_3} \log
   (N) \left(4 i {q_3}^2 \left({\omega_3}^2+6 i {\omega_3}-16\right)+{\omega_3}^2 \left(-3 i {\omega_3}^2+4 {\omega_3}-16 i\right)\right)
   \nonumber\\
   & &\hskip -0.8in -49152 \pi ^4 {g_s}^2
   M^2 N^2 {N_f} {\omega_3} \log ^2(N) \Biggl(75 {g_s}^2 M^2 {N_f} ({\omega_3}+4 i) {\omega_3}^2+8 \pi ^2 N \Biggl[32 {q_3}^4 ({\omega_3}+6 i)-48
   {q_3}^2 \left({\omega_3}^3+12 {\omega_3}+16 i\right)\nonumber\\
   & &\hskip -0.8in +{\omega_3}^2 \left(18 {\omega_3}^3-111 i {\omega_3}^2+200 {\omega_3}+16 i\right)\Biggr]\Biggr)\nonumber\\
   & & \hskip -0.8in+108 {g_s}^6
   M^6 {N_f}^3 {\omega_3} \log ^6(N) \left(-75 {g_s}^2 M^2 {N_f} {\omega_3}^3+2 \pi ^2 N \left(128 i {q_3}^4-24 {q_3}^2 {\omega_3}^2 ({\omega_3}+4
   i)+{\omega_3}^4 (19 {\omega_3}+22 i)\right)\right)\nonumber\\
   & &\hskip -0.8in +2304 \pi ^2 {g_s}^4 M^4 N {N_f}^2 \log ^4(N)\nonumber\\
   & &\hskip -0.8in \times \left(150 {g_s}^2 M^2 {N_f} ({\omega_3}+2 i)
   {\omega_3}^3+\pi ^2 N \left(256 {q_3}^4 \left({\omega_3}^2+3 i {\omega_3}+4\right)-24 {q_3}^2 {\omega_3}^2 \left(15 {\omega_3}^2-8 i
   {\omega_3}+96\right)+{\omega_3}^4 \left(125 {\omega_3}^2-636 i {\omega_3}+208\right)\right)\right)\nonumber\\
   & &\hskip -0.8in +4194304 \pi ^8 N^4 {\omega_3} \left(16 {q_3}^4 ({\omega_3}+8
   i)-24 {q_3}^2 \left({\omega_3}^3+24 {\omega_3}+64 i\right)+{\omega_3}^2 \left(9 {\omega_3}^3-74 i {\omega_3}^2+200 {\omega_3}+32
   i\right)\right)\Biggr)\Biggr\} + {\cal O}\left(\frac{1}{N^2}\right),
\end{eqnarray}
where:
\begin{eqnarray}
& & \Sigma \equiv {\omega_3}^2 \Biggl[-81 {g_s}^8 M^8 {N_f}^4 {\omega_3}^4 \log ^8(N)+41472 i \pi ^2 {g_s}^6 M^6 N {N_f}^3 {\omega_3}^3 \log
   ^6(N)+294912 \pi ^4 {g_s}^4 M^4 N^2 {N_f}^2 (26-3 i {\omega_3}) {\omega_3}^2 \log ^4(N)\nonumber\\
   & & -201326592 \pi ^6 {g_s}^2 M^2 N^3 {N_f} {\omega_3}
   ({\omega_3}+3 i) \log ^2(N)+2147483648 \pi ^8 N^4 \left({\omega_3}^2+6 i {\omega_3}-8\right)\Biggr].
   \end{eqnarray}}

The Dirichlet boundary condition $Z_v(u=0)=0$ in the hydrodynamical limit retaining therefore terms only up to ${\cal O}(\omega_3^mq_3^n):\ m+n=4$, reduces to:
$a \omega_3^4 + b \omega_3^3 + c \omega_3^2 + f \omega_3 + g = 0$ where:
\begin{eqnarray}
\label{a b c d f g-i}
& & a = 3\left(96\pi^2 + \frac{13 g_s^2 M^2 N_f (\log N)^2}{N}\right),\nonumber\\
& & b = 2 i \left(1664\pi^2 + 39 g_s^2 M^2 N_f \frac{(\log N)^2}{N}\right),\nonumber\\
& & c = 128 \pi^2\left(-70 + 3 q_3^2\right) + 78 g_s^2 M^2 N_f\left(-2 + q_3^2\right)\frac{(\log N)^2}{N},\nonumber\\
& & f = 8 i\left(64\pi^2(-16 + 7 q_3^2) - 6 g_s^2 M^2 N_f \frac{(\log N)^2}{N}q_3^2\right),\nonumber\\
& & g = 16 q_3^2\left(64\pi^2(-4 + q_3^2) + 3 g_s^2 M^2 N_f(4 - 3 q_3^2)\frac{(\log N)^2}{N}\right),
\end{eqnarray}
One of the four roots of  $Z(u=0)=0$ is:
\begin{equation}
\label{root-at-second-order}
\omega_3 = -8.18 i + \frac{0.14 i g_s^2 M^2 N_f(\log N)^2}{N} + \left(-0.005 i - \frac{0.002 i g_s^2 M^2 N_f (\log N)^2}{N}\right)q_3^2 + {\cal O}(q_3^3).
\end{equation}

(b) Using the Frobenius method and going up to ${\cal O}((u-1)^3)$ in (\ref{solution1}), one obtains:
{\footnotesize
\begin{eqnarray}
\label{NLOa3}
& & \hskip -0.6in a_3 = \frac{1}{65536 \pi ^2 N {\omega_3}^2 ({\omega_3}+2 i)^2 ({\omega_3}+4 i)^2 ({\omega_3}+6 i)^2}\Biggl\{i {g_s}^2 M^2 {N_f} \log (N) \Biggl(20 i {\omega_3} \left({\omega_3}^3+12 i {\omega_3}^2-44 {\omega_3}-48 i\right)\nonumber\\
& & \hskip -0.6in \times \left(48 {q_3}^4
   \left({\omega_3}^2+12 i {\omega_3}-48\right)-8 {q_3}^2 \left(9 {\omega_3}^4+48 i {\omega_3}^3+60 {\omega_3}^2+1472 i {\omega_3}-3840\right)+{\omega_3}^2 \left(27
   {\omega_3}^4-42 i {\omega_3}^3+1288 {\omega_3}^2+2464 i {\omega_3}-2048\right)\right)\nonumber\\
   & & \hskip -0.6in -\log (N) \Biggl[64 {q_3}^6 {\omega_3} \left(3 {\omega_3}^4+72 i
   {\omega_3}^3-652 {\omega_3}^2-2400 i {\omega_3}+2880\right)\nonumber\\
   & & \hskip -0.6in -48 {q_3}^4 \left(9 {\omega_3}^7+156 i {\omega_3}^6-668 {\omega_3}^5+3072 i {\omega_3}^4-37024
   {\omega_3}^3-124416 i {\omega_3}^2+160768 {\omega_3}+49152 i\right)\nonumber\\
   & & \hskip -0.6in +4 {q_3}^2 {\omega_3} \left(81 {\omega_3}^8+852 i {\omega_3}^7+4324 {\omega_3}^6+85824 i
   {\omega_3}^5-444320 {\omega_3}^4-1143552 i {\omega_3}^3+1270784 {\omega_3}^2-454656 i {\omega_3}+1769472\right)\nonumber\\
   & & \hskip -0.6in -{\omega_3}^3 \left(81 {\omega_3}^8+288 i
   {\omega_3}^7+13136 {\omega_3}^6+103296 i {\omega_3}^5-183440 {\omega_3}^4+289152 i {\omega_3}^3-925696 {\omega_3}^2-436224 i
   {\omega_3}+221184\right)\Biggr]\Biggr)\Biggr\}\nonumber\\
   & & \hskip -0.6in +\frac{1}{3072 {\omega_3}^2 \left({\omega_3}^3+12 i {\omega_3}^2-44 {\omega_3}-48 i\right)}\Biggl\{64 i {q_3}^6 {\omega_3}
   ({\omega_3}+12 i)\nonumber\\
   & &  \hskip -0.6in +48 {q_3}^4 \left(-3 i {\omega_3}^4+6 {\omega_3}^3-208 i {\omega_3}^2+960 {\omega_3}+512 i\right)+4 {q_3}^2 {\omega_3} \left(27 i
   {\omega_3}^5+222 {\omega_3}^4+2272 i {\omega_3}^3-7200 {\omega_3}^2+4736 i {\omega_3}-36864\right)\nonumber\\
   & & \hskip -0.6in +{\omega_3}^3 \left(-27 i {\omega_3}^5-504 {\omega_3}^4-932 i
   {\omega_3}^3-5424 {\omega_3}^2-4544 i {\omega_3}+4608\right)\Biggr\} + {\cal O}\left(\frac{1}{N^2}\right).
\end{eqnarray}}
The Dirichlet condition $Z_v(u=0)=0$ reduces to $a \omega_3^4 + b \omega_3^3 + c \omega_3^2 + f \omega_3 + g = 0$
where
\begin{eqnarray}
\label{a b c d f g-ii}
& & a = -\frac{957 {g_s}^2 M^2 {N_f} \log ^2(N)}{N}-63264 \pi ^2,\nonumber\\
& & b = -48 i \left(\frac{27 {g_s}^2 M^2 {N_f} \log ^2(N)}{N}+2240 \pi ^2\right),\nonumber\\
& & c = 8 \left(\frac{15 {g_s}^2 M^2 {N_f} {q_3}^2 \log ^2(N)}{N}+32 \pi ^2 \left(127 {q_3}^2+288\right)\right)\nonumber\\
& & f = 576 i {q_3}^2 \left(64 \pi ^2-\frac{3 {g_s}^2 M^2 {N_f} \log ^2(N)}{N}\right),\nonumber\\
& & g = 384 {q_3}^4 \left(32 \pi ^2-\frac{3 {g_s}^2 M^2 {N_f} \log ^2(N)}{N}\right).
\end{eqnarray}
One of the four roots of the quartic in $\omega_3$ is:
\begin{equation}
\label{root-at-third-order}
\omega_3 =  \left(- 0.73 i + \frac{0.003 i g_s^2 M^2 N_f (\log N)^2}{N}\right)q_3^2 + {\cal O}(q_3^3).
\end{equation}
The leading order coefficient of $q_3^2$ is not terribly far off the correct value $-\frac{i}{4}$ already at the third order in the infinite series (\ref{solution1}).

(c) Let us look at (\ref{solution1}) up to the fourth order. One finds:
{\footnotesize
\begin{eqnarray}
\label{a_4}
& & a_4 = \frac{1}{98304 {\omega_3}^4 \left({\omega_3}^4+20 i {\omega_3}^3-140 {\omega_3}^2-400 i
   {\omega_3}+384\right)}\nonumber\\
   & & \times\Biggl\{256 {q_3}^8 {\omega_3}^3 ({\omega_3}+16 i)-768 {q_3}^6 \left({\omega_3}^6+4 i {\omega_3}^5+136 {\omega_3}^4+832 i {\omega_3}^3+256 {\omega_3}^2+7168 i
   {\omega_3}-12288\right)\nonumber\\
   & &  +32 {q_3}^4 {\omega_3}^2 \left(27 {\omega_3}^6-222 i {\omega_3}^5+4880 {\omega_3}^4+18176 i {\omega_3}^3+110464 {\omega_3}^2+652288 i
   {\omega_3}-675840\right)\nonumber\\
   & & -16 {q_3}^2 {\omega_3}^3 \left(27 {\omega_3}^7-558 i {\omega_3}^6+3320 {\omega_3}^5-9232 i {\omega_3}^4+198656 {\omega_3}^3+888320 i
   {\omega_3}^2-774144 {\omega_3}+589824 i\right)\nonumber\\
   & & +3 {\omega_3}^5 \left(27 {\omega_3}^7-900 i {\omega_3}^6-1316 {\omega_3}^5-53104 i {\omega_3}^4+108800 {\omega_3}^3+147200
   i {\omega_3}^2-487424 {\omega_3}-344064 i\right)\Biggr\}\nonumber\\
   & & -\frac{1}{524288 \pi ^2 N {\omega_3}^4
   \left({\omega_3}^2+6 i {\omega_3}-8\right)^2 \left({\omega_3}^2+14 i {\omega_3}-48\right)^2}\nonumber\\
   & & \times\Biggl\{-{g_s}^2 M^2 {N_f} \log N \biggl(-\log N \Biggl[256 {q_3}^8 {\omega_3}^3
   \left({\omega_3}^5+37 i {\omega_3}^4-530 {\omega_3}^3-3500 i {\omega_3}^2+10368 {\omega_3}+10752 i\right)\nonumber\\
   & & -768 {q_3}^6 \Biggl({\omega_3}^{10}+28 i {\omega_3}^9-222
   {\omega_3}^8+848 i {\omega_3}^7-24192 {\omega_3}^6-153184 i {\omega_3}^5+399360 {\omega_3}^4\nonumber\\
   & & +133120 i {\omega_3}^3+1531904 {\omega_3}^2+3293184 i
   {\omega_3}-2359296\Biggr)\nonumber\\
   & & +16 {q_3}^4 {\omega_3}^2 \Biggl(54 {\omega_3}^{10}+1017 i {\omega_3}^9+2420 {\omega_3}^8+195388 i {\omega_3}^7-1954848 {\omega_3}^6-8216832
   i {\omega_3}^5+5373440 {\omega_3}^4\nonumber\\
   & & -87731200 i {\omega_3}^3+345751552 {\omega_3}^2+510885888 i {\omega_3}-259522560\Biggr)\nonumber\\
   & & -8 {q_3}^2 {\omega_3}^3 \Biggl(54
   {\omega_3}^{11}+513 i {\omega_3}^{10}+14300 {\omega_3}^9+252484 i {\omega_3}^8-1373088 {\omega_3}^7-588832 i {\omega_3}^6\nonumber\\
   & & -30598656 {\omega_3}^5-183382016 i
   {\omega_3}^4+519692288 {\omega_3}^3+707788800 i {\omega_3}^2-297271296 {\omega_3}+113246208 i\Biggr)\nonumber\\
   & & +3 {\omega_3}^5 \Biggl(27 {\omega_3}^{11}+11672 {\omega_3}^9+105584 i
   {\omega_3}^8+196016 {\omega_3}^7+6136320 i {\omega_3}^6-29371904 {\omega_3}^5-60586752 i {\omega_3}^4\nonumber\\
   & & +67778560 {\omega_3}^3+79093760 i {\omega_3}^2-93585408
   {\omega_3}-33030144 i\Biggr)\Biggr]+20 i \left({\omega_3}^4+20 i {\omega_3}^3-140 {\omega_3}^2-400 i {\omega_3}+384\right)\nonumber\\
   & & \times {\omega_3}^2 \biggl[64 {q_3}^6 {\omega_3}
   \left({\omega_3}^2+18 i {\omega_3}-96\right)-16 {q_3}^4 \left(9 {\omega_3}^5+84 i {\omega_3}^4+192 {\omega_3}^3+6496 i {\omega_3}^2-23296 {\omega_3}-9216 i\right)\nonumber\\
   & & +4
   {q_3}^2 {\omega_3} \left(27 {\omega_3}^6+12 i {\omega_3}^5+2756 {\omega_3}^4+22208 i {\omega_3}^3-71680 {\omega_3}^2+27136 i {\omega_3}-270336\right)\nonumber\\
   & & +{\omega_3}^4
   \left(-27 {\omega_3}^5+234 i {\omega_3}^4-3704 {\omega_3}^3-4224 i {\omega_3}^2+1408 {\omega_3}+52224 i\right)\biggr]\biggr)\Biggr\}
   + {\cal O}\left(\frac{1}{N^2}\right).
\end{eqnarray}}
In the hydrodynamical limit the Dirichlet boundary condition $Z_v(u=0)=0$ reduces to $a \omega_3^4 + b \omega_3^3 + c \omega_3^2 + f \omega_3 + g = 0$
where
\begin{eqnarray}
\label{a b c d f g-iii}
& & a = 9849372385059274752 i \pi ^2 + {\cal O}\left(q_3^2\right),\nonumber\\
& & b = \frac{19237055439568896 {q_3}^2 \left(3 {g_s}^2 {\log N} (2 {\log N}+5) M^2 {N_f}-128 \pi ^2 N\right)}{N},\nonumber\\
& & c = {\cal O}\left(q_3^4\right),\nonumber\\
& & f = {\cal O}\left(q_3^6\right),\nonumber\\
& & f = {\cal O}\left(q_3^6\right).
\end{eqnarray}

\section{$Z_t(u)$  from Tensor Mode of Metric Fluctuations}
\setcounter{equation}{0} \seceqgg

The EOM for the tensor metric perturbation mode $Z_t(u)$, inclusive of the non-conformal corrections in the metric
(\ref{Mtheory met})  was written out in equation (\ref{EOM-tensor}). Realizing that $u=1$ is a regular singular point of (\ref{EOM-tensor}), using the Frobenius method we made a double perturbative ansatz (\ref{ansatz-solution-tensor}) for the analytic part of the solution. Substituting (\ref{ansatz-solution-tensor}) into (\ref{EOM-tensor}), setting the coefficient of $\omega_3$ to zero one gets:
\begin{eqnarray}
\label{w3}
& & {z_{00}}(u) \left(-6 {g_s}^2 M^2 {N_f} \log (N) \log \left(2 \pi ^{3/2} \sqrt{{g_s}} T\right)-3 {g_s}^2 M^2 {N_f} \log ^2(N)+64 \pi ^2
   N\right)\nonumber\\
    & & \times\left(64 \pi ^2 N \left(u^2+2 u+3\right)+15 {g_s}^2 M^2 {N_f} \left(u^3+u^2+u+1\right) \log \left({N}\right)\right)\nonumber\\
    & & -128 i \pi ^2 N
   \Biggl[2 \left(z_{01}'(u) \left(-15 {g_s}^2 M^2 {N_f} \left(u^4-1\right) \log \left({N}\right)+64 \pi ^2 N \left(u^4+3\right)\right)+64 \pi ^2
   N u \left(u^4-1\right) z_{01}''(u)\right)\nonumber\\
   & & -i u \left(u^3+u^2+u+1\right) {z_{00}}'(u) \left(-6 {g_s}^2 M^2 {N_f} \log (N) \log \left(2 \pi ^{3/2}
   \sqrt{{g_s}} T\right)-3 {g_s}^2 M^2 {N_f} \log ^2(N)+64 \pi ^2 N\right)\Biggr].\nonumber\\
   & &
   \end{eqnarray}
By setting the coefficient of $q_3$ to zero:
   \begin{eqnarray}
   \label{q3}
   {z_{10}}'(u) \left(-15 {g_s}^2 M^2 {N_f} \left(u^4-1\right) \log \left({N}\right)+64 \pi ^2 N \left(u^4+3\right)\right)+64 \pi ^2 N u
   \left(u^4-1\right) {z_{10}}''(u) = 0,\nonumber\\
   & &
\end{eqnarray}
which solves to yield:
\begin{eqnarray}
\label{h10solution}
& & z_{10}(u) = c_2-\frac{1}{\left(64 \pi ^2 N+15 {g_s}^2 M^2 {N_f} \log \left({N}\right)\right) \left(128 \pi ^2 N+15
   {g_s}^2 M^2 {N_f} \log \left({N}\right)\right)}\nonumber\\
   & & \times\Biggl\{16 \pi ^2 c_1 N u^{1+\frac{15 {g_s}^2 M^2 {N_f} \log \left({N}\right)}{64 \pi ^2 N}} \Biggl(2 u \left(64 \pi ^2 N+15 {g_s}^2 M^2
   {N_f} \log \left({N}\right)\right)\nonumber\\
    & & \times \, _2F_1\left(1,1+\frac{15 {g_s}^2 M^2 {N_f} \log \left({N}\right)}{128 N \pi ^2};2+\frac{15
   {g_s}^2 M^2 {N_f} \log \left({N}\right)}{128 N \pi ^2};-u^2\right)+\left(128 \pi ^2 N+15 {g_s}^2 M^2 {N_f} \log
   \left({N}\right)\right)\nonumber\\
    & & \times\, _2F_1\left(1,1+\frac{15 {g_s}^2 M^2 {N_f} \log \left({N}\right)}{64 N \pi ^2};2+\frac{15 {g_s}^2 M^2
   {N_f} \log \left({N}\right)}{64 N \pi ^2};-u\right)\nonumber\\
   & & -\left(15 {g_s}^2 M^2 {N_f} \log \left({N}\right)-128 \pi ^2 N\right) \,
   _2F_1\left(1,1+\frac{15 {g_s}^2 M^2 {N_f} \log \left({N}\right)}{64 N \pi ^2};2+\frac{15 {g_s}^2 M^2 {N_f} \log
   \left({N}\right)}{64 N \pi ^2};u\right)\Biggr)\Biggr\}\nonumber\\
   & &  = u^{\frac{15 {g_s}^2 M^2 {N_f} \log \left({N}\right)}{64 \pi ^2 N}} \Bigg(\frac{64 N \pi ^2 c_1 u^4}{256 N \pi ^2+15 {g_s}^2 M^2 {N_f} \log
   \left({N}\right)}\nonumber\\
   & & +\frac{64 N \pi ^2 c_1 u^8}{512 N \pi ^2+15 {g_s}^2 M^2 {N_f} \log \left({N}\right)}+\frac{64 N \pi ^2 c_1 u^{12}}{768 N
   \pi ^2+15 {g_s}^2 M^2 {N_f} \log \left({N}\right)}+O\left(u^{13}\right)\Biggr)+c_2\nonumber\\
\end{eqnarray}
Setting $c_1=0$ for convenience, one obtains:
\begin{eqnarray}
\label{c10i}
& & z_{01}(u)= c_4+\frac{1}{3072}\Biggl\{u \Biggl[\frac{6 i c_2 {g_s}^2 M^2 {N_f} \left(3 u^3+4 u^2+6 u+12\right) \log (N) \log \left(2 \pi ^{3/2} \sqrt{{g_s}} T\right)}{\pi ^2
   N}\nonumber\\
   & & +\frac{3 i c_2 {g_s}^2 M^2 {N_f} \left(3 u^3+4 u^2+6 u+12\right) \log ^2(N)}{\pi ^2 N}+64 \Biggl(\frac{48 c_3 u^{3+\frac{15 {g_s}^2 M^2 {N_f}
   \log \left({N}\right)}{64 \pi ^2 N}}}{4+\frac{15 {g_s}^2 M^2 {N_f} \log \left({N}\right)}{64 \pi ^2 N}}\nonumber\\
   & & -i c_2 \left(3 u^3+4 u^2+6
   u+12\right)\Biggr)\Biggr]\Biggr\}\nonumber\\
   & & = - \left(\frac{c_5 u^4}{4}+c_3\right)\frac{15 \left(c_5 {g_s}^2 M^2 {N_f} u^4 \log \left({N}\right) (4 \log (u)-1)\right)}{1024 \pi ^2
   N}+ {\cal O}\left(\frac{1}{N^2}\right)\nonumber\\
   & & = u^3 \left(\frac{i c_2 {g_s}^2 M^2 {N_f} \log \left({N}\right) \left(\log \left({N}\right)+2 \log \left(2 \pi ^{3/2} \sqrt{{g_s}}
   T\right)\right)}{256 \pi ^2 N}-\frac{i c_2}{12}\right)\nonumber\\
   & & +u^2 \left(\frac{3 i c_2 {g_s}^2 M^2 {N_f} \log \left({N}\right) \left(\log
   \left({N}\right)+2 \log \left(2 \pi ^{3/2} \sqrt{{g_s}} T\right)\right)}{512 \pi ^2 N}-\frac{i c_2}{8}\right)\nonumber\\
   & & +u \left(\frac{3 i c_2 {g_s}^2 M^2
   {N_f} \log \left({N}\right) \left(\log \left({N}\right)+2 \log \left(2 \pi ^{3/2} \sqrt{{g_s}} T\right)\right)}{256 \pi ^2 N}-\frac{i
   c_2}{4}\right)+c_4 + {\cal O}\left(\frac{u^4}{N},\frac{1}{N^2}\right).\nonumber\\
   & &
\end{eqnarray}
Similarly,
\begin{eqnarray}
\label{phi10}
& & z_{10}(u) = \frac{15 c_5 {g_s}^2 M^2 {N_f} u^4 \log \left({N}\right) (4 \log (u)-1)}{1024 \pi ^2 N}+\frac{c_5 u^4}{4}+c_3.
\end{eqnarray}
The constant (in $\omega_3, q_3$) yields:
\begin{eqnarray}
\label{constant}
& & \frac{{z_{00}}'(u) \left(-15 {g_s}^2 M^2 {N_f} \left(u^4-1\right) \log \left({N}\right)+64 \pi ^2 N \left(u^4+3\right)\right)}{64 \pi ^2 N u
   \left(u^4-1\right)}+{z_{00}}''(u) = 0,
\end{eqnarray}
which is identical in form to the EOM of $z_{10}(u)$.
Setting $q_3=0$ in (\ref{Phi}), one  obtains:
\begin{eqnarray}
\label{Phi'overPhi}
& &\hskip -0.6in \frac{Z_t^\prime(u)}{Z_t(u)} = \left(\frac{i c_4 {\omega_3}^2}{4 \left(c_4 {\omega_3}+c_2\right)}+\frac{3 i c_4 {g_s}^2 M^2 {N_f} {\omega_3}^2 \log \left({N}\right) \left(\log
   \left({N}\right)+2 \log \left(2 \pi ^{3/2} \sqrt{{g_s}} T\right)\right)}{256 \pi ^2 N \left(c_4
   {\omega_3}+c_2\right)}+{\cal O}\left(\frac{1}{N^2}\right)\right)\nonumber\\
   & & \hskip -0.6in +u \Biggl(\frac{{\omega_3}^2 \left(c_2^2+4 i c_4 c_2+4 i {\omega_3} c_4^2\right)}{16
   \left(c_2+{\omega_3} c_4\right){}^2}\nonumber\\
   & & \hskip -0.6in -\frac{3 i {g_s}^2 M^2 {N_f} {\omega_3}^2 \left(-i c_2^2+2 c_4 c_2+2 {\omega_3} c_4^2\right) \log
   \left({N}\right) \left(\log \left({N}\right)+2 \log \left(2 \sqrt{{g_s}} \pi ^{3/2} T\right)\right)}{512 \pi ^2 \left(c_2+{\omega_3}
   c_4\right){}^2 N}\nonumber\\
   & &\hskip -0.6in  + {\cal O}\left(\frac{1}{N^2}\right)\Biggr)+u^2 \Biggl\{\frac{{\omega_3}^2 \left((i {\omega_3}+6) c_2^3+2 (3 {\omega_3}+8 i) c_4 c_2^2+32 i
   {\omega_3} c_4^2 c_2+16 i {\omega_3}^2 c_4^3\right)}{64 \left(c_2+{\omega_3} c_4\right){}^3}\nonumber\\
   & &\hskip -0.6in  -\frac{3 i {g_s}^2 M^2 {N_f} {\omega_3}^2 \left(3 ({\omega_3}-4 i)
   c_2^3+4 (4-3 i {\omega_3}) c_4 c_2^2+32 {\omega_3} c_4^2 c_2+16 {\omega_3}^2 c_4^3\right) \log \left({N}\right) \left(\log \left({N}\right)+2 \log
   \left(2 \sqrt{{g_s}} \pi ^{3/2} T\right)\right)}{4096 \pi ^2 \left(c_2+{\omega_3} c_4\right){}^3 N}\nonumber\\
   & &\hskip -0.6in  +{\cal O}\left(\frac{1}{N^2}\right)\Biggr\}+u^3\nonumber\\
   & & \hskip -0.6in \times
   \Biggl(\frac{1}{768 \left(c_2+{\omega_3} c_4\right){}^4}\Biggl\{{\omega_3} \Biggl[\left(-3 {\omega_3}^3+24 i {\omega_3}^2+88 {\omega_3}+192 i\right) c_2^4+8 {\omega_3} \left(3 i {\omega_3}^2+22 {\omega_3}+96 i\right) c_4
   c_2^3\nonumber\\
   & & \hskip -0.6in+8 {\omega_3}^2 (11 {\omega_3}+144 i) c_4^2 c_2^2+768 i {\omega_3}^3 c_4^3 c_2+192 i {\omega_3}^4 c_4^4\Biggr]\Biggr\}\nonumber\\
   & & \hskip -0.6in -\frac{1}{4096 \pi ^2 \left(c_2+{\omega_3} c_4\right){}^4
   N}\Biggl\{i
   {g_s}^2 M^2 {N_f} {\omega_3}\Biggl[\left(3 i {\omega_3}^3+18 {\omega_3}^2-44 i {\omega_3}+48\right) c_2^4+2 {\omega_3} \left(9 {\omega_3}^2-44 i
   {\omega_3}+96\right) c_4 c_2^3\nonumber\\
   & & \hskip -0.6in+4 (72-11 i {\omega_3}) {\omega_3}^2 c_4^2 c_2^2+192 {\omega_3}^3 c_4^3 c_2+48 {\omega_3}^4 c_4^4\Biggr] \log \left({N}\right)
   \left(\log \left({N}\right)+2 \log \left(2 \sqrt{{g_s}} \pi ^{3/2} T\right)\right)\Biggr\}\nonumber\\
   & & \hskip -0.6in +{\cal O}\left(\frac{1}{N^2}\right)\Biggr)+{\cal O}\left(u^4\right).\nonumber\\
   & &
   \end{eqnarray}

\section{$T$ in the $D=11$ Uplift involving a Deformed/Resolved Conifold}
\setcounter{equation}{0} \seceqhh

In this appendix we show that it is only when the resolution is larger than the deformation that the temperature from the local $D=11$ uplift of the type IIB background of \cite{metrics} goes like the horizon radius; when the deformation is larger than the resolution the temperature goes like the reciprocal of the horizon radius.

Using the exact expression for the deformed conifold metric component $g_{rr}$ \cite{metric-deformed} for appropriately redefined radial coordinate, and the type IIA dilaton $\phi^{\rm IIA}$, one obtains in the MQGP limit the following expression for the $D=11$ metric component $G^{\cal M}_{rr}$:
\begin{eqnarray}
\label{GMrr_def}
& & G^{\cal M}_{rr} = e^{-\frac{2\phi^{\rm IIA}}{3}} g_{rr}\sqrt{h} \nonumber\\
& & = \frac{3 r^8 \sqrt[3]{1-\frac{9 \varepsilon^4}{r^6}} \left(-3 {N_f} \log \left(9 a^2 r^4+r^6\right)+\frac{24 \pi }{{g_s}}+6
   {N_f} \log (N)\right)^{2/3} }{4 2^{5/6} \pi ^{7/6} \left(r^4-{r_h}^4\right) \left(\varepsilon^4 \left(\sinh \left(2 \cosh
   ^{-1}\left(\frac{r^3}{\varepsilon^2}\right)\right)-2 \cosh ^{-1}\left(\frac{r^3}{\varepsilon^2}\right)\right)\right)^{2/3}}\nonumber\\
   & & \times \sqrt{\frac{{g_s} \left(18 {g_s}^2 M^2 {N_f} \log ^2(r)+3 {g_s} M^2 \log
   (r) \left({g_s} {N_f} \log \left(\frac{1}{4 \sqrt{N}}\right)+3 {g_s} {N_f}+4 \pi \right)+8 \pi ^2
   N\right)}{r^4}},\nonumber\\
   & &
\end{eqnarray}
where $\varepsilon$ is the deformation parameter ($h_5\equiv\frac{\varepsilon^2}{r^3}$ in (\ref{RWDC})).
Using (\ref{GMrr_def}) one obtains the following expression for the temperature:
\begin{eqnarray}
\label{T-DC}
&& \hskip -0.6in T = \frac{\partial_rG^{\cal M}_{00}}{4\pi\sqrt{G^{\cal M}_{00}G^{\cal M}_{rr}}}\nonumber\\
& & \hskip -0.6in = \frac{2^{2/3} \sqrt[3]{\varepsilon^4 \left(\sinh \left(2 \cosh ^{-1}\left(\frac{{r_h}^3}{\varepsilon^2}\right)\right)-2 \cosh
   ^{-1}\left(\frac{{r_h}^3}{\varepsilon^2}\right)\right)}}{\sqrt{3 \pi } {r_h} \sqrt{{g_s} \left(18 {g_s}^2 M^2
   {N_f} \log ^2({r_h})+3 {g_s} M^2 \log ({r_h}) \left({g_s} {N_f} \log \left(\frac{1}{4
   \sqrt{N}}\right)+3 {g_s} {N_f}+4 \pi \right)+8 \pi ^2 N\right)}}\nonumber\\
   & & \hskip -0.6in = \frac{\sqrt[3]{\varepsilon^4 \left(\sinh \left(2 \cosh ^{-1}\left(\frac{{r_h}^3}{\varepsilon^2}\right)\right)-2 \cosh
   ^{-1}\left(\frac{{r_h}^3}{\varepsilon^2}\right)\right)}}{2^{5/6} \sqrt{3} \pi ^{3/2} \sqrt{{g_s}} \sqrt{N}
   {r_h}}\nonumber\\
   & & \hskip -0.6in -\frac{\sqrt{3} \sqrt{{g_s}} M^2 \log ({r_h}) \sqrt[3]{\varepsilon^4 \left(\sinh \left(2 \cosh
   ^{-1}\left(\frac{{r_h}^3}{\varepsilon^2}\right)\right)-2 \cosh ^{-1}\left(\frac{{r_h}^3}{\varepsilon^2}\right)\right)} }{32 N^{3/2} \left(2^{5/6} \pi ^{7/2}
   {r_h}\right)}\nonumber\\
   & & \hskip -0.6in \times \left({g_s}
   {N_f} \log \left(\frac{1}{N}\right)+12 {g_s} {N_f} \log ({r_h})+6 {g_s} {N_f}-2 {g_s}
   {N_f} \log (4)+8 \pi \right) + {\cal O}\left(\left(\frac{1}{N}\right)^{5/2}\right)\nonumber\\
   & & \hskip -0.6in = \frac{1}{6 \pi
   ^{3/2} \sqrt{{g_s}} \sqrt{N} {r_h}}\Biggl\{\sqrt[6]{2} \sqrt{3} \sqrt[3]{-\varepsilon^4 \left(\sinh \left(\frac{\varepsilon^4}{2 {r_h}^6}+4 \log (d)-2 \log \left(2
   {r_h}^3\right)\right)-4 \log (d)+2 \log \left(2 {r_h}^3\right)\right)}\nonumber\\
   & & \hskip -0.6in-\frac{3 \sqrt{3} \varepsilon^{4/3} {g_s} M^2
   \log ({r_h}) \sqrt[3]{\frac{\varepsilon^4}{{r_h}^6}-2 \sinh \left(\frac{\varepsilon^4}{2 {r_h}^6}+4 \log (d)-2 \log \left(2
   {r_h}^3\right)\right)+8 \log (d)-4 \log \left(2 {r_h}^3\right)} }{32 \sqrt[6]{2} \pi ^2 N}\Biggr\}\nonumber\\
   & & \hskip -0.6in \times \left(-{g_s} {N_f} \log (N)+12 {g_s}
   {N_f} \log ({r_h})+6 {g_s} {N_f}-2 {g_s} {N_f} \log (4)+8 \pi \right).
\end{eqnarray}

When the resolution is larger than the deformation, using the exact expression for the $g_{rr}$ component of the resolved conifold \cite{metric-resolved} with appropriate redefinition of the radial coordinate, the $D=11$ component
$G^{\cal M}_{rr}$ is given as under:
\begin{eqnarray}
\label{GMrr_res}
& & \hskip -0.6in G^{\cal M}_{rr} = \frac{r^4 \left(6 a^2+r^2\right) \sqrt[3]{1-\frac{9 \varepsilon^4}{r^6}} \left(-3 {N_f} \log \left(9 a^2 r^4+r^6\right)+\frac{24
   \pi }{{g_s}}+6 {N_f} \log (N)\right)^{2/3} }{4 \sqrt{2} \pi ^{7/6} \left(9 a^2+r^2\right) \left(r^4-{r_h}^4\right)}\nonumber\\
   & & \hskip -0.6in \times \sqrt{\frac{{g_s} \left(18 {g_s}^2 M^2 {N_f} \log ^2(r)+3
   {g_s} M^2 \log (r) \left({g_s} {N_f} \log \left(\frac{1}{4 \sqrt{N}}\right)+3 {g_s} {N_f}+4 \pi
   \right)+8 \pi ^2 N\right)}{r^4}}.
\end{eqnarray}
Using (\ref{GMrr_res}), one obtains the following expression for the temperature:
\begin{eqnarray}
\label{T-RC}
& & T = \frac{\partial_rG^{\cal M}_{00}}{4\pi\sqrt{G^{\cal M}_{00}G^{\cal M}_{rr}}}\nonumber\\
& & = {r_h} \left[\frac{1}{2 \pi ^{3/2} \sqrt{{g_s} N}}-\frac{3 {g_s}^{\frac{3}{2}} M^2 {N_f} \log ({r_h}) \left(-\log
   {N}+12 \log ({r_h})+\frac{8 \pi}{g_sN_f} +6-\log (16)\right)}{64 \pi ^{7/2} N^{3/2}} \right]\nonumber\\
   & & + a^2 \left(\frac{3}{4 \pi ^{3/2} \sqrt{{g_s}} \sqrt{N} {r_h}}-\frac{9 {g_s}^{3/2} M^2 {N_f} \log ({r_h})
   \left(\frac{8 \pi }{{g_s} {N_f}}-\log (N)+12 \log ({r_h})+6-2 \log (4)\right)}{128 \pi ^{7/2} N^{3/2}
   {r_h}}\right).\nonumber\\
   & &
\end{eqnarray}

\end{document}